\documentclass[final,10pt]{elsarticle}

\usepackage{amssymb}
\usepackage{subfigure}
\usepackage{amsmath}
\usepackage[margin=0.8in]{geometry}
\usepackage[monochrome]{xcolor}
\journal{Journal of Computational Physics}
\begin{document}

\begin{frontmatter}



\title{{\color{green}On energy consistency of intermediate states in HLL-type MHD Riemann solvers}}

 \author[label1,label2,label3]{Fan Zhang} 
  \author[label1]{Andrea Lani}
   \author[label1,label4]{Stefaan Poedts}
  
  \affiliation[label1]{organization={Centre for mathematical Plasma-Astrophysics, Department of Mathematics, KU Leuven},
             addressline={Celestijnenlaan 200B},
             city={Leuven},
             postcode={3001},
             country={Belgium}}

\affiliation[label2]{organization={Institute of Theoretical Astrophysics, University of Oslo},
             addressline={PO Box 1029 Blindern},
             city={Oslo},
             postcode={0315},
             country={Norway}}             
 \affiliation[label3]{organization={Rosseland Centre for Solar Physics, University of Oslo},
             addressline={PO Box 1029 Blindern},
             city={Oslo},
             postcode={0315},
             country={Norway}}

\affiliation[label4]{organization={Institute of Physics, University of Maria Curie-Sk{\l}odowska},
             addressline={Pl. M. Curie-Sk{\l}odowskiej 5},
             city={Lublin},
             postcode={20-031},
             country={Poland}}

%% Abstract
\begin{abstract}
Approximate Riemann solvers are widely used for solving hyperbolic conservation laws, including those of magnetohydrodynamics (MHD). However, due to the nonlinearity and complexity of MHD, obtaining accurate and robust numerical solutions to MHD equations is non-trivial, and it may be challenging for an approximate MHD Riemann solver to preserve the positivity of scalar variables, particularly when the plasma $\beta$ is low. As we have identified that the inconsistency between the numerically calculated magnetic field and magnetic energy may be at least partly responsible for the loss of positivity of scalar variables, 
 we propose {\color{green}a consistency condition for calculating the intermediate energies within the Riemann fan} and implement it in HLL-type MHD Riemann solvers, thereby alleviating erroneous magnetic field solutions that break scalar positivity. In addition, (I) for the HLLC-type scheme, we have designed a revised two-state approximation,  specifically reducing numerical error in magnetic field solutions, {\color{red}although sacrificing the contact-resolving capability,} and (II) for the HLLD-type scheme, we replace the constant total pressure assumption by a three-state assumption for the intermediate thermal energy, which is more consistent with our other assumptions. The proposed schemes perform better in numerical examples with low plasma $\beta$. Moreover, we explained the energy error introduced during time integration.
\end{abstract}

\setcounter{figure}{-1}

\begin{graphicalabstract}
\begin{figure}[htbp]
 \centering
 \includegraphics[width=0.6\textwidth]{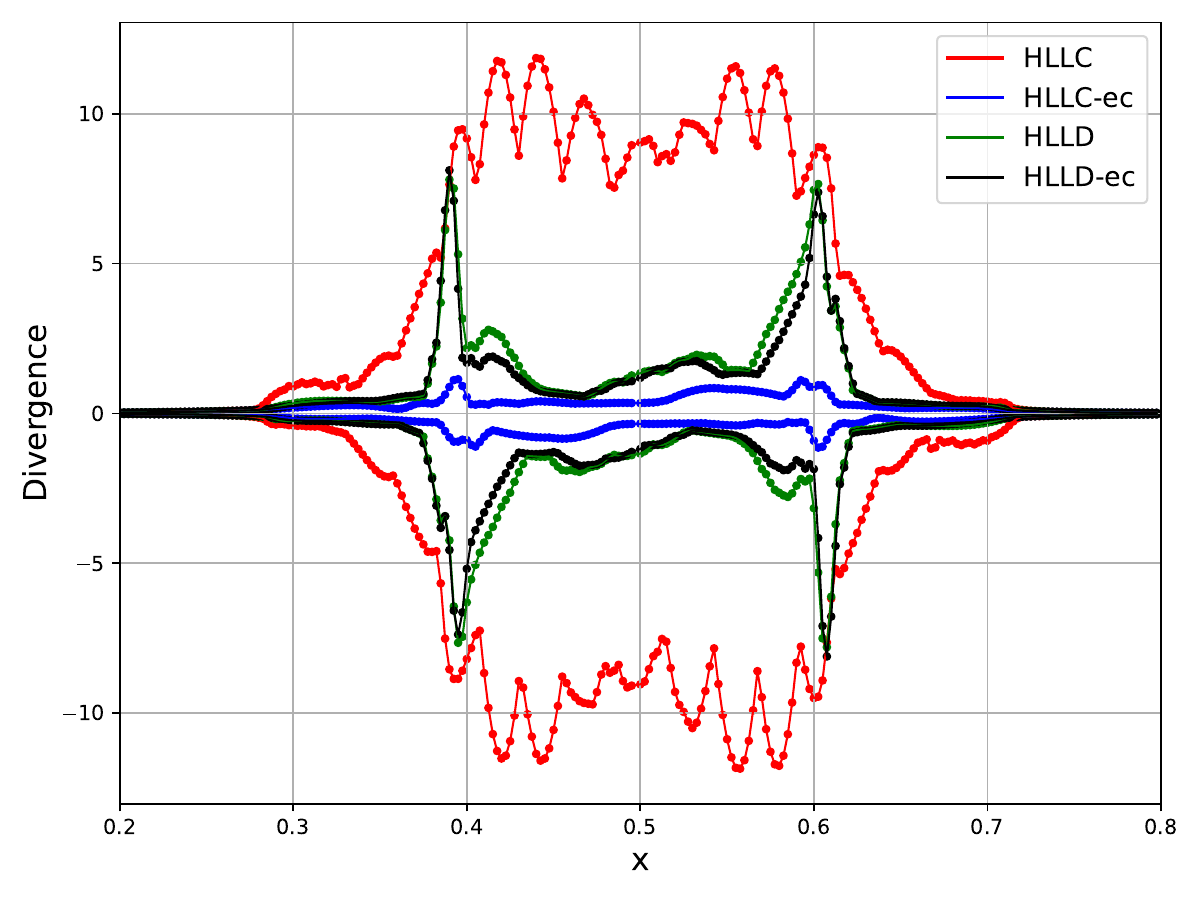}
 \caption{Envelopes of the divergence $\nabla\cdot\mathbf{B}$ of a two-dimensional low-plasma $\beta$ test. The present HLLC/D-ec schemes produce less divergence error than their conventional counterparts. For details, see Section \ref{sec:lowbetarotor}.}
\end{figure}
\end{graphicalabstract}

\setcounter{figure}{0}

%divB error production should be addressed
\begin{highlights}
\item Positivity of internal energy in  MHD Riemann solutions discussed in more realistic scenarios
\item Energy consistency investigated for the intermediate states of the MHD Riemann solutions  
\item Multi-state assumptions of HLL-type schemes revised
\item Entropy error introduced during time integration discussed
\end{highlights}

%% Keywords
\begin{keyword}
Riemann solver \sep MHD \sep positivity  \sep HLL-type scheme   \sep  plasma $\beta$


\end{keyword}

\end{frontmatter}

%\linenumbers


\section{Introduction} \label{sec:intro}

%photospheric magnetic field as the inner boundary condition

Magnetohydrodynamics (MHD) is essential for describing the macroscopic dynamics of laboratory, space, and astrophysical plasmas. Because the MHD equations are nonlinear, discontinuous solutions may arise; shock-capturing schemes are indispensable for numerical simulations of compressible MHD.
Among all shock-capturing schemes, Godunov-type schemes \cite{Godunov1959}, which utilise approximate solutions to Riemann problems, i.e.,  approximate Riemann solvers, are particularly popular for their accuracy, robustness, and efficiency. One may refer to Ref.~\cite{Toro2009} for a comprehensive introduction to such schemes for solving hydrodynamics (HD) equations. Solving an MHD Riemann problem, however, is more complex and challenging. 

While not being inclusive, several typical issues make solving MHD equations relatively challenging. Firstly, compared to the HD equations, the three-dimensional MHD equations comprise eight equations, resulting in more complex wave structures in the Riemann problem \cite{DAI1994} (Fig.~\ref{fig:MHD_modes}, further discussed later). Moreover, ensuring $\nabla\cdot\mathbf{B}\equiv 0$ (zero divergence) constraint in multi-dimensional MHD simulations is crucial, as the $\nabla \cdot\mathbf{B}$ error may cause nonphysical phenomena and even crash numerical simulations \citep{Brackbill_1980}. However, ensuring the zero divergence constraint is not a trivial task \cite{Toth_2000,Dedner_2002}. Last but not least, the positivity of scalar variables, including density, pressure, and internal energy, is challenging to preserve \cite{Wu_2018}, especially when plasma $\beta$ is low \cite{Wang2025}, which means that the ratio of gas pressure to magnetic pressure is small. {\color{green}We should also note that these issues are not isolated. For example, positivity-preservation is already essential when designing an HD Riemann solver \cite{EINFELDT1991,Batten_1997,Gallice2003}, but the $\nabla \cdot\mathbf{B}$ error makes this issue more challenging for MHD Riemann solvers.} Overall, these issues must be carefully treated for accurate and robust numerical simulations of phenomena (mainly) driven by magnetic fields. Note that the divergence constraint is not fundamental to the design of MHD Riemann solvers, and thus, it is only discussed to the extent necessary for this work.

Being relatively simple, efficient, and robust \cite{EINFELDT1991}, HLL-type Riemann solvers \cite{Harten_1983} (where HLL stands for Harten, Lax, van Leer) have been widely implemented and further developed for solving MHD equations \cite{Gurski2004,Li2005,Miyoshi2005} in various scenarios. The general idea of HLL-type schemes is to assume a spatio-temporal ($x-t$) single- or multi-state approximation for the solution to a Riemann problem, and the assumed eigenwave structure may consist of two or more waves of different speeds separating the intermediate states within the Riemann fan (Fig.~\ref{fig:MHD_modes}). The HLL scheme, being the simplest one of the HLL-type schemes, assumes two fast(est) waves and a constant state within the whole Riemann fan \cite{Harten_1983}, while there are in total seven eigenwaves in a three-dimensional MHD system. For the MHD Riemann problem, the two fastest waves are left- and right-going fast (magneto-acoustic) shocks. Having only one constant state in the Riemann fan means that the HLL scheme cannot accurately resolve Alfvén waves, slow waves, and entropy waves. The contact-resolving HLLC scheme \cite{Toro1994} has been extended to the MHD equations \cite{Gurski2004,Li2005}, resolving the entropy wave and providing better resolution than the HLL scheme. The HLLD scheme \cite{Miyoshi2005} includes two (approximated) Alfv\'en waves in the Riemann fan, resulting in a four-state approximation that further improves numerical resolution.

A more comprehensive eigenwave structure allows the HLL-type schemes to resolve (MHD) waves more accurately. In the meantime, estimates of wave speeds and calculations of intermediate states are essential for both robustness and accuracy. For example, estimating fast wave speeds helps preserve the positivity of scalar variables \cite{EINFELDT1991}. To be consistent with the integral form of the conservation laws, which is described as a consistency condition \cite{Toro2009}, the HLLC scheme of Li \cite{Li2005} (denoted as HLLC-L hereafter) uses the HLL approximation for the induction equation. In contrast, the intermediate states of the other equations are calculated based on the Rankine–Hugoniot (RH) jump relations across the fast waves.  The HLLC scheme of Gurski \cite{Gurski2004} (denoted as HLLC-G hereafter) applies the RH jump relations also to the induction equation, breaking the conservation laws within the Riemann fan, and thus extra dissipation is added via an approach similar to Linde's Riemann solver \cite{Linde2002}, to avoid nonphysical oscillations. {\color{red} Therefore, only the HLLD scheme \cite{Miyoshi2005} can satisfy the consistency condition without sacrificing the resolution for contact discontinuity}.

%% Use \subsubsection, \paragraph, \subparagraph commands to 
%% start 3rd, 4th and 5th level sections.
%% Refer following link for more details.
%% https://en.wikibooks.org/wiki/LaTeX/Document_Structure#Sectioning_commands

{\color{green} While the accuracy of Riemann solvers can be effectively improved by using more physical approximations of the MHD Riemann problem, retaining robustness in complex scenarios may still be challenging, particularly when a strong magnetic field is imposed, which} plays a crucial role in various astrophysical and space physics phenomena. For example, in numerical modelling of the solar atmosphere, plasma $\beta$ may easily go below 0.01 \cite{Gary2001,Brchnelova2023b}, which means that the plasma pressure is more than two orders of magnitude lower than the magnetic pressure. Apparently, in this kind of scenario, the accuracy of resolving the magnetic field should be prioritised, contrary to the HLLC-L scheme, which uses the relatively low-resolution HLL approximation for the induction equation. The HLLD scheme may also exhibit a similar issue when the normal component of the magnetic field is dominant, as discussed below. However, while robustness is highly desirable in numerical simulations, we have found that even the relatively more diffusive HLL scheme (alone) may not be safe from losing positivity in specific numerical simulations that have low plasma $\beta$ and strongly varying magnetic field structures, thus demanding extra processing to the magnetic boundary condition \cite{Ku_ma_2023} and proper diffusive numerical treatments \cite{Zhang2018,Perri2022}.  Solving decomposed MHD equations, in which the magnetic field is split into a time-independent background field and a time-dependent variable field, is a valuable strategy for improving robustness \cite{Wang2025}. However, this approach introduces additional complexity to algorithm implementations and may not be universally effective \cite{TANAKA1994}. 

{\color{blue}Further improvements can also be made beyond the standard conservative formulation of the MHD equations. For example, based on the so-called relaxation scheme \cite{Bouchut2007,Bouchut2010} and the (magneto-)acoustic/transport splitting strategy \cite{Chalons_2016,BOURGEOIS2024}, Ref.~\cite{TREMBLIN2024} introduced a robust finite-volume (FV) solver that includes a non-conservative but entropy-satisfying scheme, allowing robust simulations with extremely low plasma $\beta$ ($\simeq 10^{-6}$). More importantly, Ref.~\cite{TREMBLIN2024}  also addressed the fundamental reason that causes numerical stability issues under a strong magnetic field, showing how the error in magnetic field solutions breaks the entropy condition.} Although we do not provide a more extensive review of relevant approaches due to the limited scope of this work, it should be noted that numerous numerical schemes have been developed to enhance the robustness of MHD modelling. We intend to improve the Riemann solvers for solving the conservative MHD equations, and to provide further understanding regarding this issue.

In this work, we introduce a consistency condition for calculating intermediate energies between the Riemann fan and demonstrate how simple modifications can enhance the robustness of HLL-type MHD Riemann solvers, particularly in multi-dimensional simulations with strong magnetic fields. Moreover, the two-state approximation of the HLLC-type Riemann solver has been modified to enhance resolution of the magnetic field, and the assumptions used in the HLLD scheme have been revised to ensure greater consistency across different equations. 

The structure of this paper is as follows. In the next section, we briefly discuss the ideal MHD equations. In section~\ref{sec:flux}, we introduce the basics of HLL-type MHD Riemann solvers. 
Section~\ref {sec:consistency} discusses the numerical behaviours of HLL-type schemes in two typical scenarios. Then, in section~\ref{sec:newHLLC}, the energy consistency condition is proposed, and an HLLC scheme and an HLLD scheme are designed accordingly. The numerical test cases are presented in section~\ref{sec:cases}. Finally, the conclusions are given in the last section.

\section{Analytical formulations of ideal MHD equations} \label{sec:MHDEqu}
This section introduces the proposed methods and the assumptions and approximations made in the MHD equations.  More details can be found in \cite{Goedbloed2004,KULSRUD}. 
\subsection{A recap of multi-dimensional ideal MHD equations}
The  continuity equation is 
\begin{equation}
\frac{\partial \varrho}{\partial t}+\nabla\cdot\left(\varrho \mathbf{V}\right)=0, 
\end{equation}
\noindent where $\varrho$ is density and $\mathbf{V}=(u, v, w)^T$ is velocity. The equation
of motion without viscosity is 
\begin{equation}
\frac{\partial \varrho\mathbf{V}}{\partial t}+\nabla\cdot\left(\varrho\mathbf{V}\otimes\mathbf{V}+p\right)-\mathbf{J}\times\mathbf{B}=\mathbf{0}, 
\end{equation}
\noindent where  $p$ is thermal pressure, $\mathbf{J}=\nabla\times\mathbf{B}$ is  current density, and $\mathbf{B}=(B_x, B_y, B_z)^T$ is magnetic field. A factor of $\frac{1}{\sqrt{\mu_0}}$ is included in the definition of the magnetic field for simplicity, where $\mu_0=4\pi\times 10^7 \text{Hm}^{-1}$ is the magnetic permeability. 
Note that the third term, i.e., Lorentz force $\mathbf{J}\times\mathbf{B}$, does not have any component along magnetic field lines. Then, the conservation of energy is described by
\begin{equation} \label{eq:energyE}
\frac{\partial}{\partial t}\left(\varrho e+\frac{1}{2}\varrho\mathbf{V}^2 +\frac{1}{2}\mathbf{B}^2\right)+\nabla\cdot\left[\left(\varrho e+\frac{1}{2}\varrho\mathbf{V}^2+p\right)\mathbf{V}+\mathbf{B}\times\left(\mathbf{V}\times\mathbf{B}\right)\right]=0, 
\end{equation}
\noindent where  $e=\frac{p}{\varrho(\gamma-1)}$ is internal energy density, $\frac{1}{2}\varrho\mathbf{V}^2$ is kinetic energy,  $\frac{1}{2}\mathbf{B}^2$ is magnetic energy,  and $\gamma$ is the adiabatic index. Finally, the induction equation is 
\begin{equation}\label{eq:induction}
\frac{\partial \mathbf{B}}{\partial t}-\nabla\times\left(\mathbf{V}\times\mathbf{B}\right)=\mathbf{0}. 
\end{equation}

While seemingly unnecessary, we must note that in the energy equation, there should be two non-conservative terms that cancel each other, namely, 
\begin{eqnarray} \label{eq:cancelled}
\mathbf{V}\cdot\left(\mathbf{J}\times\mathbf{B}\right)  \quad\text{and} \quad -\mathbf{V}\cdot\left(\mathbf{J}\times\mathbf{B}\right).
\end{eqnarray}
\noindent These two terms represent the energy transferred from kinetic energy to magnetic energy, and vice versa. A similar scenario is the energy transfer between kinetic energy and internal energy. {\color{green}When solving the conservation equation of total energy ($E=\varrho e+\frac{1}{2}\varrho\mathbf{V}^2 +\frac{1}{2}\mathbf{B}^2$), these non-conservative terms do not (need to) appear in Eq.~(\ref{eq:energyE}).}

Moreover, we note that Eqs.~(\ref{eq:energyE}) and (\ref{eq:induction}) each have only one term containing spatial derivatives of the magnetic field. These details are taken into account in the designs of the numerical schemes. % as being analytically equivalent   does not guarantee equivalence in numerical discretizations.

 \subsection{The conservation form of  ideal MHD equations} \label{sec:conserv}

To rewrite all the equations to their conservation forms, the terms that involve the magnetic field can be rewritten as
\begin{eqnarray}
-\mathbf{J}\times\mathbf{B}= \mathbf{B}\times\left(\nabla\times\mathbf{B}\right)=\frac{1}{2}\nabla\left(\mathbf{B}^2\right)-\nabla\cdot(\mathbf{B}\otimes\mathbf{B}), \label{eq:extra}\\
\nabla\cdot\left[\mathbf{B}\times\left(\mathbf{V}\times\mathbf{B}\right)\right]=\nabla\left[\left(\mathbf{B}^2\right)\mathbf{V}-\left(\mathbf{V}\cdot\mathbf{B}\right)\mathbf{B}\right],\\ 
-\nabla\times\left(\mathbf{V}\times\mathbf{B}\right) = \nabla\left(\mathbf{V}\otimes\mathbf{B}-\mathbf{B}\otimes\mathbf{V}\right).
\end{eqnarray}
\noindent Moreover, the zero divergence constraint for the magnetic field, i.e.,  
\begin{equation} \label{eq:divB}
\nabla\cdot\mathbf{B}=0,
\end{equation}
\noindent holds; otherwise, extra terms must be added. For example, in Eq.~(\ref{eq:extra}) there would be an extra term $\mathbf{B}(\nabla\cdot\mathbf{B})$ on the right-hand side. More details can also be found, for example, in Ref.~\cite{Dedner_2002}.

Finally, the conservation form of the MHD equations can be written as first-order partial differential equations (PDEs), 
\begin{equation}  \label{eq:1stOrder}
 \frac{\partial \mathbf{U}}{\partial t}+ \nabla\cdot\mathcal{F}(\mathbf{U})=\mathbf{S},
 \end{equation}
\noindent  where $\mathbf{U}$ includes all the (conservative) variables, the tensor $\mathcal{F}$ includes all the flux terms, and $\mathbf{S}$ includes all the source terms.
Specifically, for the ideal MHD equations without gravitational acceleration, we usually have 
\begin{eqnarray}\label{eq:govern} 
\mathbf{U}  =  \left(
\begin{array}{cc}
\varrho  \\
\varrho \mathbf{V}\\
E \\
\mathbf{B}
\end{array}     \right),
\quad \mathcal{F}(\mathbf{U}) =
 \left(
\begin{array}{cc}
\varrho\mathbf{V} \\
\varrho\mathbf{V}\otimes\mathbf{V}+P\mathbf{I}-\mathbf{B} \otimes\mathbf{B}\\
(E+P)\mathbf{V}-\mathbf{B}(\mathbf{V}\cdot\mathbf{B}) \\
\mathbf{V}\otimes\mathbf{B}-\mathbf{B}\otimes\mathbf{V}
\end{array}     \right), ~\text{and}~ 
\mathbf{S} =
 \mathbf{0},
\end{eqnarray}
\noindent in which the total pressure $P=p+\frac{1}{2}\mathbf{B}^2$ includes plasma thermal pressure $p$ and magnetic pressure $\frac{1}{2}\mathbf{B}^2$.

Thus far, no approximation has been applied, and the conservation form of the equations is equivalent to the non-conservation form, as long as Eq.~(\ref{eq:divB}) holds. FV discretisation can then be applied to the conservation form of the MHD equations. However, we may realise that several original terms have been transformed into separate terms, which need to be numerically approximated consistently, as being analytically equivalent does not guarantee equivalence in numerical discretisations. For example, in the flux term of the energy equation, the original $\mathbf{B}^2$ is now separated into total energy and total pressure, respectively being $\frac{1}{2}\mathbf{B}^2$. Therefore, the approximations of these two separated terms shall be equivalent, but this is not always the case in specific numerical approaches.

\subsection{One-dimensional numerical flux of ideal MHD equations}
With FV discretisation, the numerical flux functions are typically designed based on a one-dimensional assumption, and thus, we write the one-dimensional MHD equations as 
\begin{eqnarray}  \label{eq:1DMHD}
 \frac{\partial \mathbf{U}}{\partial t}+  \frac{\partial\mathbf{F}(\mathbf{U})}{\partial {r}_{\parallel}}=\mathbf{0},
 \end{eqnarray}
 \noindent where  $\mathbf{F}(\mathbf{U})=\mathcal{F}_{\parallel}(\mathbf{U})$, the subscript ${\parallel}$ denotes the longitudinal direction of flux transportation, i.e., the normal direction $\mathbf{n}=(n_x,n_y,n_z)^{\text{T}}$ of an interface between two neighbouring cells in FV discretisation, and thus $r_{\parallel}=xn_x+yn_y+zn_z$. Correspondingly, the components perpendicular to the direction of the flux are denoted by the subscript $\perp$. Specifically, we have the one-dimensional MHD flux
 \begin{eqnarray} \label{eq:ExactFlux}
\mathbf{F}(\mathbf{U}) =
 \left(
\begin{array}{cc}
\varrho V_{\parallel}\\
\varrho V_{\parallel}u+P_x-B_{\parallel}B_{x} \\
\varrho V_{\parallel}v+P_y-B_{\parallel}B_{y} \\
\varrho V_{\parallel}w+P_z-B_{\parallel}B_{z} \\
V_{\parallel}(E+P)-B_{\parallel}(\mathbf{B}\cdot\mathbf{V}) \\
V_{\parallel}B_{x} - B_{\parallel}u \\
V_{\parallel}B_{y}-B_{\parallel}v  \\
V_{\parallel}B_{z}-B_{\parallel}w  
\end{array}     \right),
\end{eqnarray}
\noindent where $B_{\parallel}=B_xn_x+B_yn_y+B_zn_z$ and $V_{\parallel}=un_x+vn_y+wn_z$. Note that $\mathbf{F}(\mathbf{U})$ is written as $\mathbf{F}$ hereafter for simplicity. The eigen-wave modes shown in Fig.~\ref{fig:MHD_modes} can be derived from the present equations. However, the one-dimensional equations are no longer equivalent to the original three-dimensional equations.

In addition, the divergence constraint of the magnetic field becomes  
  \begin{eqnarray} \label{eq:Bn}
 \frac{\partial B_{\parallel}}{\partial r_{\parallel}}=0.
   \end{eqnarray}
\noindent  While $B_{\parallel}$  naturally being constant in a one-dimensional space, it is not the case in multi-dimensional simulations, as Eq.~(\ref{eq:Bn}) is only a sufficient but not necessary condition of Eq.~(\ref{eq:divB}).  This difference needs to be treated carefully in numerical solutions \cite{Fuchs2011a}.

\section{Numerical fluxes of ideal MHD equations} \label{sec:flux}
\subsection{Hyperbolic divergence cleaning} \label{sec:HDC}
To solve the MHD equations, the divergence constraint on the magnetic field must typically be imposed in numerical solutions. However, it is not always necessary, depending on the numerical formulae \cite{DAO2024,TREMBLIN2024}. Here, we use the hyperbolic divergence cleaning (HDC) approach \cite{Dedner_2002,MIGNONE2010} that leads to two extra hyperbolic PDEs
  \begin{eqnarray}  \label{eq:HDC}
  \left\{\begin{array}{cl}
 \frac{\partial B_{\parallel}}{\partial t}+\frac{\partial  \psi}{\partial r_{\parallel}}=0,  \\
 \frac{\partial  \psi}{\partial t}+  c_{\text{h}}^2\frac{\partial B_{\parallel}}{\partial r_{\parallel}}=0, 
 \end{array}    \right.
 \end{eqnarray}
 \noindent where $\psi$ is a Lagrange multiplier and $c_{\text{h}}$ is a constant that is expected to be sufficiently large in comparison to the fastest physical wave speeds, thus allowing the divergence error to efficiently propagate away from the computational domain \cite{Dedner_2002}.

\begin{figure}[htbp]
 \centering
 \includegraphics[width=0.5\textwidth]{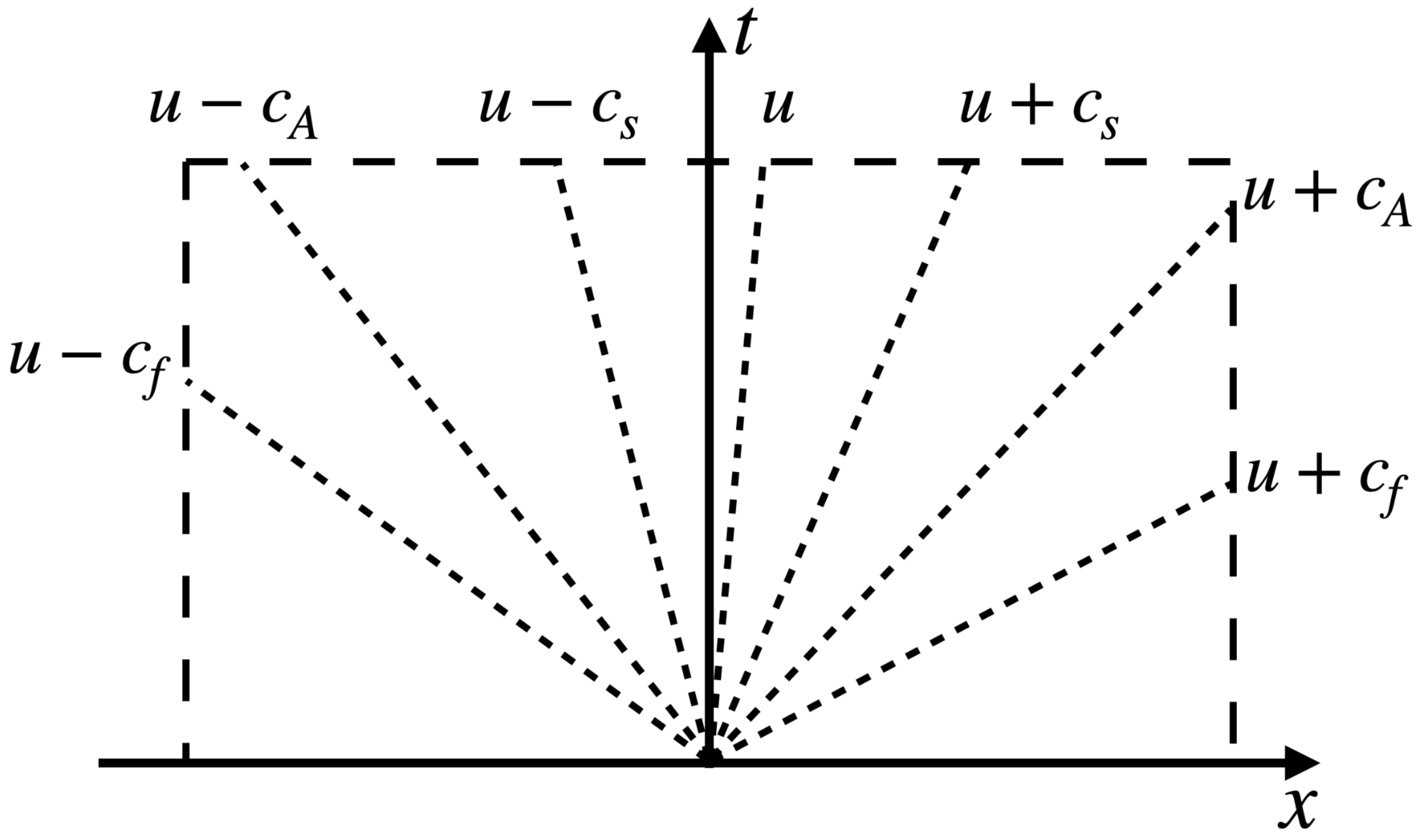}
 \caption{A $x-t$ spatio-temporal schematic of eigen-wave propagation in the MHD Riemann problem. Here, the subscripts $f$, $A$, and $s$ denote fast, Alfv\'en, and slow modes and the corresponding eigen-speeds.}
 \label{fig:MHD_modes}
\end{figure}

 Note that these two equations can be combined with the one-dimensional MHD equations and solved altogether. Then, the original seven-wave system (Fig.~\ref{fig:MHD_modes}) becomes a nine-wave, entirely hyperbolic system, including two non-physical eigenwaves with speeds $\pm c_{\text{h}}$  \cite{Yalim2011}. 
 However, separately solving this extra hyperbolic system is theoretically and numerically more convenient. Without going into details, the numerical flux of the hyperbolic system in Eq.~(\ref{eq:HDC}) is given as
\begin{eqnarray}\label{eq:HDCSolution}
\hat{\mathbf{F}}^{\text{hdc}}=\frac{1}{2}\left[
\begin{array}{cc}
 \psi^{\text{l}}+ \psi^{\text{r}}  \\ 
c_{\text{h}}^2\left(B_{\parallel}^{\text{l}}+ B_{\parallel}^{\text{r}}\right)
\end{array}     \right]-\frac{c_{\text{h}}}{2}\left(
\begin{array}{cc}
B_{\parallel}^{\text{r}}- B_{\parallel}^{\text{l}} \\ 
 \psi^{\text{r}}- \psi^{\text{l}} 
\end{array}     \right),  
\end{eqnarray}
\noindent where the superscripts l and r respectively denote the left and right states of a FV interface, and the  $\hat{\cdot}$ over the numerical flux $\mathbf{F}$ indicates that the variable is approximated, which is omitted hereafter unless mentioned otherwise. Then, this numerical flux can be combined with the one-dimensional MHD numerical flux and is used in the FV solver introduced in \ref{sec:CF}. 

As mentioned above, the condition in Eq.~(\ref{eq:Bn}) is usually not proper in a multi-dimensional space. However, the one-dimensional MHD flux function is only valid and accurate when   $\frac{\partial B_{\parallel}}{\partial r_{\parallel}}\equiv 0$, as in the general MHD jump relations.  Specifically, in the collocated FV discretisation, which we use, $B_{\parallel}$ is provided using approximated solutions in two neighbouring cells of a numerical interface, thus typically resulting in $\frac{\partial B_{\parallel}}{\partial r_{\parallel}}\ne 0$ in multi-dimensional cases, even if the zero divergence constraint is satisfied. Further discussion is provided later. {\color{red}Note that this numerical issue can be formally 
avoided if $B_{\parallel}$ is defined at the midpoint or interface between two grid points or cells, for example, by using a staggered-grid system \cite{BALSARA1999,GARDINER2005}.} We however do not further compare the differences between implementations on the collocated-grid system and the staggered-grid system, because the numerical nuances would need a rather extensive discussion. 
By introducing and numerically solving non-conservative source terms (i.e., the terms omitted in section \ref{sec:conserv}), Riemann solvers can be designed with varying $B_{\parallel}$ \cite{Fuchs2011a}, which is beyond the scope of this work. More importantly, it is shown later that even if  $B_{\parallel}$ is indeed constant, problems may still appear.

\textbf{Remark 3.1\label{re:first}}: As it is not always addressed, we note that the numerical flux and dissipation terms of the induction equation and energy equation should, {\color{green}in theory}, have no (non-zero) magnetic component perpendicular to the interface (of a FV cell). Thus numerical fluxes should not affect the corresponding magnetic field components ${B}^{\text{l,r}}_{\parallel}$, which may only be changed by the HDC solution for satisfying the zero divergence constraint. Otherwise, more divergence errors may be generated in the magnetic field. 
Correspondingly, a part of magnetic energy, i.e., $\frac{1}{2}{B}_{\parallel}^2$, should also not be affected by the numerical flux of MHD equations.
%Consequently, instead of calculating all the components of the magnetic field using the same formula \cite{Yalim2011}, here we only use the HLL average to calculate the transversal components of the magnetic field,  $\mathbf{B}_{\parallel}$, using Eq.~(\ref{eq:consistentB}). 
In fact, in the (exact) flux terms, the contributions of the longitudinal component of the magnetic field may cancel each other. Therefore, whenever possible,  the numerical dissipation terms below are explicitly without ${B}_{\parallel}$. On the other hand, as the HDC process changes  ${B}_{\parallel}$, magnetic energy should be changed correspondingly, meaning that when combining with HDC, the energy equation should have a flux term calculated based on the solution of Eq.~(\ref{eq:HDCSolution}). However, the relevant discussion is beyond the scope of this work. {\color{red}It is important to note that these issues do not occur when the local ${B}_{\parallel}$ is indeed constant.} {\color{green}For the present solver, they are relevant to the numerical performance, although not as significant as the issues addressed in the following sections}. $\square$

\subsection{Basics of HLL-type schemes}

After obtaining the HDC solution, we can solve the one-dimensional MHD equations. As introduced in Section \ref{sec:intro}, several multi-state HLL-type approximate Riemann solvers are available for solving the conservative MHD equations.

The classic HLL scheme \cite{Harten_1983} can be written as  
\begin{eqnarray}  \label{eq:HLLCoupled}
\mathbf{F}^{\text{hll}} =
\left\{\begin{array}{cl}
\mathbf{F}^{\text{l}},~&\text{if}~S^{\text{l}}>0,\\
\frac{S^{\text{r}}\mathbf{F}^{\text{l}}-S^{\text{l}}\mathbf{F}^{\text{r}}+S^{\text{l}}S^{\text{r}}(\mathbf{U}^{\text{r}}-\mathbf{U}^{\text{l}})}{S^{\text{r}}-S^{\text{l}}},~&\text{if}~S^{\text{l}}\le 0\le S^{\text{r}}, \\
\mathbf{F}^{\text{r}},~&\text{if}~S^{\text{r}}<0, 
\end{array}    \right.
\end{eqnarray}
\noindent where $S^{\text{l,r}}$ are the fastest left and right propagating wave speeds, and ${B}_{\parallel}$ can be easily removed from $\mathbf{U}^{\text{l,r}}$ (for both the induction equation and energy equation). It is known that there are different ways of (numerically) calculating the fastest wave speeds \citep{Einfeldt_1988}. Here, we use 
  \begin{eqnarray}   \label{eq:SLSR}
  \left\{\begin{array}{cl}
S^{\text{l}}=\min\left(V_{\parallel}^{\text{l}}-c_{\text{f}}^{\text{l}}, V_{\parallel}^{\text{r}}-c_{\text{f}}^{\text{r}}\right),  \\
S^{\text{r}}=\max\left(V_{\parallel}^{\text{l}}+c_{\text{f}}^{\text{l}}, V_{\parallel}^{\text{r}}+c_{\text{f}}^{\text{r}}\right), 
 \end{array}    \right.
 \end{eqnarray}
 \noindent where $c^{\text{l,r}}_{\text{f}}$ are the fast magneto-acoustic speeds calculated respectively using the left and right states $\mathbf{U}^{\text{l,r}}$.
Note that the HLL flux naturally satisfies the consistency condition \cite{Toro2009}, which is further discussed below, but the resulting constant intermediate state within the whole Riemann fan, i.e.,
 \begin{equation} \label{eq:U_HLL}
\mathbf{U}^{\text{hll}} =\frac{S^{\text{r}}\mathbf{U}^{\text{r}}-S^{\text{l}}\mathbf{U}^{\text{l}}-\mathbf{F}^{\text{r}}+\mathbf{F}^{\text{l}}}{S^{\text{r}}-S^{\text{l}}},
 \end{equation}
 \noindent {\color{red}causes more numerical diffusion. Note that $\mathbf{F}^{\text{hll}}\ne\mathbf{F}(\mathbf{U}^{\text{hll}})$ in the subsonic scenario.}

%  \begin{eqnarray}   
%%  \left\{\begin{array}{cl}
%S^{\text{L}}S^{\text{R}}(B^{\text{R}}-B^{\text{L}}),  \\
%S^{\text{L}}S^{\text{R}}\left(\frac{1}{2}(B^{\text{R}})^2-\frac{1}{2}(B^{\text{L}})^2\right), 
% \end{array}    \right.
% \end{eqnarray}

Knowing that the HLL scheme is based on the two-wave and one-state approximation, we arrive at the more accurate HLLC scheme, which is based on a three-wave, two-state assumption \cite{Toro1994}, i.e., two fast-moving nonlinear waves and one linear middle wave (entropy wave). The general form of the HLLC scheme reads
\begin{eqnarray}  \label{eq:HLLC}
\mathbf{F}^{\text{hllc}} =
\left\{\begin{array}{ll}
\mathbf{F}^{\text{l}},~&\text{if}~S^{\text{l}}>0,\\
\mathbf{F}^{\text{l}}+S^{\text{l}}(\mathbf{U}^{\text{l*}}-\mathbf{U}^{\text{l}}),~&\text{if}~S^{\text{l}}\le 0\le S^{\text{m}}, \\ 
\mathbf{F}^{\text{r}}+S^{\text{r}}(\mathbf{U}^{\text{r*}}-\mathbf{U}^{\text{r}}),~&\text{if}~S^{\text{m}}< 0\le S^{\text{r}}, \\
\mathbf{F}^{\text{r}},~&\text{if}~S^{\text{r}}<0,
\end{array}    \right.
\end{eqnarray}
\noindent where the superscripts l* and r* denote the intermediate states on the left and right sides of the middle wave, which propagates at speed $S^{\text{m}}$. As $\mathbf{F}^{\text{l},\text{r}}$ and $\mathbf{U}^{\text{l},\text{r}}$ are already given, the only unknowns are the $\mathbf{U}^{\text{l*},\text{r*}}$, which can be calculated based the jump relations, written as
\begin{equation} \label{eq:RH}
S^{\text{l,r}}\mathbf{U}^{\text{l,r}}-\mathbf{F}^{\text{l,r}}=S^{\text{l,r}}\mathbf{U}^{\text{l*,r*}}-\mathbf{F}^{\text{l*,r*}},
\end{equation}
\noindent where $\mathbf{F}^{\text{l*,r*}}$ are given as the functions of $\mathbf{U}^{\text{l*,r*}}$. As mentioned above, Eq.~(\ref{eq:RH}) may not be used for certain equations to ensure the consistency condition in a HLLC-type scheme. Further details of a typical HLLC scheme are provided in \ref{sec:HLLC_Li}, which is essentially the same as the HLLC-L scheme, except that the longitudinal component of the magnetic field and the corresponding numerical flux are explicitly treated following the discussion given in the last subsection.

\textbf{Remark 3.2\label{re:second}}: 
One of the major differences between the HLLC-L scheme and the HLLC-G scheme is that the HLLC-L scheme uses the HLL approximation to provide the numerical flux of the induction equation, which means that only the one-state approximation is used for the magnetic field. Such a solution is used  because the consistency condition \cite{Toro2009}
 \begin{equation} \label{eq:consistency}
\frac{S^{\text{m}}-S^{\text{l}}}{S^{\text{r}}-S^{\text{l}}}\mathbf{U}^{\text{l*}}+\frac{S^{\text{r}}-S^{\text{m}}}{S^{\text{r}}-S^{\text{l}}}\mathbf{U}^{\text{r*}}=\frac{S^{\text{r}}\mathbf{U}^{\text{r}}-S^{\text{l}}\mathbf{U}^{\text{l}}-\mathbf{F}^{\text{r}}+\mathbf{F}^{\text{l}}}{S^{\text{r}}-S^{\text{l}}},
 \end{equation}
\noindent cannot be satisfied when using the two-state approximation for the fluid and magnetic field solutions simultaneously \cite{Li2005}, and violating consistency conditions may result in numerical oscillations \cite{Gurski2004}. 
Therefore, the HLLC-G scheme, which uses the RH jump relations to calculate all the components in $\mathbf{U}^{\text{l*},\text{r*}}$, needs an extra "smoothing" approach to suppress the oscillations. 
To distinguish it from the {\color{green}\textit{energy consistency} condition proposed below for the intermediate energies}, Eq.~(\ref{eq:consistency}) is hereafter referred to as the \textit{integral consistency} condition. Note that this condition has also been formulated for the four-state approximation of the HLLD scheme \cite{Miyoshi2005}, which will be further specified later. $\square$

Similar to the HLLC scheme, the HLLD scheme can be written as 
\begin{eqnarray}  \label{eq:HLLD}
\mathbf{F}^{\text{hlld}} =
\left\{\begin{array}{ll}
\mathbf{F}^{\text{l}},~&\text{if}~S^{\text{l}}>0,\\
\mathbf{F}^{\text{l}}+S^{\text{l}}(\mathbf{U}^{\text{l*}}-\mathbf{U}^{\text{l}}),~&\text{if}~S^{\text{l}}\le 0< S^{\text{l*}}, \\ 
\mathbf{F}^{\text{l}}+S^{\text{l}}(\mathbf{U}^{\text{l*}}-\mathbf{U}^{\text{l}})+S^{\text{l*}}(\mathbf{U}^{\text{l**}}-\mathbf{U}^{\text{l*}}),~&\text{if}~S^{\text{l*}}\le 0< S^{\text{m}}, \\ 
\mathbf{F}^{\text{r}}+S^{\text{r}}(\mathbf{U}^{\text{r*}}-\mathbf{U}^{\text{r}})+S^{\text{r*}}(\mathbf{U}^{\text{r**}}-\mathbf{U}^{\text{r*}}),~&\text{if}~S^{\text{m}}\le 0 < S^{\text{r*}}, \\
\mathbf{F}^{\text{r}}+S^{\text{r}}(\mathbf{U}^{\text{r*}}-\mathbf{U}^{\text{r}}),~&\text{if}~S^{\text{r*}}\le 0< S^{\text{r}}, \\
\mathbf{F}^{\text{r}},~&\text{if}~S^{\text{r}}\le 0.
\end{array}    \right.
\end{eqnarray}
\noindent Here the superscript $\text{l*},\text{r*}$ indicate the states behind the left and right going fast waves, respectively, and the superscript $\text{l**},\text{r**}$ indicate the states on the left and right sides of the middle wave (i.e., behind the left and right going Alfv\'en waves). Two more eigenwaves are included in the multi-state approximation of HLLD compared to the HLLC schemes. However, as constant total pressure is assumed in the original HLLD scheme \cite{Miyoshi2005} between the fastest left and right-moving eigenwaves, only the Alfv\'en mode can be included, but not the slow mode. Details of the HLLD scheme are introduced in \ref{sec:HLLD}, essentially the same as introduced in Ref.~\cite{Miyoshi2005}, but with minor differences.

\section{Positivity and monotonicity: on the HLL-type schemes} 
\label{sec:consistency}

As mentioned, the HLLC-L scheme still uses the HLL approximation for magnetic fields. When plasma $\beta$ is low and $B_{\parallel}$ is dominant, Alfv\'en speed approaches fast magnetoacoustic speed. In this case, the HLLD scheme's approximation of the magnetic field (but not the magnetic energy, which is a significant difference) becomes closer to that of the HLL scheme. Therefore, as an example, we discuss how the HLL-type approximation may encounter challenges in preserving positivity or monotonicity, {\color{green}inspiring the efforts introduced in the next section.} {\color{red}Note that the example cannot be \textit{directly} used to explain the multi-state HLL-type schemes.}

A Riemann solver provides the numerical flux between two sets of physical states, which are expected to be physically admissible. This means that the density and (internal) energy must be positive.  However, internal energy  is usually not directly calculated by the Riemann solver but is given by 
\begin{equation} \label{eq:posi}
{e}=\frac{1}{\varrho}\left(E-\frac{1}{2}\mathbf{B}^2\right)-\frac{1}{2}\mathbf{V}^2.
\end{equation}
\noindent   Obviously, the total energy being positive does not guarantee the internal energy being positive \cite{EINFELDT1991,Batten_1997,JANHUNEN2000}, since kinetic energy and (especially) magnetic energy could be significantly larger than internal energy, such that a small numerical error may overwhelm the internal energy.
Discussions about positivity preservation typically follow the one-dimensional assumption, assuming that (I) Eq.~(\ref{eq:Bn}) holds, and (II) the tangential components of the velocity and magnetic field are negligible. However, neither assumption can be guaranteed in multi-dimensional simulations.

Moreover, when high(er)-order approximations are used,  Total Variation Diminishing \cite{HARTEN1983,Harten1984}   or monotonicity preserving \cite{Balsara2000}  methods are frequently needed for suppressing numerical oscillations, even for scalar equations \cite{Zhang2024}. Below, we demonstrate that numerical oscillations can occur due to Riemann solvers, even when using first-order approximations, which hinders monotonicity and positivity preservation.

\subsection{An example with varying $B_{\parallel}$} \label{sec:varyingBx}
From a simple yet multi-dimensional example below, we can see that the positivity of scalar variables (pressure or internal energy) can be easily compromised when plasma $\beta$ is low and $B_{\parallel}$ is not constant, which is common in realistic physical problems.
We consider calculating an HLL average within a supposed three-dimensional FV discretisation, i.e., Eq.~(\ref{eq:U_HLL}).   To simplify the discussion, we consider an interface between two FV cells of which the normal direction is $\mathbf{n}=(1, 0, 0)^{\text{T}}$. At this specific location, we have $B_x\ne 0$ (i.e., $B_{\parallel}\ne 0$) and more importantly, $\frac{\partial B_x}{\partial x}\ne 0$. Moreover, while $\frac{\partial B_{y}}{\partial y}+\frac{\partial B_{z}}{\partial z}\ne 0$ is needed for the zero divergence constraint, we assume that at this location we have $B_y=B_z=\frac{\partial B_{y}}{\partial x}=\frac{\partial B_z}{\partial x}=0$ (or sufficiently small to be negligible). The magnetic field is assumed to be sufficiently strong, and the plasma $\beta$ and temperature are  sufficiently low such that the internal energy is small in comparison to the magnetic energy, i.e.
 \begin{equation} 
\frac{1}{2}\mathbf{B}^2=\frac{1}{2}B_x^2\gg{\varrho}e.
 \end{equation}
 \noindent In addition, density $\varrho$, plasma pressure $p$, and one of the tangential components of the velocity field, $v$, are all constant across the interface (or their variations are negligible). Yet, the longitudinal component and another tangential component of the velocity field, i.e., $u$ and $w$, are set to zero to simplify the discussion. Similar scenarios may exist, for example, in coronal streamers when plasma outflows cross closed magnetic field lines.

 To calculate the HLL average of conservative variables, first, the left and right fast wave speeds {\color{green}$c^{\text{l,r}}_{\text{f}}$}  are needed. While considering that plasma $\beta$ (thus also plasma pressure) is sufficiently low such that sound speed can be omitted, we have:
\begin{equation}  
c^{\text{l,r}}_{\text{f}}= \frac{|B_x|^{\text{l,r}}}{\sqrt{\varrho}}, 
 \end{equation}
 \noindent which become equal to the local Alfv\'en speeds. Since $u$ (i.e., $V_{\parallel}$) is zero, we may simplify Eq.~(\ref{eq:SLSR}) to  
     \begin{eqnarray}  \label{eq:SL_SR}
  \left\{\begin{array}{cl}
 S^{\text{l}}=-\max(c^{\text{l}}_{\text{f}}, c^{\text{r}}_{\text{f}}),\\
S^{\text{r}}=\max(c^{\text{l}}_{\text{f}},  c^{\text{r}}_{\text{f}}),
 \end{array}    \right.
 \end{eqnarray}
 \noindent and $\max(c^{\text{l}}_{\text{f}}, c^{\text{r}}_{\text{f}})$ is denoted as $c^{\text{max}}_{\text{f}}$ hereafter for simplicity.
The longitudinal component of the magnetic field within the Riemann fan may be calculated in different ways \cite{Li2005,MIGNONE2010}. Ideally, if there is no divergence error ($\nabla\cdot\mathbf{B}=0$), which means that in an HDC solution \cite{Dedner_2002,MIGNONE2010}, we have $\psi=0$, then we can easily derive the following approximation 
   \begin{equation} \label{eq:Ideal}
B^{\text{m}}_{\parallel}={B}^{\text{m}}_x=\frac{1}{2}\left( B_x^{\text{l}}+ B_x^{\text{r}}\right).
    \end{equation}
    \noindent  Li \cite{Li2005}  used the HLL average,  which may also result in the same formula as Eq.~(\ref{eq:Ideal}) with the assumptions given for the example here. However, we expect a more accurate approximation of $B_{\parallel}$ at the interface that is not affected by either the HDC solution or other physical quantities, and thus we use Eq.~(\ref{eq:Ideal}) for other scenarios as well.

Then, the HLL average of the tangential components of the magnetic field is  
  \begin{equation} 
\mathbf{{B}}^{\text{hll}}_{\perp}=\left({B}^{\text{hll}}_y, {B}^{\text{hll}}_z\right)^{\text{T}},
 \end{equation}
 \noindent where
 \begin{equation} \label{eq:errorA}
{B}_y^{\text{hll}}=\frac{v(B_x^{\text{r}}-B_x^{\text{l}})}{2c^{\text{max}}_{\text{f}}}, \quad \text{and}  \quad {B}^{\text{hll}}_z=0.
  \end{equation}
  \noindent 
\noindent Similarly, to calculate the HLL average of total energy, we first simplify the left and right total energies under the assumptions above. We thus have 
\begin{eqnarray}   
{E}^{\text{l,r}}=\frac{1}{2}\left[\varrho v^2+(B^{\text{l,r}}_x)^2\right], 
 \end{eqnarray}
 \noindent where plasma pressure is neglected, and the energy fluxes in $\mathbf{F}^{\text{l,r}}$ are zero. Therefore, the HLL average of total energies becomes 
  \begin{equation} 
{E}^{\text{hll}}=\frac{1}{2}\varrho v^2+
\frac{(B_x^{\text{l}})^2+(B_x^{\text{r}})^2}{4}.
  \end{equation}  

\noindent  
Since density is assumed to be constant, using the HLL average on the left and right densities also yields the same value, i.e., ${\varrho}^{\text{hll}}=\varrho$.  We then have the momentum components
\begin{equation}  
({\varrho u})^{\text{hll}}= \frac{(B^{\text{r}}_x)^2-(B^{\text{l}}_x)^2}{4c^{\text{max}}_{\text{f}}}, \quad \text{and}\quad
({\varrho v})^{\text{hll}}=\varrho v.
 \end{equation}
\noindent Finally, the error between (I) the HLL average of left and right total energies and (II) the total energy calculated using the HLL averages of other quantities is
    \begin{equation} \label{eq:errorBxV}
 \text{Error}({E})^{\text{hll}}={E}^{\text{hll}}-\frac{1}{2}({B}_x^{\text{m}})^2-\frac{1}{2}({B}_y^{\text{hll}})^2-\frac{[({\varrho u})^{\text{hll}}]^2}{2\varrho}-\frac{[({\varrho v})^{\text{hll}}]^2}{2\varrho}.
  \end{equation}  

All the terms on the right-hand side of  Eq.~(\ref{eq:errorBxV}) are not below zero, and the last term in Eq.~(\ref{eq:errorBxV}) cancels the kinetic energy in ${E}^{\text{hll}}$. We notice that when $B_x^{\text{l}}=B_x^{\text{r}}$, we have $\text{Error}({E})^{\text{hll}}=0$. However, when $B_x^{\text{l}}\ne B_x^{\text{r}}$, the third term on the right hand side, $\frac{1}{2}({B}_y^{\text{hll}})^2$, particularly, is unbounded and proportional to the squares of $v$ and of the absolute difference between $B_x^{\text{l}}$ and $B_x^{\text{r}}$. Consequently, when $B_x$ varies and the tangential velocity is significant, the positivity of the internal energy calculated from the HLL averages of the conservative variables cannot be guaranteed. {\color{red}In short, we have an inconsistency ${E}^{\text{hll}}\ne\frac{1}{2}({B}_x^{\text{m}})^2+\frac{1}{2}({B}_y^{\text{hll}})^2+\frac{1}{2}\varrho^{\text{hll}}(\mathbf{V}^{\text{hll}})^2$, which may lead to a negative internal energy, once the magnetic energy and kinetic energy need to be subtracted from the total energy.}
As discussed above, when plasma $\beta$ is small, the approximation of the HLLC and HLLD schemes for the magnetic field may be close to that of the HLL scheme. Although the energy equation is solved differently in the HLLC and HLLD schemes, issues may still arise because they do not account for varying $B_{\parallel}$ \cite{Fuchs2011a}. {\color{blue}One may also find a more general discussion on an entropy-satisfying solution that does not necessarily need the zero divergence constraint, gaining further insights \cite{TREMBLIN2024}.}

  %In case the flux function is changed to Eq.~(\ref{eq:ExactFlux_Bx}), it is easy to find that the error becomes 
   %   \begin{equation}  
 %\text{Error}({E})^{\text{HLL}}_{B_{\parallel}}={E}^{\text{HLL}}_{B_{\parallel}}-\frac{1}{2}({B}_y^{\text{HLL}})^2-\frac{[({\varrho v})^{\text{HLL}}]^2}{2\varrho}=-\frac{1}{2}({B}_y^{\text{HLL}})^2,
 % \end{equation}  
 %\noindent where ${E}^{\text{HLL}}_{B_{\parallel}}=\frac{1}{2}\varrho v^2$, 
  %\noindent since the perpendicular component of the magnetic field is excluded corresponding to the change of the flux function, as has been addressed in the last section.    

\subsection{An example with constant $B_{\parallel}$}\label{sec:example}

As has been mentioned, varying $B_{\parallel}$  can be formally avoided if this component is defined at the interface (or midpoint) between two grid cells (or grid points). However, the effective values at cell centres may still not be constant. More importantly, even if $B_{\parallel}$ is indeed constant, or $\nabla\cdot\mathbf{B}$ is indeed zero,  the positivity of numerical solutions may still be in question during time integration. To the authors' knowledge, only the Lax-Friedrichs flux has been proven to be positivity-preserving when solving multi-dimensional ideal MHD equations \cite{Wu_2018}. Here, we do not intend to prove or analyse the positivity-preserving property of  {\color{red}any HLL-type schemes. Instead, using the HLL approximation as an example, we show that numerical issues may still appear in a MHD Riemann solver when  $B_{\parallel}$ is constant.}

A simple example with the same geometrical conditions (location, coordinate, etc.) as in the last subsection is given. 
We assume that $B_x$, density $\varrho$, plasma pressure $p$, and one of the tangential velocity components, $v$, are constant. In addition, we have $u=w=B_z=0$. More importantly, one of the tangential components of the magnetic field has a discontinuity 
\begin{eqnarray}  
  \left\{\begin{array}{cl}
B_y^{\text{l}}=B_{y0},  \\
 B_y^{\text{r}}=-B_{y0},
 \end{array}    \right.
 \end{eqnarray}
 \noindent where $B_{y0}$ is a constant. Again, plasma $\beta$ is assumed to be small, so the thermal pressure/energy can be omitted when directly compared with the magnetic pressure/energy. Therefore, the fast magneto-acoustic speeds are 
 \begin{equation}  
c^{\text{l,r}}_{\text{f}}= \frac{|\mathbf{B}|^{\text{l,r}}}{\sqrt{\varrho}},
 \end{equation}
\noindent and thus Eq.~(\ref{eq:SL_SR}) and the definition of {\color{green}$c^{\text{max}}_{\text{f}}$} can also be \textit{formally} used here.

Then  the (HLL-type) intermediate states of $\varrho u$ and $\varrho v$ are  
\begin{equation}     \label{eq:rhov_hll_pp}
({\varrho u})^{\text{hll}}=0,\quad \text{and} \quad
({\varrho v})^{\text{hll}}=\varrho v-B_x\frac{B_{y0}}{c^{\text{max}}_{\text{f}}}. 
 \end{equation}
\noindent Then, the intermediate state of $B_y$ is 
\begin{equation} \label{eq:by_hll_pp}
{B}_y^{\text{hll}}=\frac{B_y^{\text{r}}+B_y^{\text{l}}}{2}=0.
  \end{equation}
  After having omitted the (small) internal energy, the intermediate state of the total energy is 
    \begin{equation} \label{eq:E_hll_pp}
{E}^{\text{hll}}=\frac{\varrho v^2}{2}+
\frac{B_{y0}^2}{2}-vB_x\frac{B_{y0}}{c^{\text{max}}_{\text{f}}}.
  \end{equation}  
 Note that in both the induction and energy equations, the contributions of $B_x$ are excluded since $B_x$ is constant.
 Consequently, the error between (I) the HLL average of left and right total energies and (II) the total energy calculated using the HLL averages of other quantities is
    \begin{equation}  \label{eq:errorBp}
 \text{Error}({E})^{\text{hll}}={E}^{\text{hll}}-\frac{[({\varrho v})^{\text{hll}}]^2}{2\varrho}=
 \frac{B_{y0}^2}{2}\left(1-\frac{B_x^2}{\mathbf{B}^2}\right)\ge 0.
  \end{equation}  
\noindent Therefore, the positivity of the \textit{intermediate} internal energy can be preserved in this scenario. However, this also means that, after subtracting magnetic energy (or kinetic energy) from the total energy, the resulting internal energy (thus also thermal pressure) may change by orders of magnitude when plasma $\beta$ is small. Although it may be argued that entropy (thus internal energy)  should be produced in weak solutions of conservation laws, the issue can be more complicated during time integration, {\color{green} and needs to be further investigated.}

{\color{green} Here, we only examine this one-dimensional example to show simple effects that may appear when using an HLL-type solver}. After one forward Euler step, the  $i-$th cell centre variables at the $n+1$ step are
\begin{equation}
\mathbf{U}_i^{n+1}=\mathbf{U}_i^{n}-\frac{\Delta t}{\Delta x}\left(\mathbf{\hat{F}}^{n}_{i+\frac{1}{2}}-\mathbf{\hat{F}}^{n}_{i-\frac{1}{2}}\right),
\end{equation}
\noindent where $\Delta t=t^{n+1}-t^{n}$ is assumed to be within the CFL limitation of the forward Euler scheme, and $\Delta x=x_{i+\frac{1}{2}}-x_{i-\frac{1}{2}}$ is the grid cell size. The magnetic field discontinuity is located at $i+\frac{1}{2}$, but across $i-\frac{1}{2}$ all variables are constant.  Assuming that the HLL flux is used, we note that the HLL scheme can be formally written as Eq.~(\ref{eq:HLLC}), as long as the intermediate states on both sides of the middle wave are equal and given using the HLL average. Consequently, assuming $S^{\text{m}}\gtrsim 0$, we can specify the updated conservative variables  after one forward Euler step
%%% be careful that SL=0-Smax
  \begin{eqnarray}  \label{eq:posi_Euler}
  \left\{\begin{array}{cl}  
(\varrho v)^{n+1}_i=\varrho v-\frac{\Delta t}{\Delta x} \left(B_xB_{y0}\right), \\ 
(B_y)^{n+1}_i=B_{y0}-\frac{\Delta t}{\Delta x}\left(c^{\text{max}}_{\text{f}} B_{y0}\right), \\
(E)^{n+1}_i=\frac{\varrho v^2}{2}+\frac{B_{y0}^2}{2}-\frac{\Delta t}{\Delta x}\left(vB_xB_{y0}\right),  
 \end{array}    \right.
 \end{eqnarray}
 \noindent while the other variables remain unchanged and are therefore not discussed. It is easy to notice that when the tangential velocity component $v$ is zero, the total energy remains constant. However, imposing a constant, nonzero tangential velocity may reduce the total energy.  This issue arises due to the one-dimensional simplification of the flux function. Specifically, in the original multi-dimensional energy flux in Eq.~(\ref{eq:govern}), $v$ contributes to both terms that involve the magnetic field, but in the one-dimensional energy flux in Eq.~(\ref{eq:ExactFlux}), $v$ only affects the second term $B_{\parallel}(\mathbf{B}\cdot\mathbf{V})$. This type of error thus cannot be cured by any of the Riemann solvers discussed in this work.

Excluding the error due to the one-dimensional simplification, the result appears to be as expected for weak solutions of conservation laws: the internal energy increases after the forward Euler step, accompanied by reductions in kinetic and magnetic energies. This type of numerical error indeed does not threaten positivity preservation and might be beneficial for numerical stability.  However, we will further explain that this type of error should not be overlooked, as it may lead to spurious numerical behaviours. In particular, the discussion only involves the first-order spatial and temporal approximations. Thus, any non-physical oscillation in the present scenario may potentially hinder monotonicity preservation \cite{Balsara2000} when higher-order schemes are used \cite{Zhang2024}.

  \section{Consistency of intermediate energies within the Riemann fan} 
  \label{sec:newHLLC}
 \subsection{An energy consistency condition} \label{sec:PC_E}
In the two scenarios discussed in the previous section, the intermediate solutions of the magnetic field and magnetic energy become inconsistent, resulting in negative, e.g., Eq~(\ref{eq:errorBxV}), or positive but erroneous, e.g., Eq.~(\ref{eq:errorBp}), intermediate internal energy, {\color{red}after subtracting the (intermediate) kinetic and magnetic energies from the total energy}. To avoid or alleviate this issue, it is straightforward to split the computations of energy terms/fluxes, or even to use different variables \cite{Ryu_1993,BALSARA1999a,Popovas2025}, thus avoiding erroneous magnetic energy or kinetic energy that affects the calculation of internal energy. Splitting the equations may also bring other conveniences, both numerically and theoretically \cite{FUCHS2009,DAO2024}, but further discussion is beyond the scope of this work.  

Here, to design a Riemann solver without  actually changing the energy conservation equation, we first \textit{formally} split the total energy as follows:
 \begin{equation} \label{eq:energySplit}
E=E^{\text{e}}+E^{\text{v}}+E^{\text{b}}=\varrho e+\frac{1}{2}\varrho\mathbf{V}^2+\frac{1}{2}\mathbf{B}^2,
 \end{equation}
\noindent and define
 \begin{equation} 
E^{\text{f}}=E^{\text{e}}+E^{\text{v}}=\varrho e+\frac{1}{2}\varrho\mathbf{V}^2,
 \end{equation}
 \noindent as fluid energy for convenience. Correspondingly,  we may also \textit{formally} split the total energy flux into
  \begin{eqnarray}   \label{eq:splitting}
  \left\{\begin{array}{ll} 
F^{\text{e}}=(\varrho e+p)V_{\parallel}=\gamma\varrho eV_{\parallel},  \\
F^{\text{v}}=\frac{1}{2}\varrho\mathbf{V}^2V_{\parallel},  \\
 F^{\text{b}}=\mathbf{B}^2V_{\parallel}-B_{\parallel}(\mathbf{V}\cdot\mathbf{B}), 
 \end{array}    \right.
 \end{eqnarray} 
\noindent which are respectively the internal energy flux, kinetic energy flux, and magnetic energy flux. Similar to defining the fluid energy term, we may also define fluid energy flux $F^{\text{f}}=F^{\text{e}}+F^{\text{v}}$. Then, in the next subsections, numerical solutions will be designed accordingly. 

If the fluid or the magnetic energy flux is solved separately, one of the two non-conservative terms in Eq.~(\ref{eq:cancelled}) should be added into the corresponding energy conservation equation. 
{\color{red}However, we do not actually separate the numerical fluxes; we simply use a formal splitting to facilitate the design of the numerical solutions later. Eventually, we calculate the total energy flux in $\mathbf{F}^{\text{l,r}}$ in Eq.~(\ref{eq:HLLC}) or (\ref{eq:HLLD}).}
We assume that the (hypothetical) numerical approximations of $\pm\mathbf{V}\cdot\left(\mathbf{J}\times\mathbf{B}\right)$ are equivalent, as their analytical counterparts. Thus they can be {\color{red}formally cancelled in the total energy flux function. Correspondingly,  omitting the non-conservative terms in the split flux functions does not change the total energy flux function.} Similarly, the (hypothetical) approximations of the exchange terms between the internal energy and kinetic energy ($\pm\mathbf{V}\cdot\nabla p$, or  $\pm p\nabla\cdot\mathbf{V}$ if   $pV_{\parallel}$  is moved from $F^{\text{e}}$ to $F^{\text{v}}$) would be assumed to be equivalent and then cancelled if we split the fluid energy flux to internal energy flux and kinetic energy flux, but note that they are typically orders of magnitude smaller than magnetic energy flux in, for example, global coronal modelling \cite{Brchnelova2023b}.
 
\textbf{Remark 5.1}: Seemingly, Eq.~(\ref{eq:splitting}) is not the only option for formally splitting the total energy flux. For example, the magnetic energy flux, which is written as a single term (the last term on the left-hand side) in Eq.~(\ref{eq:energyE}), includes $\mathbf{B}^2V_{\parallel}$, which may be formally considered as magnetic energy and magnetic pressure "propagating" at the local advection speed.
However, in numerical approximations, we emphasise the "physical origin" of magnetic pressure, although it leads to consequences similar to those of plasma pressure. More specifically, the two terms of $F^{\text{b}}$ respectively include $B_{\parallel}^2V_{\parallel}$ and $-B_{\parallel}^2V_{\parallel}$, which should cancel each other also in numerical solutions, as they do analytically. This issue is sometimes ignored in certain numerical approaches. Here, a consistent numerical approximation is suggested for both $B_{\parallel}^2V_{\parallel}$ and $-B_{\parallel}^2V_{\parallel}$, and thus they are both included in $F^{\text{b}}$. A less obvious choice is the $pV_{\parallel}$ term, which could be added to either the internal energy flux or the kinetic energy flux \cite{Qu2014}. Considering that thermal pressure contributes to the momentum equation, but not to the continuity equation, one may suggest moving the term $pV_{\parallel}$ to the kinetic energy flux. On the other hand, thermal pressure is physically connected to internal energy, and thus {\color{green}we may add $pV_{\parallel}$ to the internal energy flux}, which may be physically consistent. $\square$

Note that the above splitting process does not yet introduce any numerical approximation. 
To constrain the numerical solutions later, we propose {\color{green}an \textit{energy consistency} condition for calculating the intermediate energies} of the (multi-state) HLL-type schemes
  \begin{eqnarray}   \label{eq:phyConsistent}
  \left\{\begin{array}{ll} 
\hat{E}^{\text{e}}=\hat{(\varrho e)},  \\
\hat{E}^{\text{v}}=\frac{1}{2}\hat{\varrho}\hat{\mathbf{V}}^2,  \\
\hat{E}^{\text{b}}=\frac{1}{2}\hat{\mathbf{B}}^2, 
 \end{array}    \right. 
 \end{eqnarray} 
 \noindent or 
  \begin{eqnarray}
\hat{E}^{\alpha}=E^{\alpha}(\hat{\mathbf{U}}), \quad \alpha = \text{e}, \text{v}, \text{b},
  \end{eqnarray} 
\noindent where $\hat{\varrho}$, $\hat{\mathbf{V}}$, and $\hat{\mathbf{B}}$ are results of solving the continuity equation, momentum equation, and induction equation, and  {\color{green}$\hat{(\varrho e)}$ can be one independent variable, without explicitly calculating $\hat{e}$. Note that Eq.~(\ref{eq:phyConsistent}) is not expected to be valid at FV cell centres or FD (finite difference) grid points.} {\color{red}The resulting $\hat{E}^{\text{e}}$, $\hat{E}^{\text{v}}$ and $\hat{E}^{\text{b}}$ 
are the intermediate energies within the Riemann fan. Since we actually solve the conservation law of total energy, these intermediate  energies
can be used to calculate the numerical dissipation term of the conservation equation of total energy.} As shown in the last section, when directly calculating the intermediate total energy, the HLL average may directly violate the energy consistency condition for the magnetic and kinetic energies, {\color{green}and the error must be compensated for by the internal energy, whose consistency is then also broken, and which may become negative or erroneous.} 
 Under this consistency condition, however, the numerically calculated internal energy may remain positive, even if the kinetic or magnetic energy is incorrect. {\color{red}More specifically, satisfying the energy consistency condition means that, (I) when calculating the total energy or (II) when subtracting kinetic or magnetic energy from the total energy, the approximated energies are the same.}
  The energy consistency condition can be satisfied automatically if internal energy (or entropy) rather than total energy is used as one of the conservative variables \cite{Popovas2025}, but extra source terms are then necessary (which are not needed in our solutions, since the conservation of total energy is solved).

In summary, we have realised that magnetic energy, $\frac{1}{2}\mathbf{B}^2$, and kinetic energy, $\frac{1}{2}\varrho\mathbf{V}^2$, are not independent variables, but should be respectively consistent with the solutions of the induction equation, and the momentum and continuity equations. However, suppose the energy consistency is to be fully respected. In that case, the integral consistency may be broken, not only for the HLL scheme, as can be easily seen from the last section, but also for the HLLC scheme. 
More discussion will be given later, along with the numerical results.
 
\subsection{An (intermediate-)energy-consistent HLLC scheme: HLLC-ec}

As has been introduced (including in the Appendices), HLL-type schemes are expected to satisfy the integral consistency condition. Thus, the intermediate states in the Riemann fan are sometimes not calculated following the jump relations across eigenwaves. Even the HLLD scheme cannot accurately resolve the parallel components of the magnetic field when $|B_{\parallel}|\approx |\mathbf{B}|$ and plasma $\beta$ is small, thus being almost equivalent to the HLL scheme.  Only when $|B_{\parallel}|$ is significantly smaller than $|\mathbf{B}|$,  the excessive diffusion due to the constant intermediate magnetic field between the Alfv\'en waves, i.e., Eq.~(\ref{eq:B_HLLD}),  may disappear.
Here, we propose a new HLLC-type scheme, which is explicitly enhanced for resolving magnetic fields {\color{green}(as the HLLC-L scheme assumes a constant magnetic field within the Riemann fan)} while {\color{red}sacrificing the capability of resolving density and velocity variations}.  

For the present HLLC scheme, the solutions to the continuity and momentum equations are the same as those of the HLL scheme. To be more specific, we have 
  \begin{eqnarray}     
\varrho^{\text{l*,r*}}=\varrho^{\text{hll}} ,  \quad \text{and} \quad
 (\varrho\mathbf{V})^{\text{l*,r*}}=(\varrho\mathbf{V})^{\text{hll}} , 
 \end{eqnarray}
\noindent and thus
\begin{equation} \label{eq:middleV}
\mathbf{V}^{\text{m}}=(\varrho\mathbf{V})^{\text{hll}}/\varrho^{\text{hll}},
\end{equation}
\noindent in which the longitudinal component $V^{\text{m}}_{\parallel}$ is assumed to be the middle wave speed $S^{\text{m}}$. However, as we split the total energy and the energy flux, we first use the HLL approximation to calculate the intermediate fluid energy, which is again constant over the whole Riemann fan 
 \begin{equation} 
\Biggl \{({E}^{\text{f}})^{\text{l*,r*}}=({E}^{\text{f}})^{\text{hll}}=\frac{S^{\text{r}}({E}^{\text{f}})^{\text{r}}-S^{\text{l}}({E}^{\text{f}})^{\text{l}}-(F^{\text{f}})^{\text{r}}+(F^{\text{f}})^{\text{l}}}{S^{\text{r}}-S^{\text{l}}} \Biggr \}_{\text{HLLC-ec}}.
 \end{equation}
\noindent Note that the brackets $\bigl \{  \bigr \}$ are used to highlight that a formula different from the original HLLC/D schemes is used by the energy consistent scheme, having no further mathematical implication.  Then, the tangential components of the intermediate magnetic field can be provided by 
  \begin{equation}    \label{sec:Bparallel} \Biggl \{\mathbf{B}_{\perp}^{\text{l*,r*}}=\frac{\mathbf{B}_{\perp}^{\text{l,r}}\left(S^{\text{l,r}}-V_{\parallel}^{\text{l,r}}
 \right)-\mathbf{V}_{\perp}^{\text{m}}B_{\parallel}^{\text{m}}+\mathbf{V}_{\perp}^{\text{l,r}}B_{\parallel}^{\text{l,r}}}{S^{\text{l,r}}-S^{\text{m}}}\Biggr \}_{\text{HLLC-ec}},
 \end{equation}
 \noindent in which the longitudinal component of the magnetic field, $B_{\parallel}^{\text{m}}$,  is given by Eq.~(\ref{eq:Ideal}). The last, but important, step is to calculate the intermediate magnetic energy. To satisfy the energy consistency, the intermediate  magnetic energy is not calculated using the jump relation but directly using the results of Eq.~(\ref{sec:Bparallel}):
  \begin{equation} \label{eq:Eb_2}
\biggl \{(E^{\text{b}})^{\text{l*,r*}}=\frac{1}{2}(\mathbf{B}_{\perp}^{\text{l*,r*}})^2\biggr \}_{\text{HLLC-ec}}.
 \end{equation}
 \noindent Nevertheless, {\color{red}compared to using the jump relation, Eq.~(\ref{eq:Eb_2}) will not introduce error when resolving a "real" fast wave,} in which case the energy consistency should be satisfied automatically. Therefore, the error and its consequences may be more significant when other wave modes are involved that do not propagate at the fast magnetoacoustic speed. We, however, do not intend to provide a theoretical analysis; instead, we will present numerical tests in the next section. 

 Finally, using the \textit{intermediate} states obtained above to replace the ones in Eq.~(\ref{eq:HLLC}), having
   \begin{equation}  
\biggl \{E^{\text{l*,r*}}= ({E}^{\text{f}})^{\text{hll}}+(E^{\text{b}})^{\text{l*,r*}}\biggr \}_{\text{HLLC-ec}},
 \end{equation}
 \noindent we then have a new HLLC-type scheme that uses a new two-state approximation for better resolving the magnetic field and, more importantly, satisfies the energy consistency condition to avoid breaking the positivity of internal energy {\color{red}when subtracting magnetic energy, i.e., $(E^{\text{b}})^{\text{l*,r*}}$, from total energy, i.e., $E^{\text{l*,r*}}$.} Therefore, the present HLLC scheme is referred to as the HLLC-ec (energy consistent) scheme hereafter. It is trivial to prove that the integral consistency condition is satisfied for the continuity, momentum, and induction equations, but not the energy equation. However, the intermediate fluid energy of the HLLC-ec scheme satisfies the integral consistency. Note that here the \textit{intermediate} momentum and density need to be constant for the integral consistency to be respected.  
 
 As the solutions of the fluid equations of the present schemes are given using the HLL approximation, the positivity-preserving property of the present scheme may be analysed (partly) following the discussion regarding the  HLL scheme for MHD \cite{Gurski2004}.  Lastly, using the HLL approximation for the momentum equation may alleviate numerical shock instabilities \cite{Zhang2017,Xie2017}, which are not further examined in this work.

\subsection{An (intermediate-)energy-consistent HLLD scheme: HLLD-ec}

The HLLD scheme \cite{Miyoshi2005} includes more eigenwaves, and more importantly, it satisfies the integral consistency condition without sacrificing accuracy for certain equations (compared to the HLLC-type schemes). Here, we present a new HLLD scheme that satisfies the energy-consistency condition. Moreover, some of the assumptions used to calculate the intermediate states have been revised. 

For the states behind the fast waves, denoted by superscript l* and r*, we use the same density, velocity/momentum, and magnetic field components, exactly as those of the original HLLD scheme introduced in \ref{sec:HLLD}. However, to calculate the intermediate total energy behind the fast waves, we follow
 \begin{equation} \label{eq:totalE_PC}
\Biggl \{E^{\text{l*,r*}}=\left(\varrho e\right)^{\text{l*,r*}}+\left(\frac{1}{2}\varrho\mathbf{V}^2\right)^{\text{l*,r*}}+\frac{1}{2}(\mathbf{B}^{\text{l*,r*}}_{\perp})^2\Biggr \}_{\text{HLLD-ec}},
 \end{equation}
\noindent and thus we only need to provide each of these terms on the right-hand side. Similar to the HLLC-ec scheme, here only the tangential components of the magnetic field are included in the magnetic energy, as discussed in Section~\ref{sec:HDC}. While the intermediate magnetic energies $\frac{1}{2}(\mathbf{B}^{\text{l*,r*}}_{\perp})^2$  are calculated using Eqs.~(\ref{eq:Eb_2}) and (\ref{eq:B_perp}), the intermediate internal energy and kinetic energy are provided differently.

In HLL-type schemes $V_{\parallel}=S^{\text{m}}$ is always calculated using the HLL average, but in the HLLD scheme (and the HLLC-L scheme) $\mathbf{V}_{\perp}$ behind the fast waves is calculated following the RH jump relation. If the intermediate kinetic energy is calculated using the jump relation as well, it may be inconsistent with the velocity/momentum, thereby breaking the energy consistency condition. Therefore, for the new HLLD scheme, the intermediate kinetic energies $\left(\frac{1}{2}\varrho\mathbf{V}^2\right)^{\text{l*,r*}}$ are calculated following Eq.~(\ref{eq:phyConsistent}), using the density and velocity given by Eqs.~(\ref{eq:HLLDSM})--(\ref{eq:B_perp}), thus satisfying the energy consistency condition.

To calculate the intermediate internal energy, we cannot assume it is constant over the whole Riemann fan. Therefore, based on Eq.~(\ref{eq:splitting}), we define a formal jump relation across the fast wave
\begin{equation}
\biggl \{S^{\text{l,r}}(E^{\text{e}})^{\text{l,r}}-(F^{\text{e}})^{\text{l,r}}=S^{\text{l,r}}(E^{\text{e}})^{\text{l*,r*}}-(F^{\text{e}})^{\text{l*,r*}}\biggr \}_{\text{HLLD-ec}},
\end{equation}
\noindent where, however, $(\varrho e)$ is calculated as an independent variable without directly considering the solution of the continuity equation. Therefore, the intermediate internal energies behind the fast waves are given as {\color{green}
 \begin{equation}
\Biggl \{(\varrho e)^{\text{l*,r*}}=(\varrho e)^{\text{l,r}}\frac{S^{\text{l,r}}-\gamma V_{\parallel}^{\text{l,r}}}{S^{\text{l,r}}-\gamma S^{\text{m}}}\Biggr \}_{\text{HLLD-ec}}.
\end{equation}
 \noindent}{\color{green}Then, all the terms needed for Eq.~(\ref{eq:totalE_PC}) are available, and the resulting intermediate total energies can be used in (the diffusion terms of) the HLLD flux Eq.~(\ref{eq:HLLD}).}
 
 Similarly, we may calculate the intermediate energies between two Alfv\'en waves, denoted by superscripts  $\text{l**}$ and $\text{r**}$. We also assume that density does not change across the  Alfv\'en waves, and thus Eq.~(\ref{eq:HLLDRho}) can be reused. As has been explained in Ref.~\cite{Miyoshi2005}, within the Riemann fan, the jump relations across the Alfv\'en waves are not solvable, and thus Eqs.~(\ref{eq:V_HLLD}) and (\ref{eq:B_HLLD}) are also reused. As the intermediate magnetic energy and kinetic energy are  calculated using the intermediate magnetic field and velocity, the intermediate internal energy is separately given following the integral consistency condition  Eq.~(\ref{eq:consistency4}), resulting in  {\color{green}
 \begin{equation} \label{eq:ee}
\Biggl \{(\varrho e)^{\text{l**,r**}}=\frac{\left[S^{\text{m}}(1-\gamma)+c^{\text{r*}}_{\text{a}}\right](\varrho e)^{\text{r*}}+\left[S^{\text{m}}(\gamma-1)+c^{\text{l*}}_{\text{a}}\right](\varrho e)^{\text{l*}}}{c^{\text{l*}}_{\text{a}}+c^{\text{r*}}_{\text{a}}}\Biggr \}_{\text{HLLD-ec}}.
\end{equation}
\noindent } where the Alfv\'en speeds are
\begin{equation}  
c^{\text{l*,r*}}_{\text{a}}= \frac{|B^{\text{m}}_{\parallel}|}{\sqrt{\varrho^{\text{l*,r*}}}}. 
 \end{equation}
\noindent Finally, Eq.~(\ref{eq:totalE_PC}) can be reused (except that the superscripts $\text{l*}$ and $\text{r*}$ are changed to $\text{l**}$ and $\text{r**}$) to calculate the intermediate total energy between two  Alfv\'en waves, and then all the intermediate states are available for the HLLD flux Eq.~(\ref{eq:HLLD}).

The resulting scheme is denoted as the HLLD-ec scheme hereafter. Similar to the HLLC-ec scheme, the major change introduced into the  HLLD-ec scheme is the solution to the energy conservation equation, or more specifically, the calculation of the intermediate (total) energies. {\color{red}In particular, the intermediate energies behind the fast waves are not calculated following the jump relations, and this difference affects how the fast waves are captured.}  Moreover, we do not use the jump relations across Alfvén waves, nor do we assume a constant total pressure over the whole Riemann fan. The following remark provides further clarification of the rationale.

\textbf{Remark 5.2\label{re:totalP}}: 
In the original HLLD scheme, the intermediate total energies are calculated using the jump relations across the fast and Alfv\'en waves. In the meantime, it also assumes a constant total pressure over the whole Riemann fan, which is {\color{red} reasonable when considering the given incompressible eigen-modes (Alfv\'en wave and entropy wave) within the Riemann fan.}  However, when the intermediate total energy Eq.~(\ref{eq:LMVars}) is calculated using tangential magnetic field components from Eq.~(\ref{eq:B_perp}), the contributions of magnetic pressure and magnetic tension are in fact calculated differently. As suggested previously, we allow different numerical processes to be used for $F^{\text{f}}$ and $F^{\text{b}}$, respectively, but a consistent approximation should be applied for the terms in $F^{\text{b}}$. Moreover, in the HLLD scheme, the intermediate velocity/momentum and magnetic field {\color{red}between two Alfv\'en  waves are not calculated following the jump relations across the Alfv\'en waves \cite{Miyoshi2005},} {\color{green}but following the integral consistency condition.} {\color{red}Thus, a constant total pressure is not always consistent with the fact that the magnetic pressure may change within the Riemann fan, particularly when plasma $\beta$ is low.}
  $\square$

Overall, we conclude that strictly limiting the assumptions made for the intermediate states is not necessary; thus, we may attempt to relax, for example, the constant total pressure assumption. However, the HLLD-ec scheme effectively has constant total pressure \textit{between} the Alfv\'en waves.
% \textbf{Remark 5.3}: Ref.~\cite{Miyoshi2005} explains that slow shocks cannot form within the Riemann fan. However, even if slow shocks are assumed to exist within the Riemann fan, the HLLD scheme would be essentially the same, as long as the four-state approximation is still used. A brief explanation is as follows: (I) with a four-state approximation, the jump relations across the two intermediate eigenwaves behind the fast waves are either not solvable, or not physically meaningful (constant tangential velocity and magnetic field components); and consequently, (II) when following the integral consistency condition Eq.~(\ref{eq:consistency4}), slower $S^{\text{l*,r*}}$ would be compensated by correspondingly changed $\mathbf{U}^{\text{l**,r**}}$.   $\square$

\subsection{How (much) is it consistent?} \label{sec:How}
 In the present HLLC/D-ec schemes, the positivity of the intermediate internal energy is no longer affected by the inconsistency between the intermediate magnetic field and energy. More specifically, any numerical error introduced into the intermediate magnetic energy directly contributes to the intermediate total energy without needing to be "compensated" by the internal energy or vice versa. 
However, during time integration, the \textit{effects} of the numerical error in the intermediate states are more difficult to define. 

%Here, we further simplify the example in subsection \ref{sec:example}, by setting $v_0$ to zero.

To provide more specifics, we reuse most conditions in subsection \ref{sec:example}, particularly the wave speeds. However, to simplify the discussion, we again assume a discontinuity of $B_y$ at $i+\frac{1}{2}$, while also assuming that the tangential velocity components are all zero. The cell centre variable and the intermediate variable at time step $n$ are 
\begin{equation}
(B_{y})_i^n=B_{y0}, \quad \text{and} \quad (B_y)^n_{i+\frac{1}{2}}=B_y^*,
\end{equation}
\noindent in which $B_y^*$ may be calculated by approximating the induction equation using different HLL-type numerical schemes (unlike what is assumed in subsection \ref{sec:example}). 
 Energy-consistent schemes can readily provide the intermediate magnetic energy. The cell centre magnetic energy and intermediate magnetic energy at time step $n$ are
\begin{equation} \label{eq:En}
(E^{\text{b}})^n_i=\frac{1}{2}B_{y0}^2,  \quad \text{and} \quad  (E^{\text{b}})^n_{i+\frac{1}{2}}=\frac{1}{2}(B_y^*)^2.
\end{equation}
\noindent Then, we may define 
\begin{equation} \label{eq:deltaBE}
(\Delta B_{y})^n_{i+\frac{1}{2}}=B_y^*-B_{y0}, \quad \text{and} \quad (\Delta E^{\text{b}})^n_{i+\frac{1}{2}}=\frac{1}{2}(B_y^*)^2-\frac{1}{2}B_{y0}^2,
\end{equation}
\noindent to simplify the notations hereafter. In the meantime, both $(\Delta B_{y})^n_{i-\frac{1}{2}}$ and $(\Delta E^{\text{b}})^n_{i-\frac{1}{2}}$ are zero and not discussed, and thus the subscript $i+\frac{1}{2}$ in Eq.~(\ref{eq:deltaBE}) can be omitted, without causing confusion. As we only discuss one forward Euler step, the superscript $n$ of $\Delta E^{\text{b}}$ and $\Delta B_y$ is also omitted.

Using the assumption $S^{\text{m}}\gtrsim 0$ again, we can calculate the cell centre variables after one forward Euler step. Ideally, the updated magnetic energy should be consistent with the updated magnetic field. However, while we have 
\begin{equation} \label{eq:updatedB}
(B_{y})_i^{n+1}=(B_{y})_i^n-S^{\text{l}}\frac{\Delta t}{\Delta x}\Delta B_{y}=(B_{y})_i^n+C\Delta B_{y},
\end{equation}
\noindent the updated magnetic energy is 
\begin{equation}\label{eq:updatedEB}
(E^{\text{b}})_i^{n+1}=(E^{\text{b}})_i^n-S^{\text{l}}\frac{\Delta t}{\Delta x}\Delta E^{\text{b}}=(E^{\text{b}})_i^n+C\Delta E^{\text{b}},
\end{equation}
 \noindent where $C=c^{\text{max}}_{\text{f}}\frac{\Delta t}{\Delta x}$ is used hereafter for simplicity.  Note that, here, the same left-going fast wave speed $S^{\text{l}}=-c^{\text{max}}_{\text{f}}$ is used to simplify the discussion, but this choice is not necessary for concluding below.

 Obviously, since the total energy is a conservative variable and the kinetic energy is calculated based on the solutions of the momentum equation and the continuity equation,  the internal energy will increase when $(E^{b})_i^{n+1}> \frac{1}{2}\left[(B_{y})_i^{n+1}\right]^2$. When subtracting magnetic energy calculated using the result of Eq.~(\ref{eq:updatedB}) from the updated total energy Eq.~(\ref{eq:updatedEB}), we have
 \begin{equation} \label{eq:SpaceTimeError}
\text{Error}({E^{\text{b}}})_i^{n+1}=C\left[\Delta E^{\text{b}}-(B_{y})_i^n\Delta B_{y}-\frac{1}{2}C(\Delta B_{y})^2\right], 
\end{equation}
\noindent for which we assume that there is no error at time step $n$, given by the first formula in Eq.~(\ref{eq:En}). Therefore, the error term is not proportional to the time step.
We may further expand the first two terms on the right-hand side of Eq.~(\ref{eq:SpaceTimeError}), written as
\begin{equation}
\Delta E^{\text{b}}-(B_{y})_i^n\Delta B_{y}=\frac{1
}{2}(B_y^*+B_{y0})(B_y^*-B_{y0})-B_{y0}(B_y^*-B_{y0}).
\end{equation}
\noindent While it cannot be guaranteed, we may expect these two terms to be in the same order of magnitude in many scenarios, thus reducing the error term. To be complete for this specific case, we may further write the error as 
\begin{equation} \label{eq:TimeError}
 \text{Error}({E}^{\text{b}})^{n+1}_i=\frac{1}{2}(C-C^2)\left(B_y^{\text{*}}-B_{y0}\right)^2,
\end{equation}
\noindent   which is positive when $C<1$. Therefore, although the intermediate magnetic field and magnetic energy are consistent, the updated cell centre solutions are not. 

This error term arises when the energy-consistency condition is satisfied. However, the error term may continue to increase without satisfying the consistency condition. A simple example is when the tangential components rotate across the discontinuity, while the magnetic pressure remains constant. In this scenario, $\Delta E^{\text{b}}$ becomes zero; thus, the error term may be significantly larger than the energy-consistent solution. Therefore, the energy consistency limits the inconsistency between the magnetic field and magnetic energy. On the other hand, increasing the estimated eigenwave speed, which increases numerical diffusion, does not help suppress this error. 

Here, we use this simple example only to convey a general idea of potential errors, {\color{green}without intending to be exhaustive.} We will need to discuss more detailed consequences, especially when using implicit time-stepping.
Completely eliminating this type of error can be achieved by using internal energy rather than total energy as one of the conservative variables. Still, the corresponding weak solution may need to be discussed \cite{Tadmor_2003}, and the error may be necessary, at least to some extent. More discussion, however, is beyond the scope of this work.

  %\begin{equation}
%\mathcal{E}=\left(S-V_{\perp}\right)\left(\varrho e+\frac{1}{2}\varrho\mathbf{V}^2+ \frac{1}{2}\mathbf{B}^2\right)-\left(p+\frac{1}{2}\mathbf{B}^2\right)V_{\perp}+B_{\perp}(\mathbf{V}\cdot\mathbf{B}),
 %\end{equation}
 %\noindent is only defined to simplify the formula.

  \section{Numerical tests} \label{sec:cases}
%\footnote{While the results shown were generated by \texttt{COOLFluiD}, some of them were cross-examinated using an opensource package  \texttt{MLAU} \cite{Minoshima2020,Minoshima2021} (available at {https://github.com/minoshim/MLAU}). Details are not further discussed.}

We use the fully-implicit MHD solver in \texttt{COOLFluiD}, which is introduced in \ref{sec:CF}, to run the numerical test cases shown here. 
Specifically, the HLL, HLLC(-L), HLLD, and the proposed HLLC/D-ec schemes are compared. {\color{red}The conventional HLLC/D schemes are further introduced in the Appendices, which consider the discussion in Section \ref{sec:HDC} to be fairly compared with the proposed schemes.} Moreover,  using typical unstructured FV limiters may still produce non-physical oscillations as they do not guarantee local monotonicity near strong discontinuities \cite{Zhang2018}. Therefore, the first-order spatial approximation is discussed to focus on the effects of Riemann solvers, which is a necessary step before introducing higher-order solutions.

To verify the results, we examine some of the tests using an explicit FV solver\footnote{Opensource package  \texttt{MLAU} \cite{Minoshima2020,Minoshima2021} available at {https://github.com/minoshim/MLAU}.}, also solving the conservative MHD equations. {\color{red}Instead of the HDC, this explicit solver uses constrained transport to remove the $\nabla\cdot\mathbf{B}$ error and a staggered-grid system to discretise the computational domains in multi-dimensional simulations. Consequently, its Riemann solvers do not consider the numerical nuances related to varying $B_{\parallel}$, and the divergence error does not develop in the simulations. Otherwise, the explicit simulations, whose results are not shown, confirm the numerical resolution of the proposed schemes compared with the conventional schemes.}

%Below, two types of test cases are compared. One type is MHD numerical simulations, which include canonical one- and two-dimensional shock wave problems. However, before showing the numerical simulations we introduce a (semi-)numerical test for the flux functions of the HLL average, Eq.~(\ref{eq:U_HLL}). This test simply provides a series of admissible inputs to the numerical flux functions and then compares (the positivity of). This simple test is designed because: (I) nonlinear flux functions are difficult to analytically compare, and (II) it is difficult to isolate important behaviours of numerical fluxes from complex numerical simulation results. 

%\subsection{Positivity in the HLL average: a numerical test}

\subsection{One-dimensional MHD shock-tube} \label{sec:BW}

\begin{figure}[ht]
 \centering
 \subfigure[\label{fig:HLLC_density}{}]{
 \includegraphics[width=0.48\textwidth]{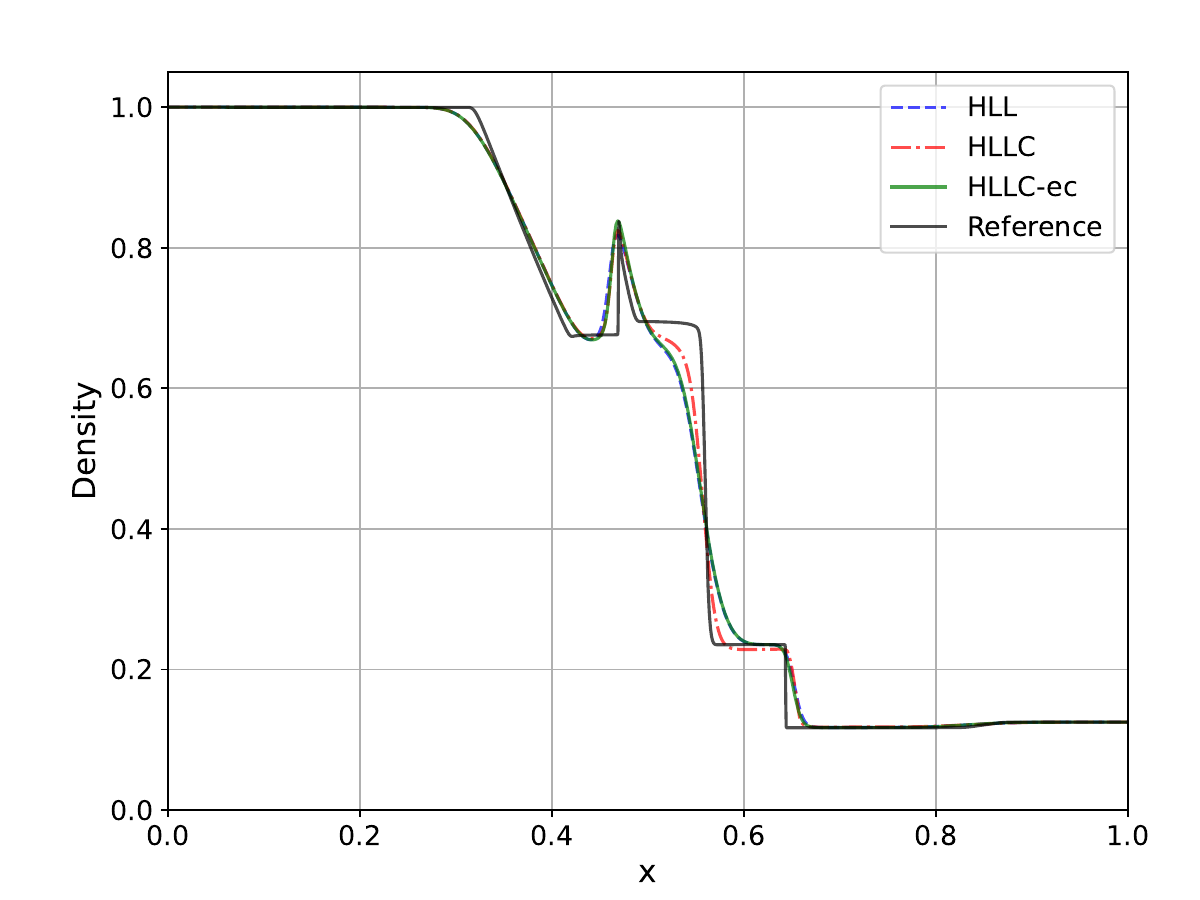}
 }
 \subfigure[\label{fig:HLLC_density_zoomin}{}]{
 \includegraphics[width=0.48\textwidth]{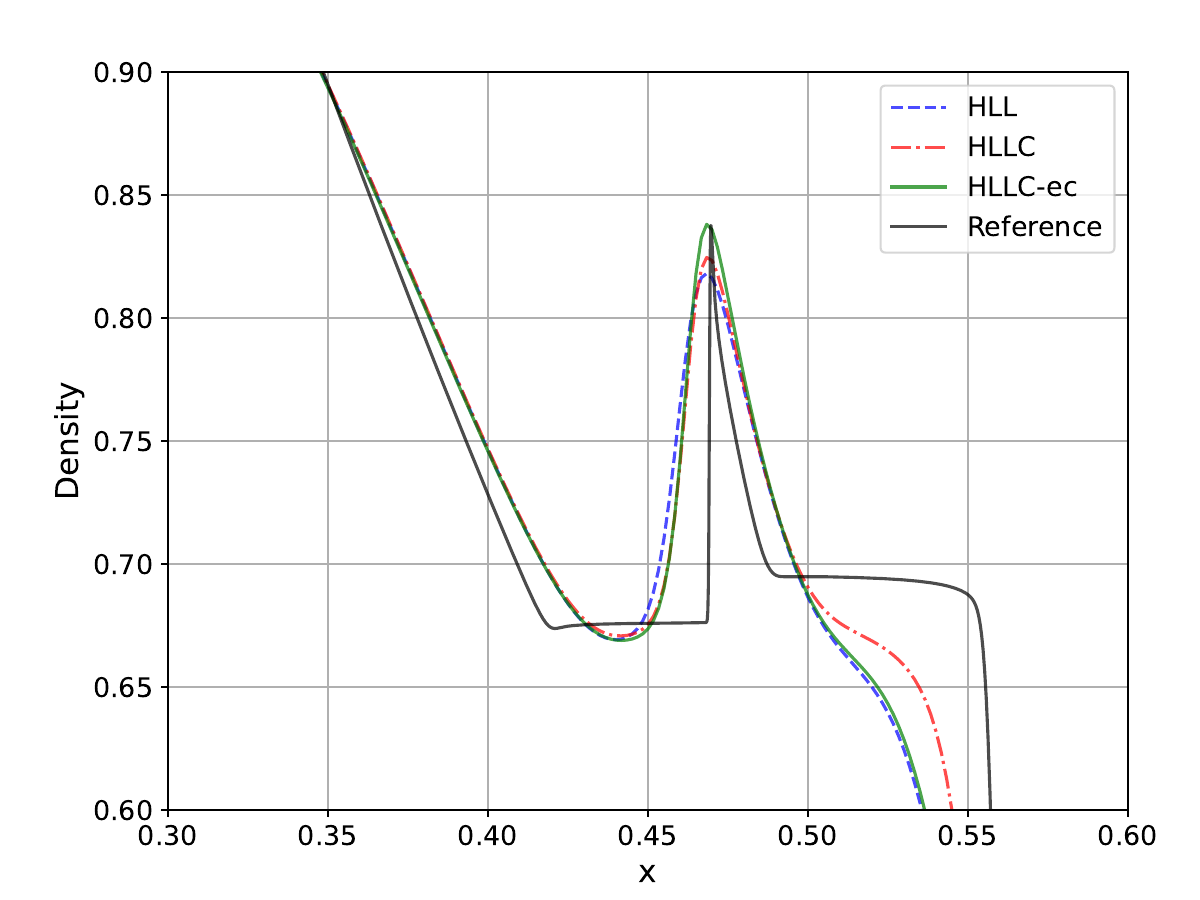}
 } 
\subfigure[\label{fig:HLLC_x_velocity}{}]{
 \includegraphics[width=0.48\textwidth]{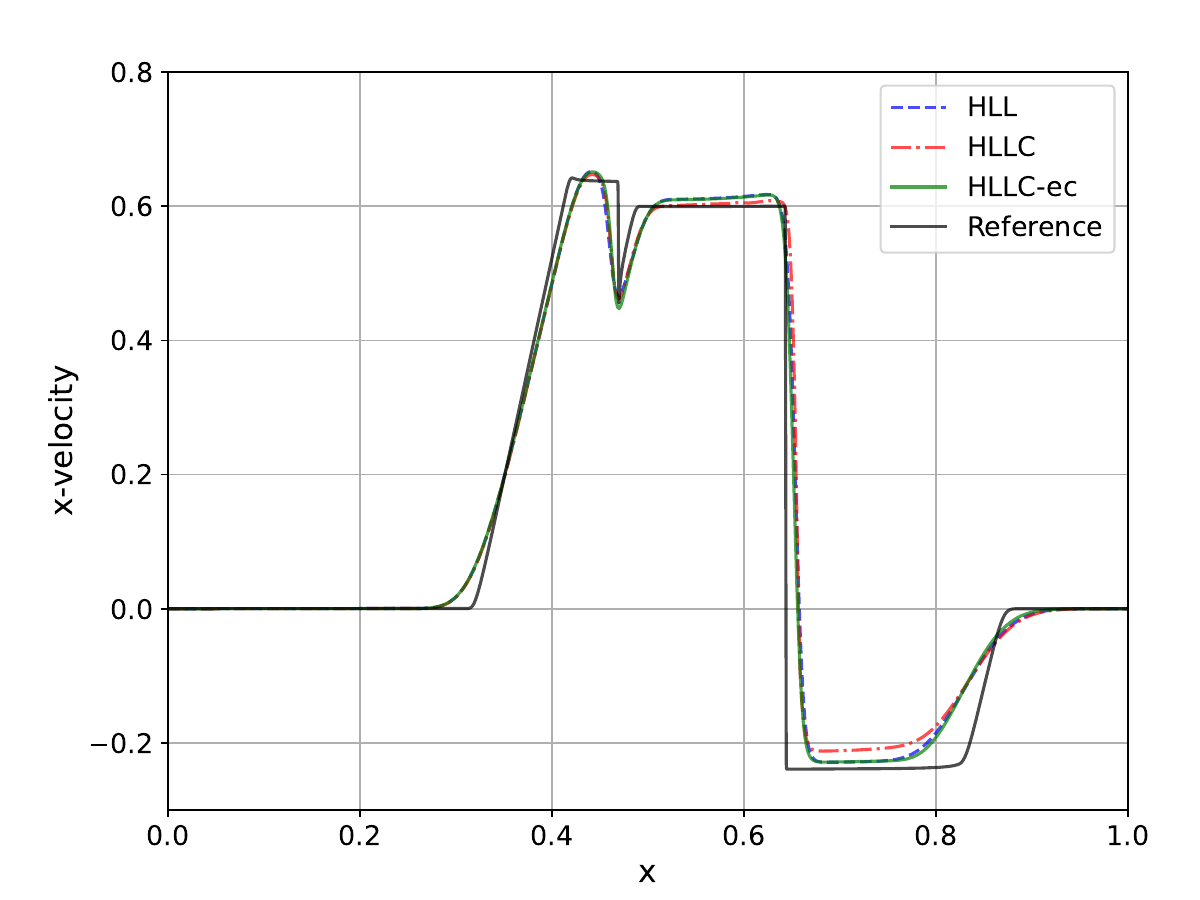}
 }
 \subfigure[\label{fig:HLLC_x_velocity_zoomin}{}]{
 \includegraphics[width=0.48\textwidth]{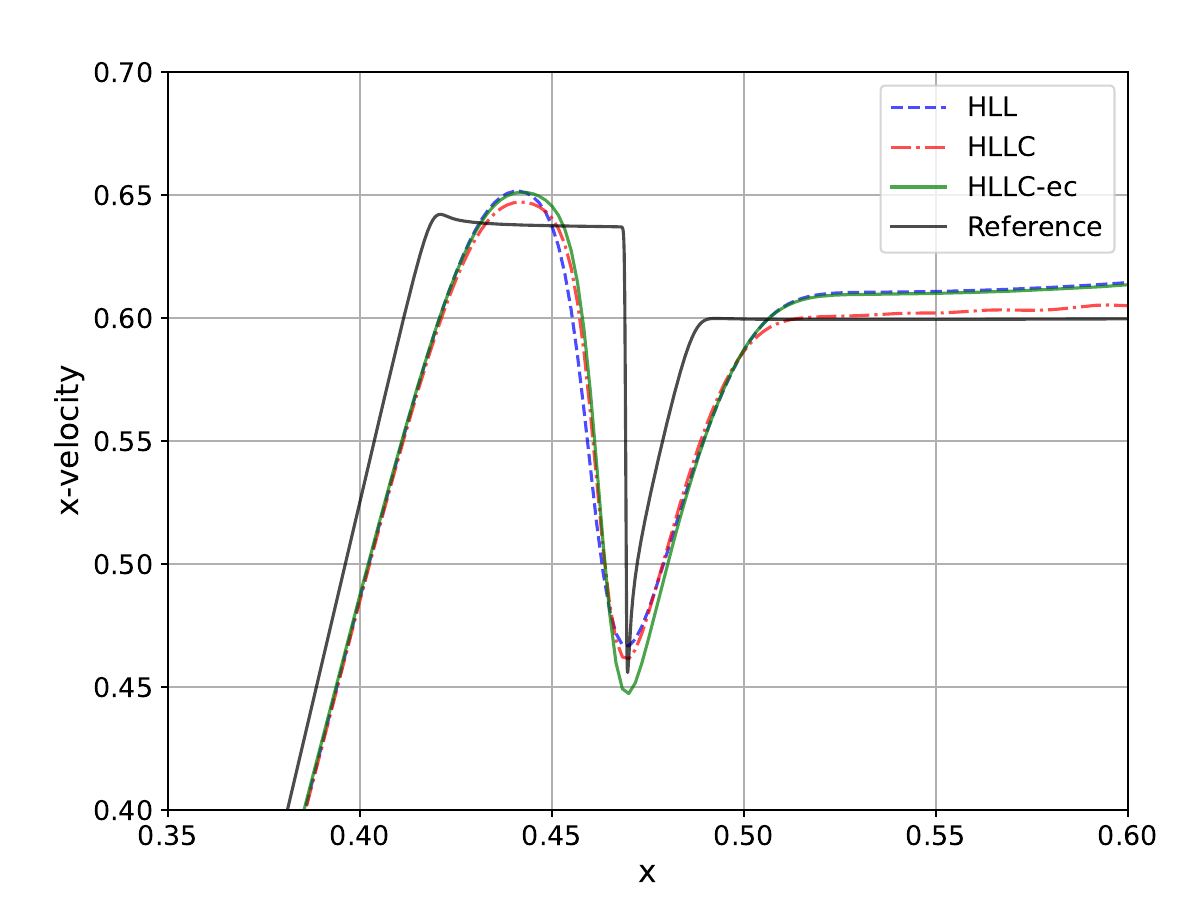}
 } 
 \caption{Density and velocity ($x$-component) distributions of HLLC-type schemes.}
 \label{fig:HLLC_BW}
\end{figure}

The MHD shock-tube problem of Brio and Wu \cite{BRIO1988} is first used to show the basic behaviours of the proposed schemes. While differences will be observed, no significant advantage is expected from the results because of the relatively high plasma $\beta$. This quasi-one-dimensional problem within $x\in[0,1]$ has two sets of states separated by a discontinuity:  
  \begin{eqnarray}   
  \left\{\begin{array}{ll}
 (\varrho, u, v, p, B_x, B_y)^{\text{l}}=(1, 0, 0, 1, 0.75, 1), & \text{if} \quad x<0.5, \\
 (\varrho, u, v, p, B_x, B_y)^{\text{r}}=(0.125, 0, 0, 0.1, 0.75, -1),& \text{if} \quad x\ge0.5, 
 \end{array}    \right.
 \end{eqnarray} 
\noindent with an adiabatic index $\gamma=2$.  A two-dimensional mesh with $600\times 4$ FV cells is used to model this one-dimensional problem, and the initial solutions change along the $x$-direction but are constant along the $y$-direction. The results at nondimensional time $t = 0.1$ are presented after 100 time steps {\color{red}(thus CFL $\simeq 2.3$)} using the fully implicit method. However, we do not consider the effects of changing the time steps further, as it is not the focus. A reference solution is provided using a mesh with 20,000 FV cells along the $x$-direction, with the difference between MHD waves captured by the HLL and HLLD schemes being minor and thus not discussed.

\begin{figure}[ht]
 \centering
 \subfigure[\label{fig:HLLD_density}{}]{
 \includegraphics[width=0.48\textwidth]{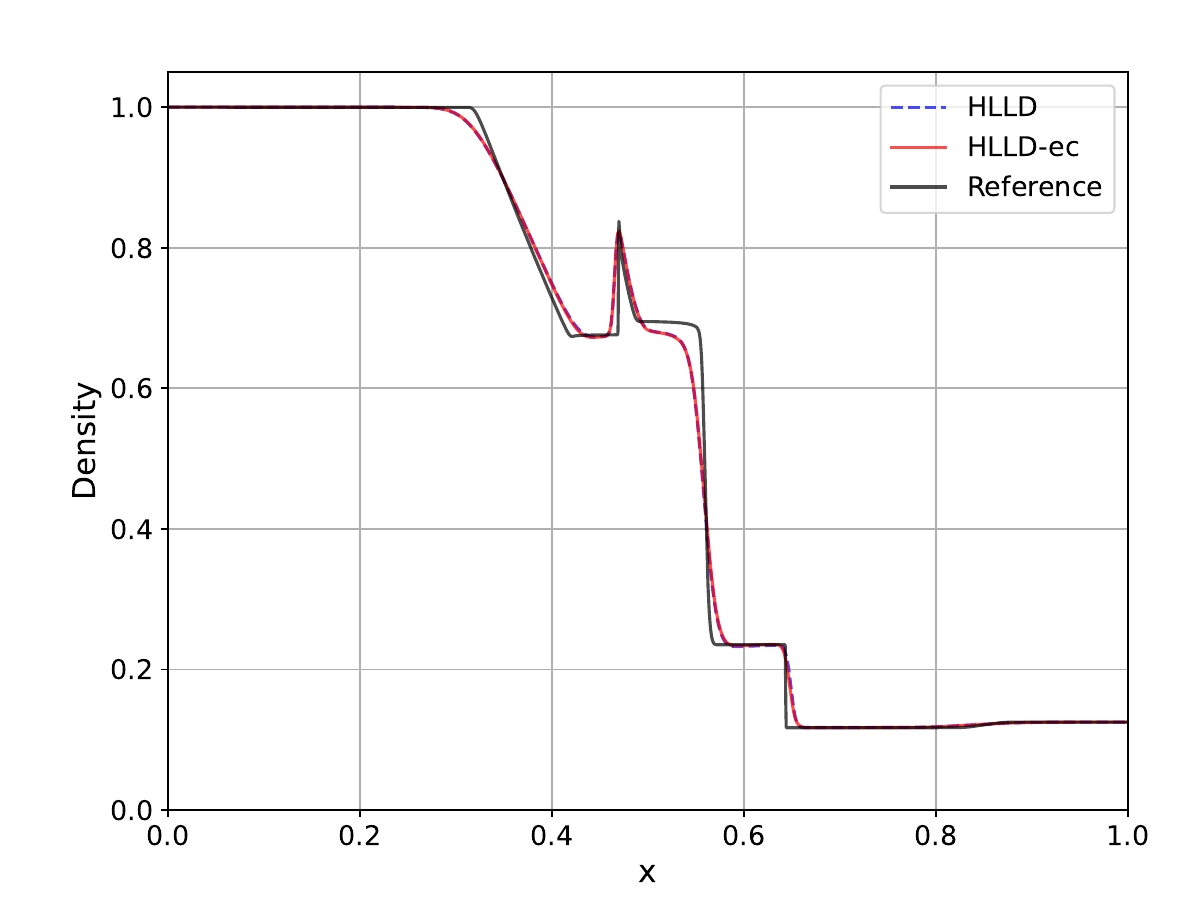}
 }
 \subfigure[\label{fig:HLLD_density_zoomin}{}]{
 \includegraphics[width=0.48\textwidth]{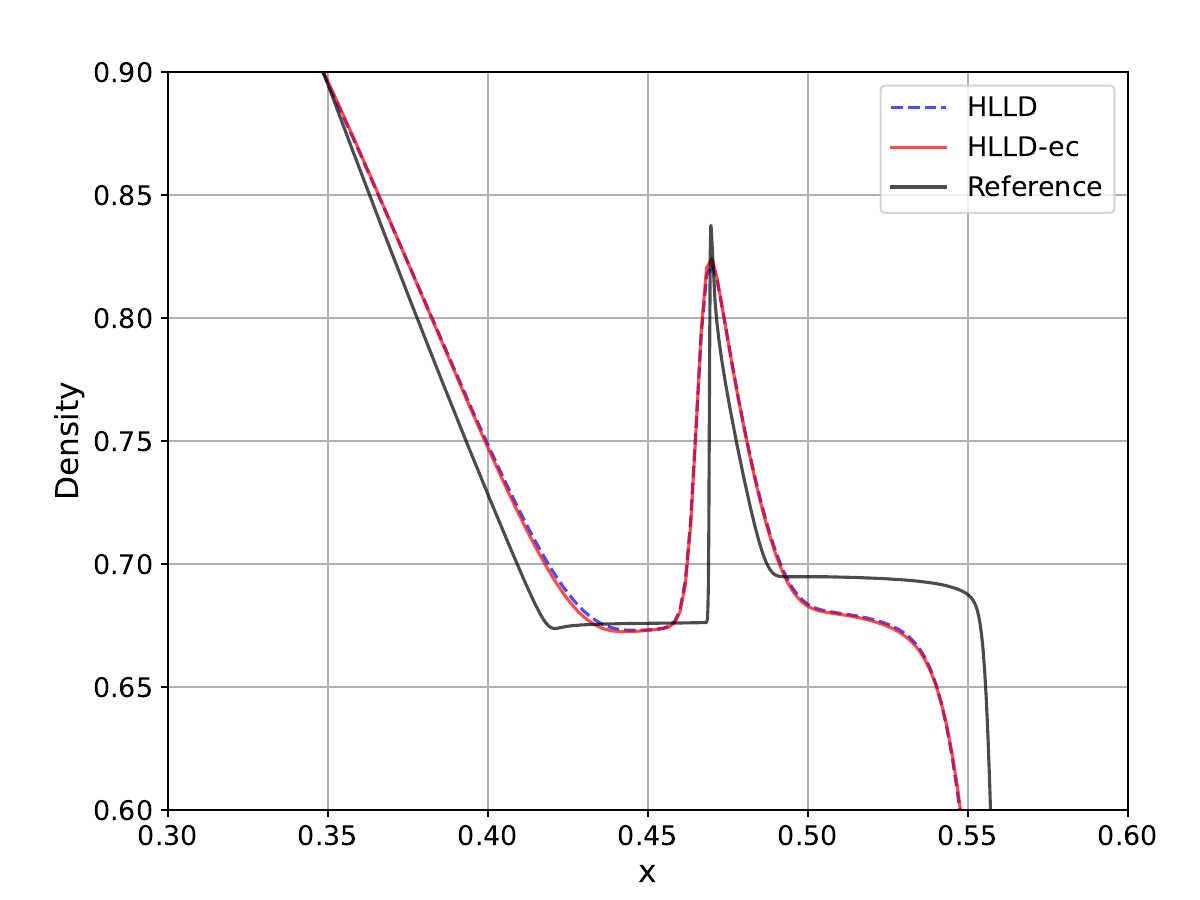}
 } 
\subfigure[\label{fig:HLLD_x_velocity}{}]{
 \includegraphics[width=0.48\textwidth]{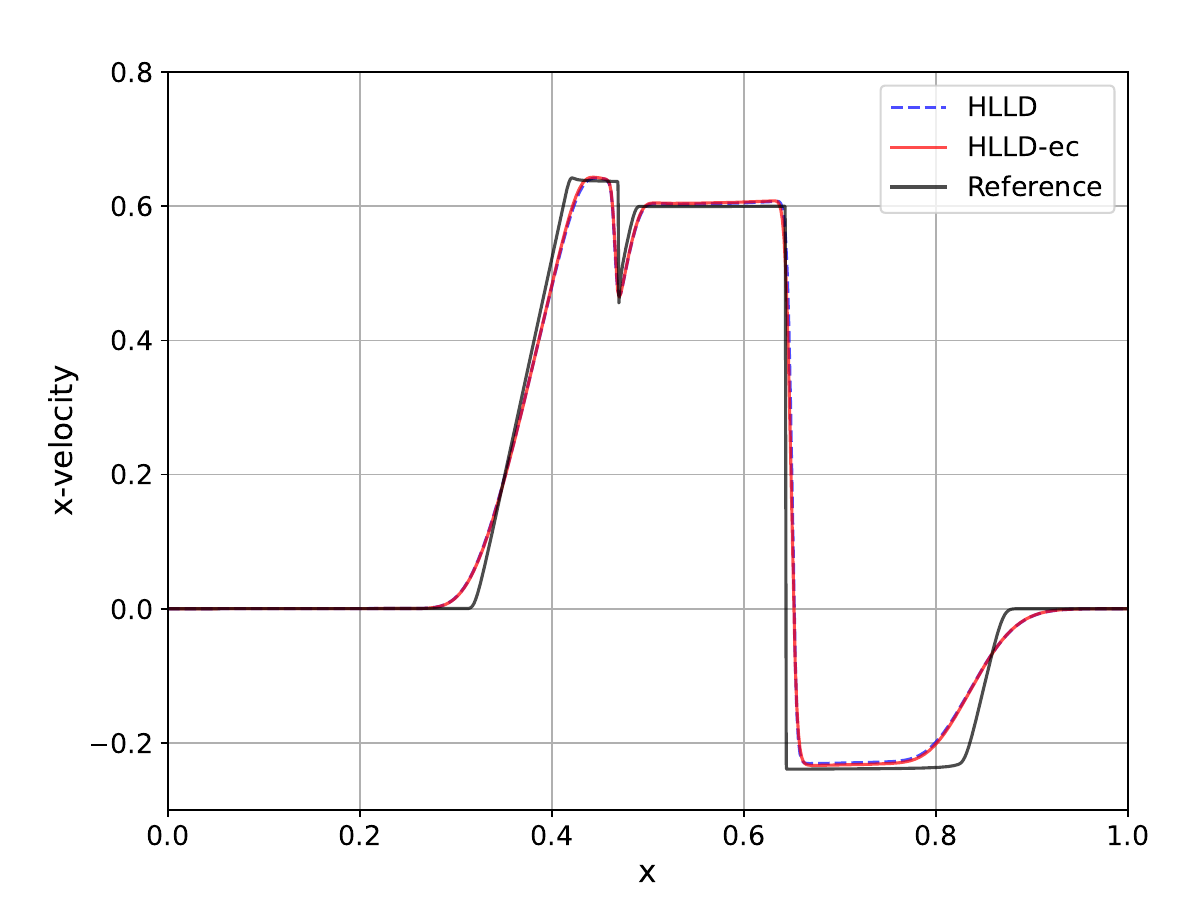}
 }
 \subfigure[\label{fig:HLLD_x_velocity_zoomin}{}]{
 \includegraphics[width=0.48\textwidth]{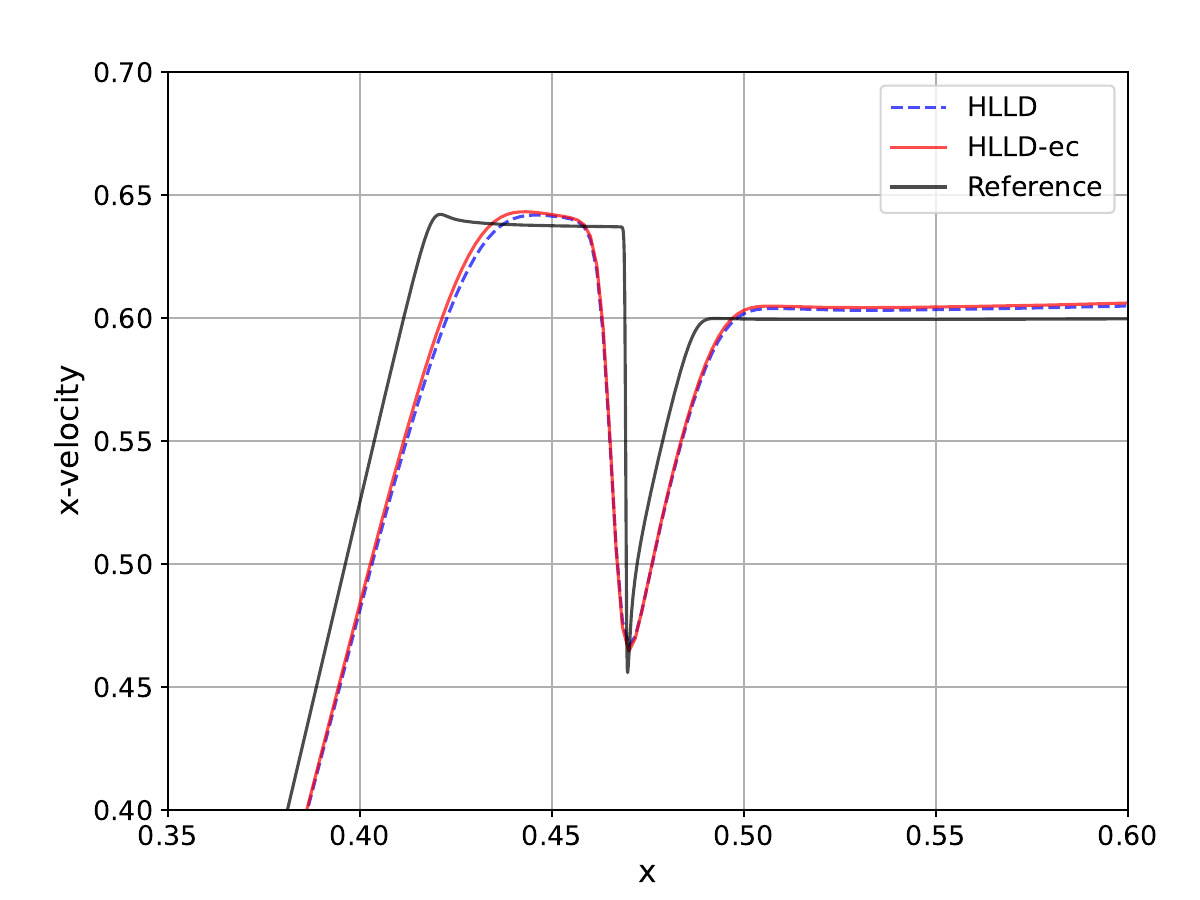}
 } 
 \caption{Density and velocity ($x$-component) distributions of HLLD-type schemes.}
 \label{fig:HLLD_BW}
\end{figure}

Figs.~\ref{fig:HLLC_BW} and \ref{fig:HLLD_BW} show the results of HLLC-type and HLLD-type schemes. In addition, Fig.~\ref{fig:HLLC_BW} also includes the results of the HLL scheme. The energy-consistent schemes generally show resolutions similar to those of the original HLLC/D schemes.
More specifically, the HLLC-ec scheme is more diffusive for the contact discontinuity, showing a resolution similar to that of the HLL scheme. {\color{red}This is expected since the HLLC-ec scheme reduces its resolution for density and momentum.} On the other hand, the HLLC-ec scheme shows the best resolution for the slow compound wave since its resolution for the magnetic field is improved. As this case has a relatively high plasma $\beta$, the density difference is more pronounced. Minor differences are observed between the results of the HLLD-type schemes. Specifically, the HLLD-ec scheme shows slightly better resolution for the rarefaction wave and the slow compound wave, but somewhat more diffusive resolution for the contact discontinuity and shock.

%{
%\begin{figure}[htbp]
% \centering
% \includegraphics[width=0.48\textwidth]{Bx.pdf}
%% \caption{Numerical errors of hyperbolic divergence cleaning.}
% \label{fig:HDC_error}
%\end{figure}
%}

Overall, we may conclude that the numerical behaviours of the present schemes are as expected when capturing isolated MHD waves under a relatively weak magnetic field.  

\subsection{One-dimensional discontinuities with varying $B_{\parallel}$}

\begin{figure}[ht]
 \centering
 \subfigure[\label{fig:IE_Bx}{}]{
 \includegraphics[width=0.48\textwidth]{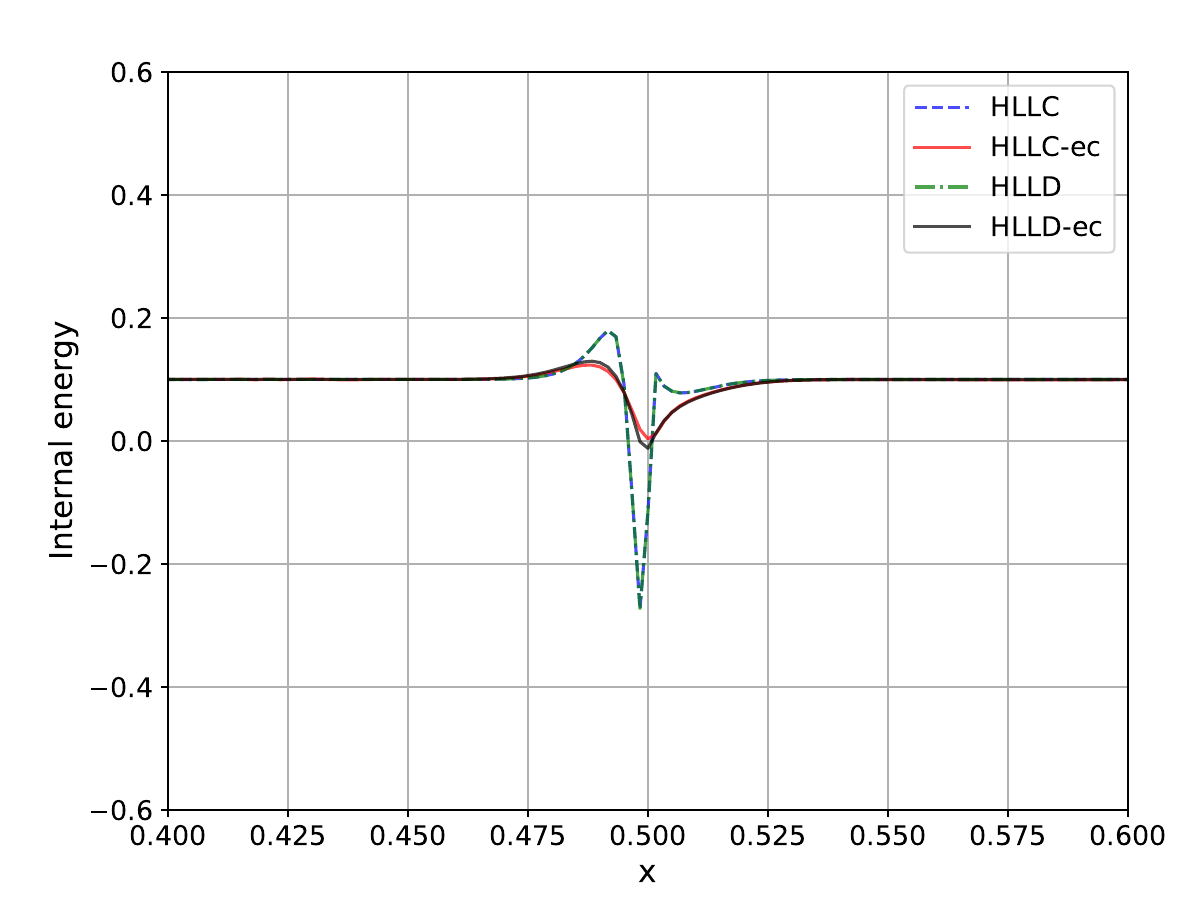}
 }
 \subfigure[\label{fig:Density_Bx}{}]{
 \includegraphics[width=0.48\textwidth]{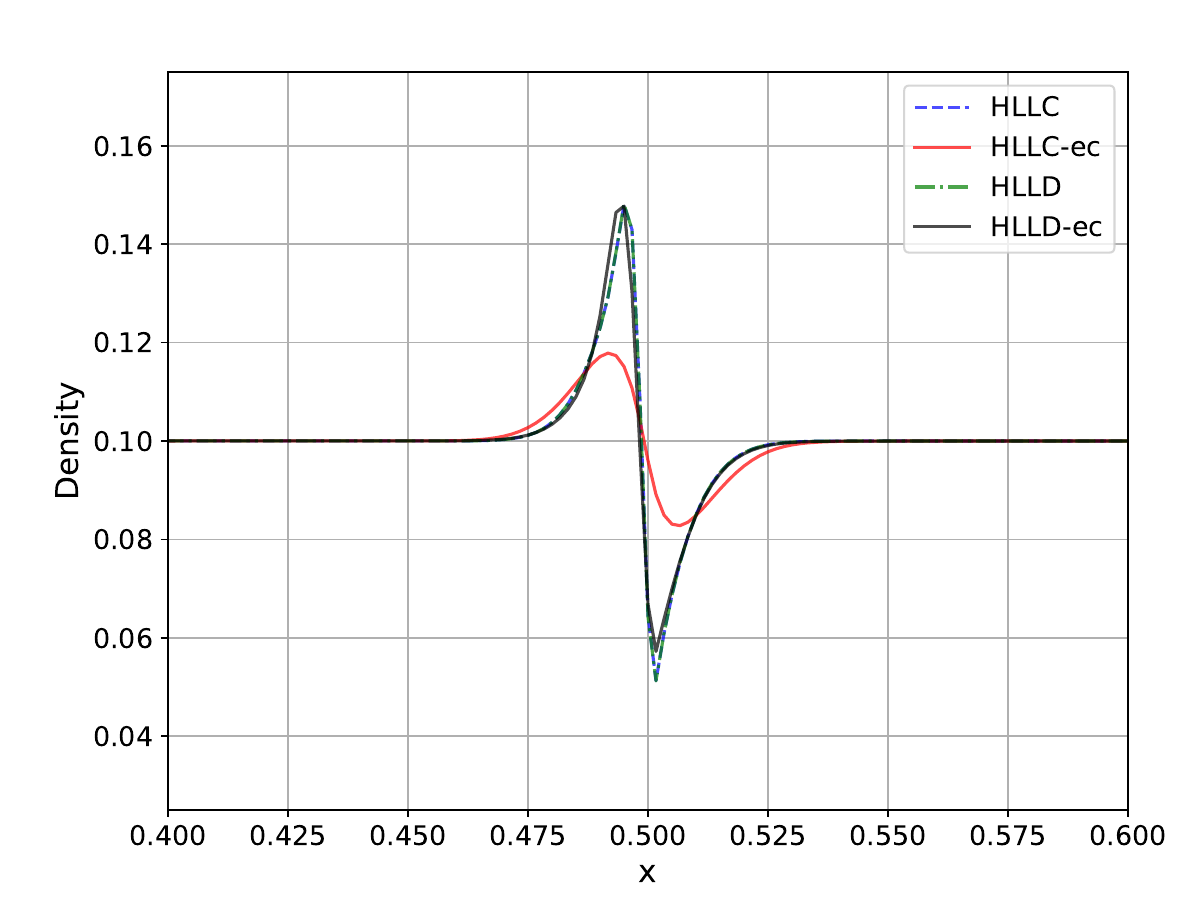}
 } 
\subfigure[\label{fig:Vy_Bx}{}]{
 \includegraphics[width=0.48\textwidth]{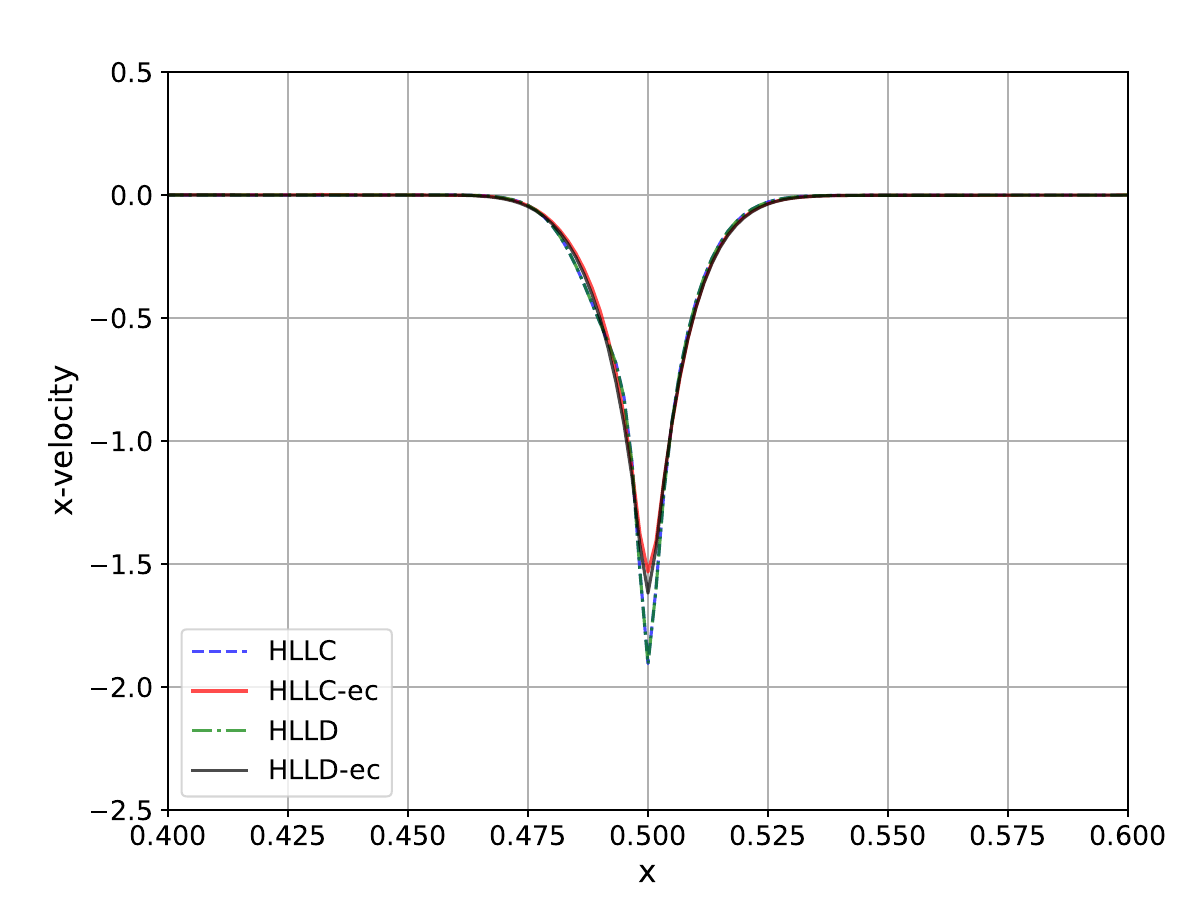}
 }
 \subfigure[\label{fig:By_Bx}{}]{
 \includegraphics[width=0.48\textwidth]{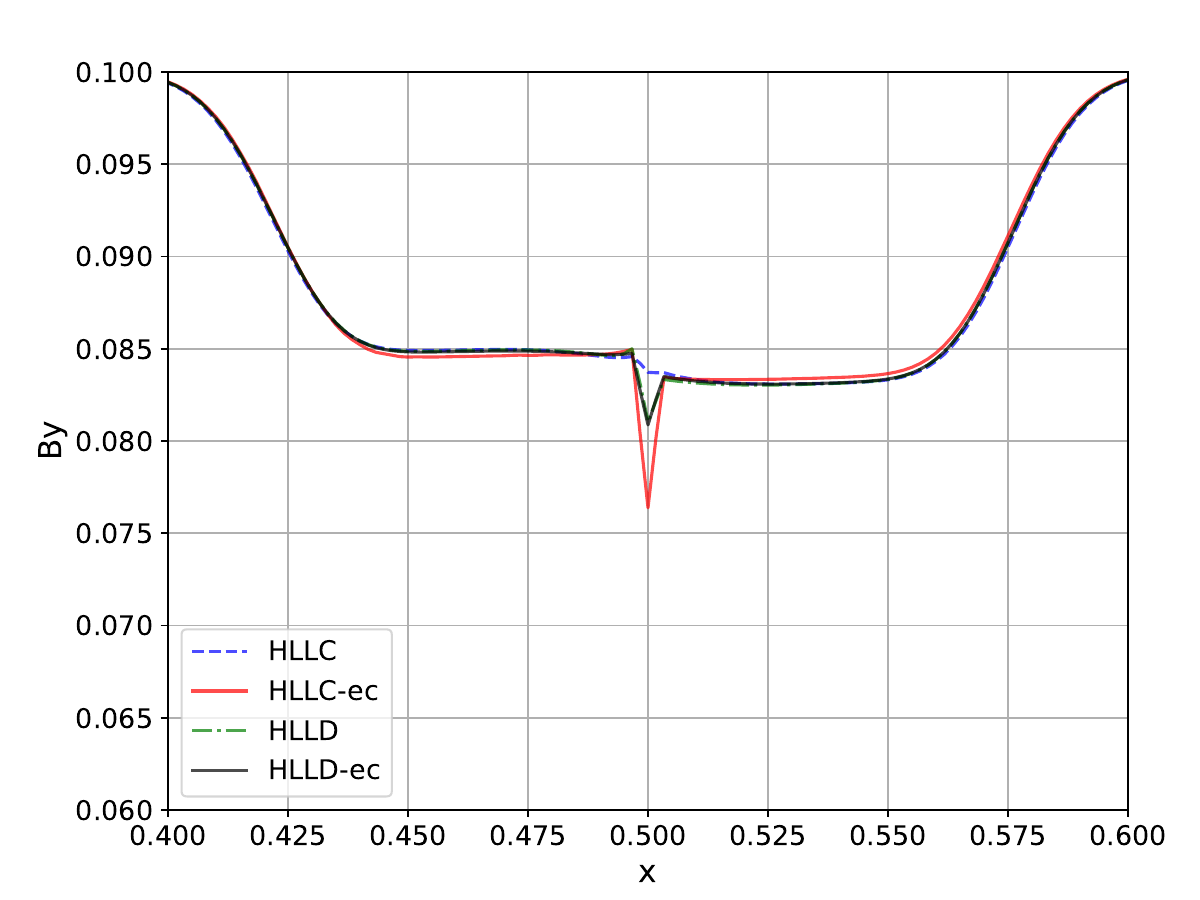}
 } 
 \caption{Results of 1D simulations with varying $B_x$.}
 \label{fig:VaryingBx}
\end{figure}

To show the performance of the present schemes with varying $B_{\parallel}$, we design the present test case. As discussed, a one-dimensional problem always has a constant $B_{\parallel}$, but this is usually not true in three-dimensional scenarios. Especially in astrophysical MHD simulations, the amplitude of the magnetic field may vary significantly simply due to areal expansion \cite{Pinto_2017}.
Therefore, varying $B_{\parallel}$ may be assumed in one-dimensional simulations to model multi-dimensional effects \cite{Sykora_2023}. 
 
To obtain numerical  behaviours that are specifically related to varying $B_{\parallel}$, we use the following initial solution for the one-dimensional problem along the $x$-axis
  \begin{eqnarray}   
  \left\{\begin{array}{ll}
 (\varrho, u, v, p, B_x, B_y)^{\text{l}}=(0.1, 0, 5, 0.1, 5, 0.1), & \text{if} \quad x<0.5, \\
 (\varrho, u, v, p, B_x, B_y)^{\text{r}}=(0.1, 0, 5, 0.1, 4.9, 0.1),& \text{if} \quad x\ge0.5, 
 \end{array}    \right.
 \end{eqnarray} 
 \noindent  and again an adiabatic index $\gamma=2$ is used. There is a relatively small discontinuity in $B_x$, but  $B_y$ is constant.  The previous mesh is reused; thus, we have $\Delta x=1/600$. Since we use an initial condition that causes numerical oscillations, only a short period of temporal evolution is sufficient for the oscillations to develop. The results at $t=0.005$ are shown in Fig.~\ref{fig:VaryingBx}. All simulations use the same constant time step $\Delta t=10^{-5}$ {\color{red}(thus CFL $\simeq 0.1$)}. Note that $B_x$ remains the initial value during the simulations, as the HDC is not activated.

 As discussed in subsection \ref{sec:varyingBx}, 
 tangential velocity components are essential in causing negative pressure, even if they are continuous. 
 More importantly, tangential velocity components do not contribute to the values of longitudinal eigenwave speeds. Thus, numerical methods cannot adjust the numerical diffusion or the time step to account for a more significant tangential velocity. Therefore, a relatively sizeable tangential velocity component is imposed to challenge the numerical schemes. However, the resulting kinetic energy is still significantly smaller than the magnetic energy and is one order of magnitude larger than the internal energy.  
The minimum plasma $\beta$ is below $0.01$.

In general, Fig.~\ref{fig:VaryingBx} shows that the proposed HLLC/D-ec schemes are less prone to producing oscillations compared to the original HLLC/D schemes.  More importantly, as shown in Fig.~\ref{fig:IE_Bx}, the HLLC/D schemes have already caused more significant negative internal energy, and the computations can continue only because the total energy is still positive. The HLLC-ec scheme produces a smoother density distribution but a more significant magnetic field variation, as expected. We should note that the oscillations of the HLLC/D-ec schemes will continue to develop and eventually crash the simulations; however, we focus on the relative differences/advantages in the results. Note that a constant $B_x$ or a zero $v$ can remove such oscillations. Therefore, we may assume that mesh refinement in regions where the amplitude of the magnetic field component along field lines varies significantly would be beneficial for preserving positivity, even if all the physical variables are smooth.

Since $B_x$ is not constant, there is no exact Riemann solution to this problem. It is also difficult to quantify the specific contribution of each algorithmic change. Still, the results suggest that the present energy-consistent schemes are significantly more robust in this non-ideal scenario with changing $B_x$ (which may result from numerical error or physical variation).  
 
\subsection{One-dimensional false wave structures  with constant  $B_{\parallel}$} \label{sec:false}
This subsection introduces a one-dimensional test case with constant $B_{\parallel}$, again along the $x$ direction. This problem is initialised with low plasma $\beta$, and the amplitude of the longitudinal magnetic field component is significantly larger than that of the tangential components. Specifically, we have 
 \begin{eqnarray}   
  \left\{\begin{array}{ll}
 (\varrho, u, v, p, B_x, B_y)^{\text{l}}=(0.1, 0, 0, 0.1, 5, 0.1), & \text{if} \quad x<0.5, \\
 (\varrho, u, v, p, B_x, B_y)^{\text{r}}=(0.1, 0, 0, 0.1, 5, -0.1),& \text{if} \quad x\ge0.5, 
 \end{array}    \right.
 \end{eqnarray} 
\noindent and again an adiabatic index $\gamma=2$ is used. Therefore, the initial condition has both constant plasma thermal and magnetic pressure. The only initial discontinuity is in the tangential magnetic field component $B_y$. This small discontinuity should not cause strong nonlinearity.

\begin{figure}[ht]
 \centering
 \subfigure[\label{fig:density_velocity}{}]{
 \includegraphics[width=0.48\textwidth]{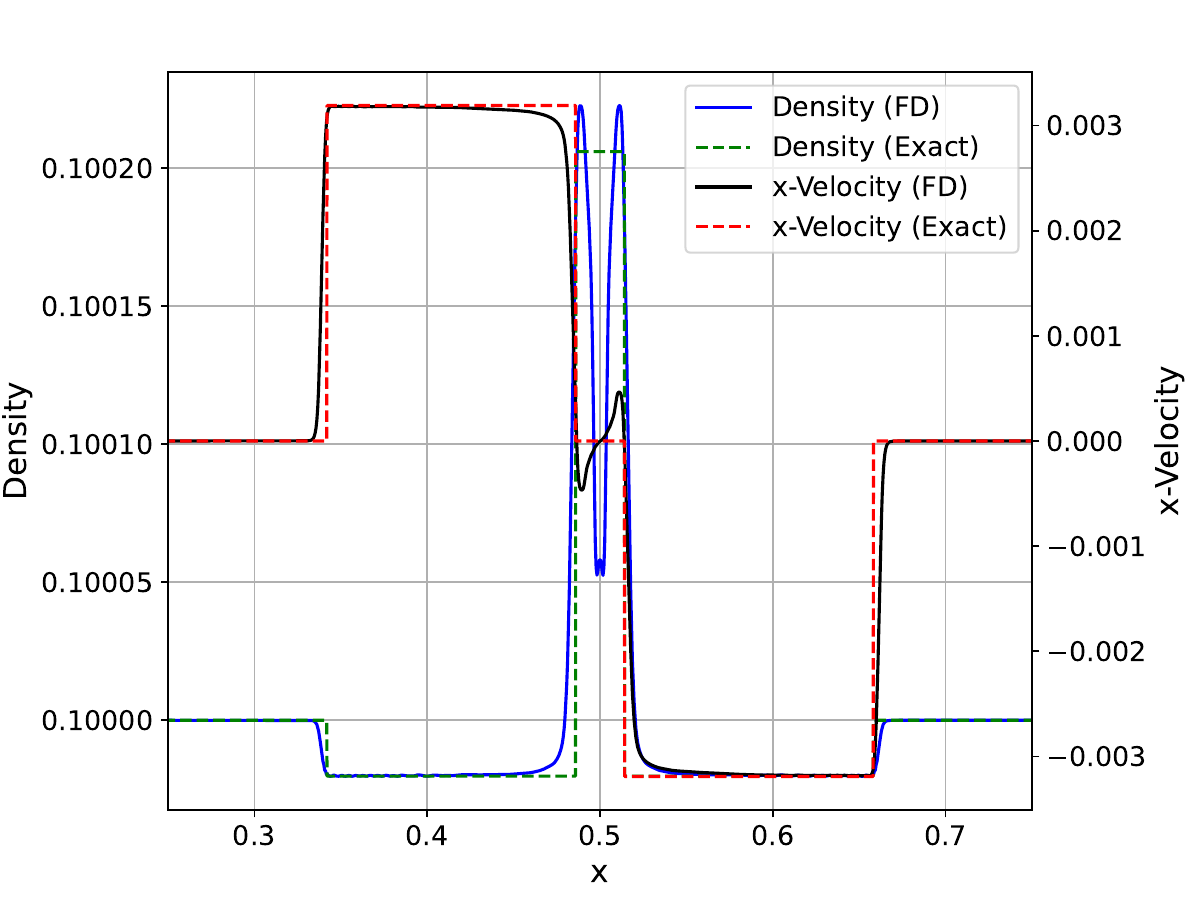}
 }
 \subfigure[\label{fig:pressure_By}{}]{
 \includegraphics[width=0.48\textwidth]{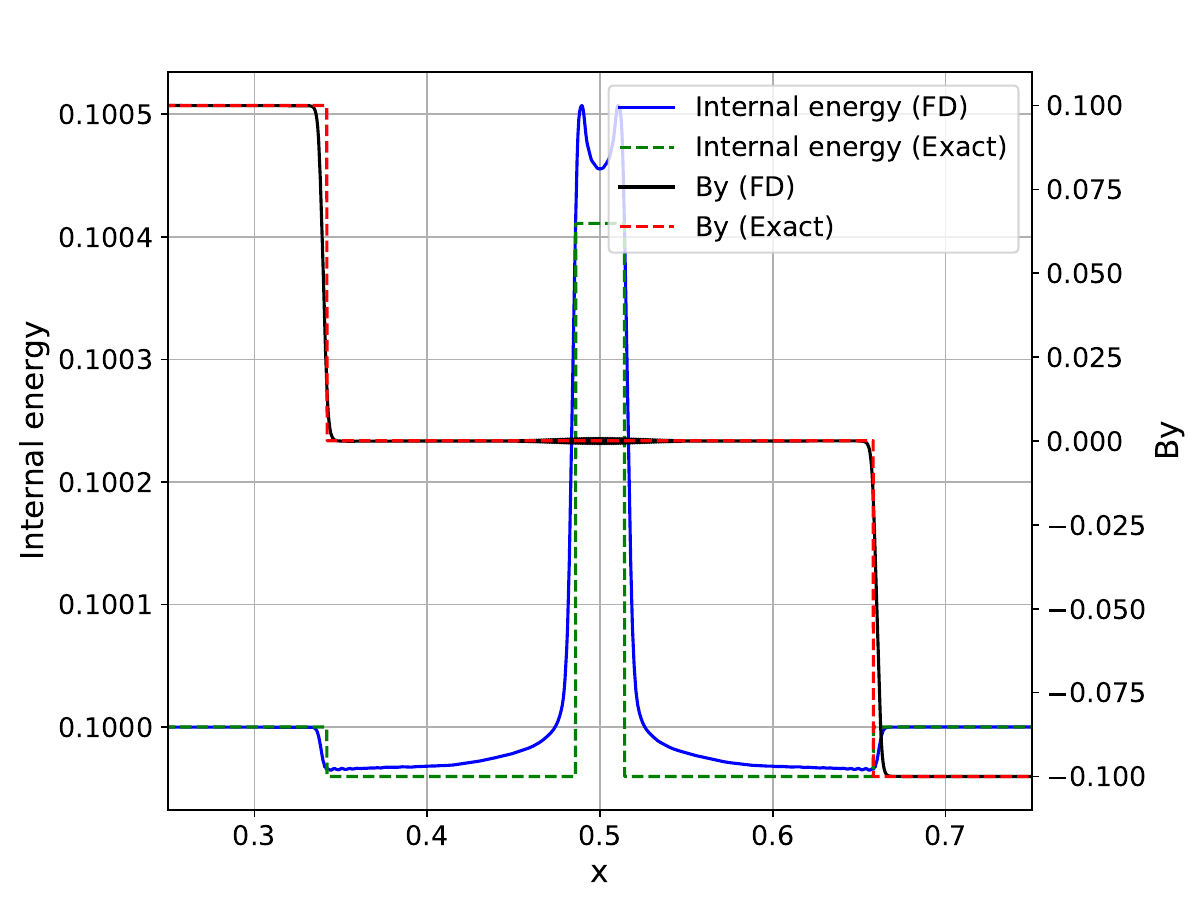}
 }  
 \caption{Reference solutions at $t=0.01$,  given by a high-order FD solver and an exact Riemann solver.}
 \label{fig:Exact_FD}
\end{figure}

As references, an exact solution given by an exact MHD Riemann solver \cite{Xu_2024} and a high-resolution numerical solution (2000 grid points) provided by a sixth-order FD solver with hyper-diffusivity \cite{Sykora_2023} are both shown in Fig.~\ref{fig:Exact_FD}. The FD  solver uses internal energy as the variable when solving the energy conservation equation. Since magnetic energy is not directly involved, there is no inconsistency between the calculated magnetic field and magnetic energy. It is well known that artificial diffusivity may need to be adjusted to suppress numerical oscillations adequately. Indeed, small oscillations are observed in both internal energy and $B_y$ in the FD solution. However, such oscillations are not further discussed, as they relate only to the FD method. A significant numerical phenomenon is the density dip at approximately $x=0.5$ in Fig.~\ref{fig:density_velocity}, which is not associated with any physical eigenwave structure and whose amplitude remains unchanged with a finer mesh. The FD solution generally agrees with the exact solution, and the exact wave structures are also relatively weak. For example, the relative density or pressure variation is more minor than $1\%$.

\begin{figure}[ht]
 \centering
 \subfigure[\label{fig:HLLCBxCDensity}{}]{
 \includegraphics[width=0.48\textwidth]{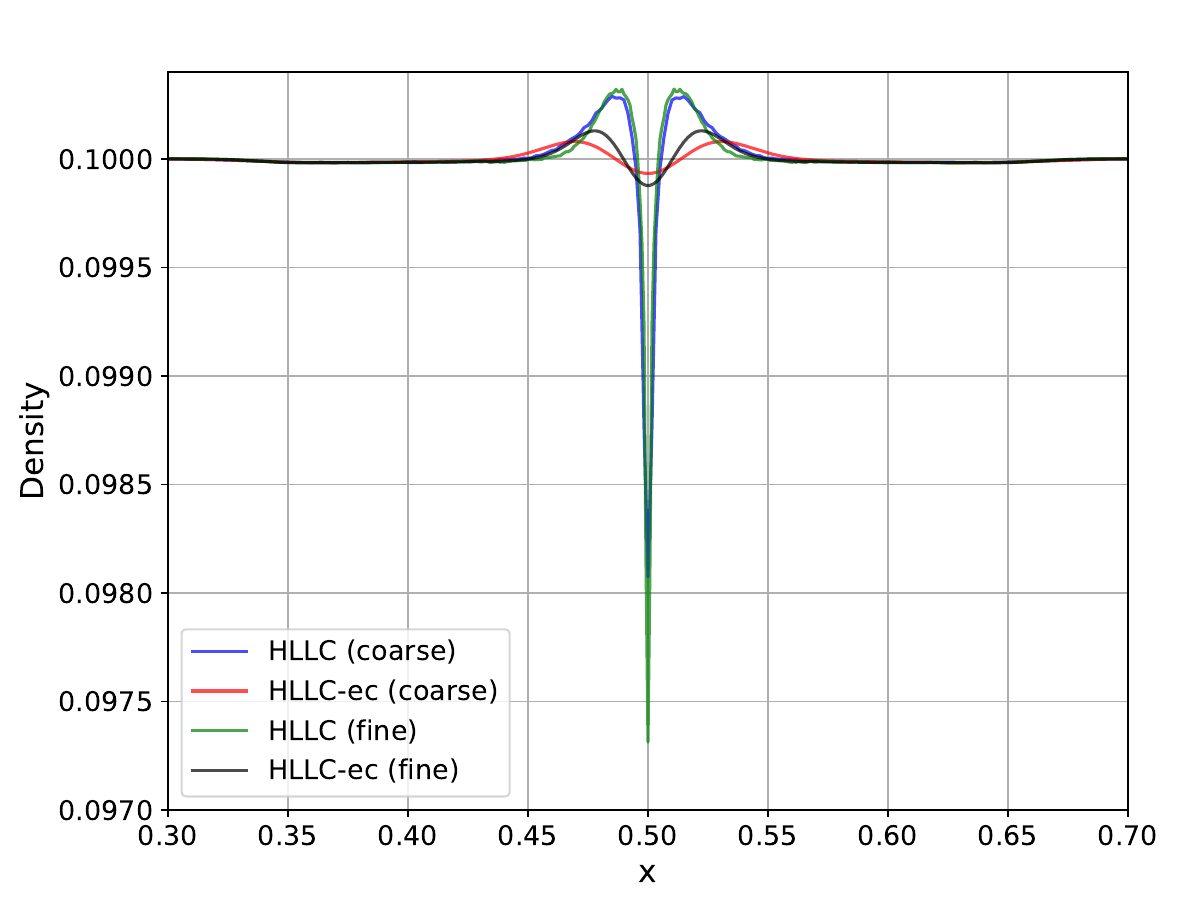}
 }
 \subfigure[\label{fig:HLLCBxCEnergy}{}]{
 \includegraphics[width=0.48\textwidth]{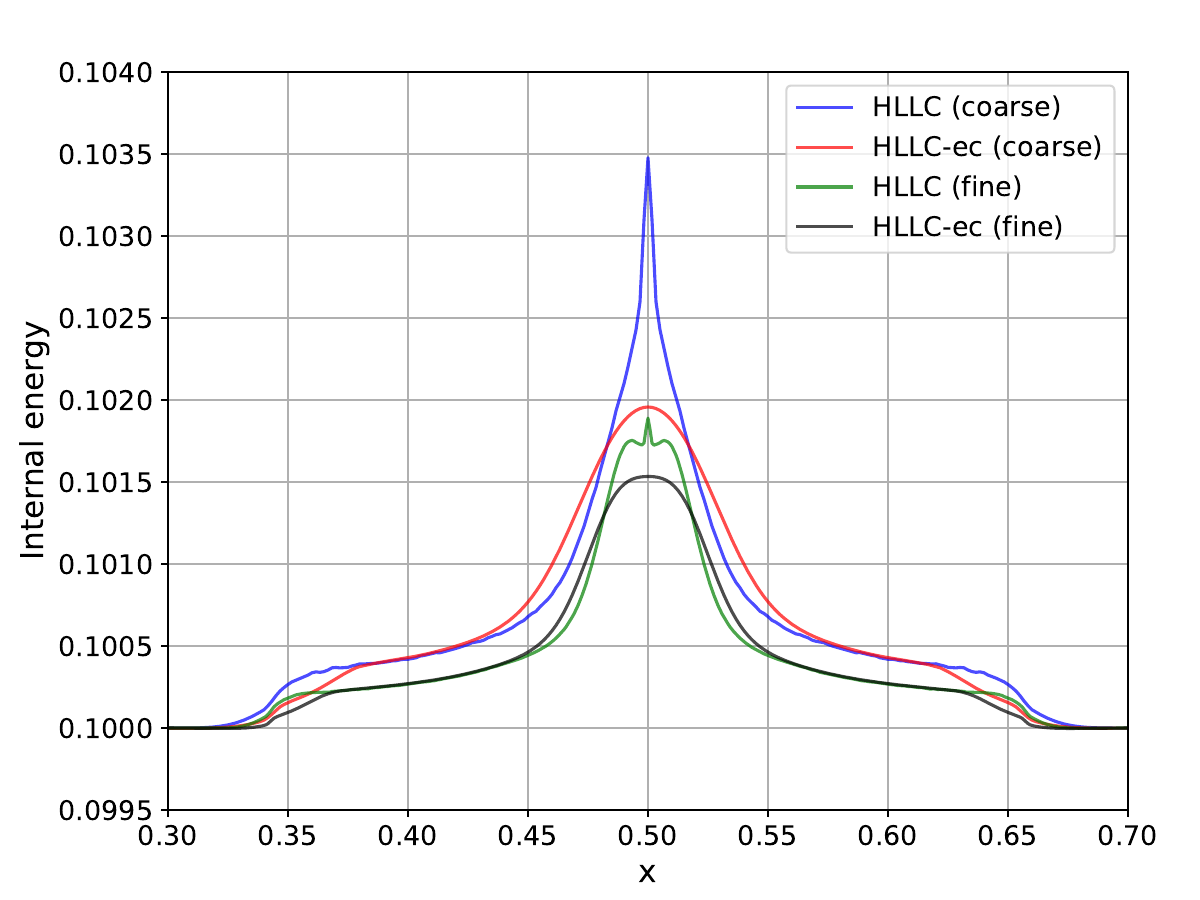}
 }  
 \caption{Numerical solutions at $t=0.01$, given by the HLLC-type schemes on two different meshes.}
 \label{fig:HLLCBxC}
\end{figure}

\begin{figure}[ht]
 \centering
 \subfigure[\label{fig:HLLDBxCDensity}{}]{
 \includegraphics[width=0.48\textwidth]{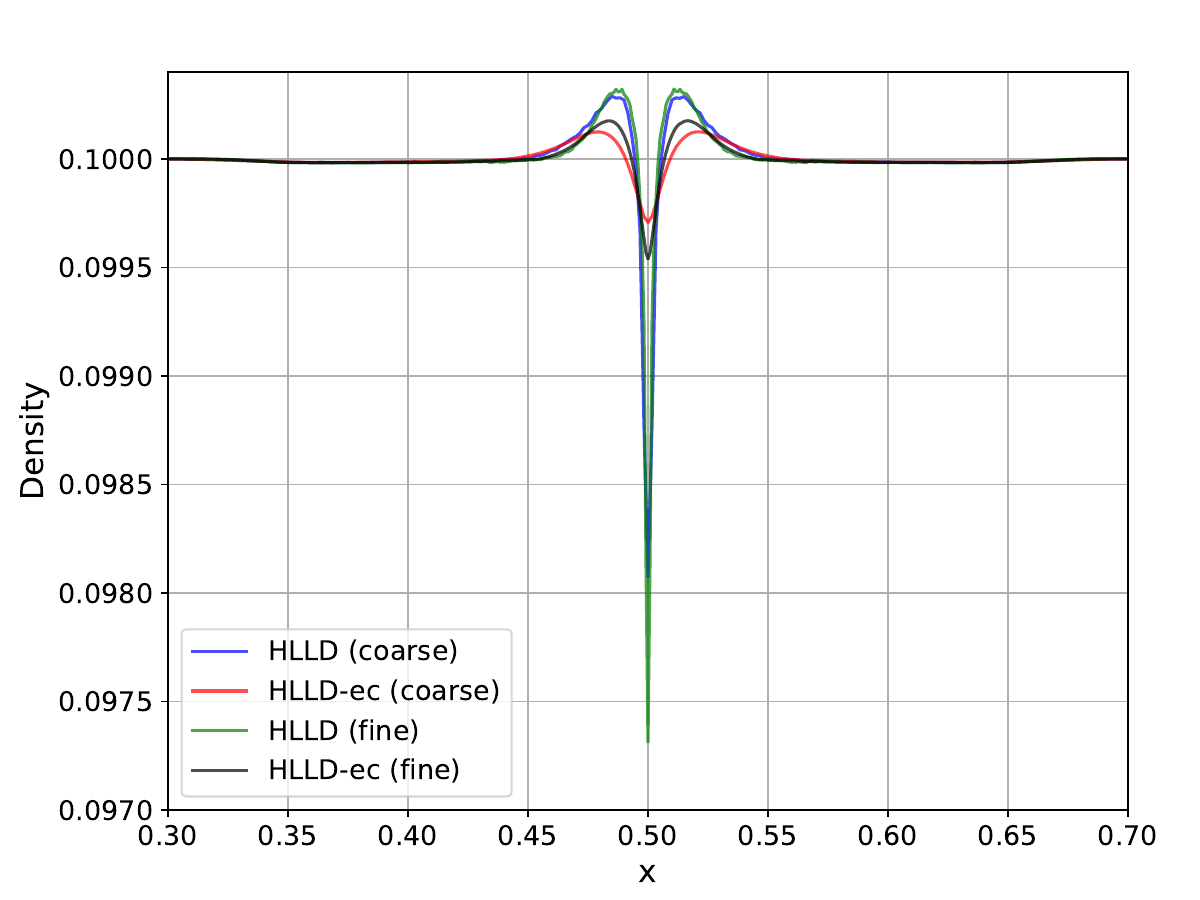}
 }
 \subfigure[\label{fig:HLLDBxCEnergy}{}]{
 \includegraphics[width=0.48\textwidth]{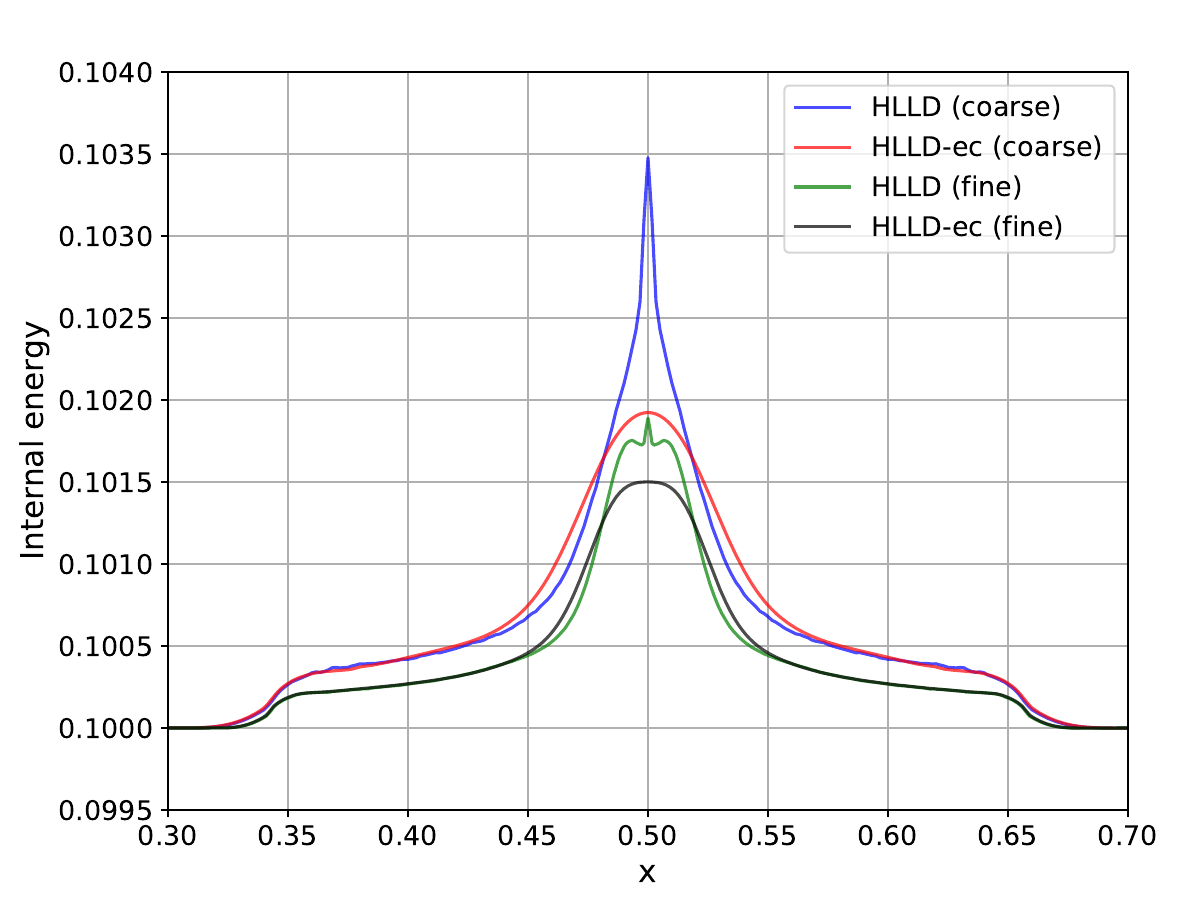}
 }  
 \caption{Numerical solutions at $t=0.01$, obtained using the HLLD-type schemes on two different meshes.}
 \label{fig:HLLDBxC}
\end{figure}

Numerical solutions of the HLLC and HLLD-type schemes are shown in Figs.~\ref{fig:HLLCBxC} and \ref{fig:HLLDBxC}, respectively. Two meshes are used. The coarse and fine meshes, respectively, have 600 and 1200 cells along the $x$-direction. Correspondingly, the implicit time step used for the coarse mesh is $\Delta t= 0.0002$, and the one used for the fine mesh is $\Delta t= 0.0001$ {\color{red}(thus CFL $\simeq 1.9$)}. Note again that only the first-order FV discretisation is used.  The FV results are also confirmed by the previously mentioned explicit FV MHD solver.

In the present numerical solutions, two phenomena are particularly worth attention. Firstly, the HLLC and HLLD schemes exhibit relatively more substantial density dips, particularly on the finer mesh. Secondly, all the Riemann solvers exhibit internal energy increments behind the fast waves, which is the opposite of the FD solution, where internal energy rightfully decreases along with density. The density dips caused by the HLLC/D-ec schemes are significantly weaker. It is less interesting regarding the HLLC-ec scheme, since it is more diffusive in density space. In realistic simulations, the HLLD-ec scheme also shows only a relatively slight density dip, while its high resolution has already been demonstrated. These two results suggest that numerical resolution is not the essential reason for this phenomenon. Regarding the second phenomenon, we believe that the internal energy increments are caused by the erroneous thermal pressure resulting from excessive entropy, because the thermal pressure is calculated by subtracting the erroneous magnetic pressure from the total pressure, as discussed in Section \ref{sec:How}. More essentially, this may be explained by the fact that Riemann solvers may produce entropy even for isentropic processes.
In this specific case, refining the mesh shows that such increments decrease, potentially due to lower numerical diffusion. Overall, all Riemann solvers tested exhibit clear errors compared to the exact solution; however, this numerical test case focuses on the relative differences between the numerical results.

The oscillation amplitudes shown here are relatively small. However, this may change in more complex scenarios, and some issues here are already of concern. Firstly, the simulations here (except the reference numerical simulation) use first-order FV discretisation, and thus, when higher-order schemes are used, the effects may be more complicated. It is well known that monotonicity is also crucial for numerical stability, and small-scale high-frequency oscillations can interact with nonlinear discontinuities, further complicating the preservation of positivity and monotonicity \cite{Zhao2019,Zhang2024}.  Secondly, the oscillations do \textit{not necessarily} decrease with higher spatial and temporal resolutions, which relates to the fact that the error introduced in subsection \ref{sec:How} depends on the resolutions and the relative change of the magnetic field. This issue may thus introduce additional complexity into realistic simulations.
Lastly, the false wave structures, albeit small, may affect small-scale flow phenomena. 

Overall, although the present schemes cannot remove all the issues mentioned, they can significantly reduce the false oscillations compared to the original HLLC/D schemes. {\color{red}The results also indicate that simply ensuring the positivity of scalar variables does not guarantee correct numerical results.}

\begin{figure}[!ht]
 \centering
 \subfigure[\label{fig:HLLC_Rotor}{HLLC}]{
 \includegraphics[width=0.48\textwidth]{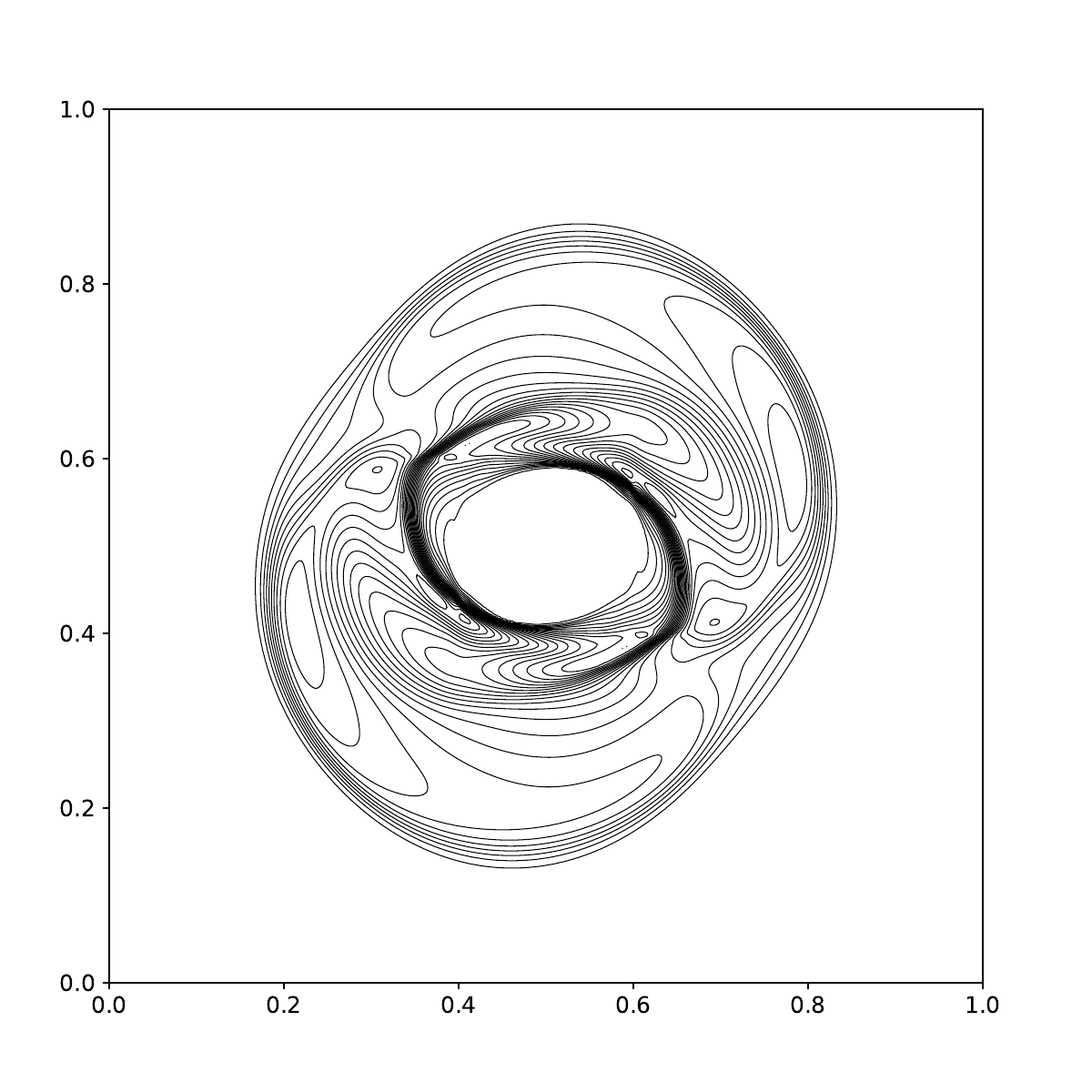}
 }
 \subfigure[\label{fig:HLLCPC_Rotor}{HLLC-ec}]{
 \includegraphics[width=0.48\textwidth]{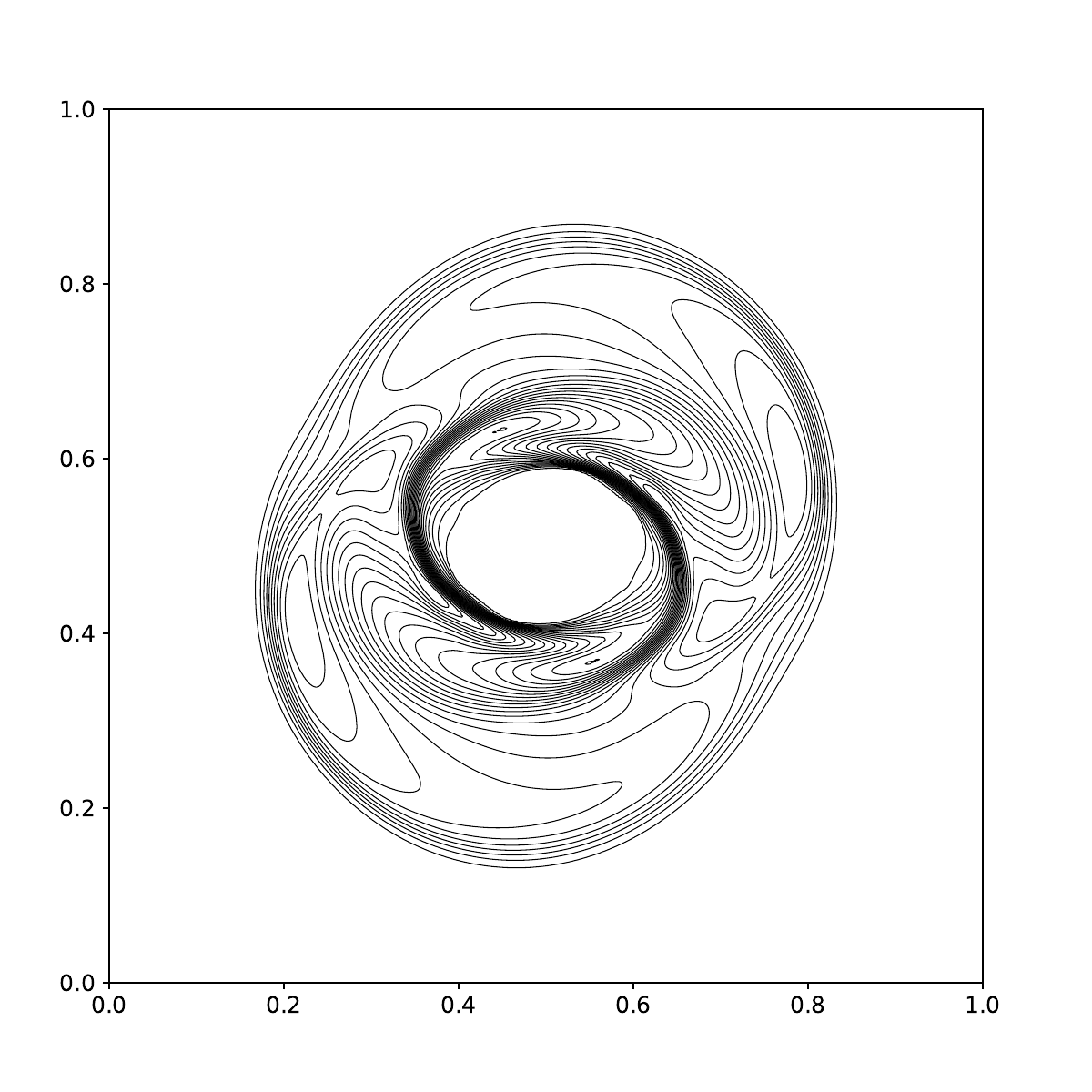}
 }  
  \subfigure[\label{fig:HLLD_Rotor}{HLLD}]{
 \includegraphics[width=0.48\textwidth]{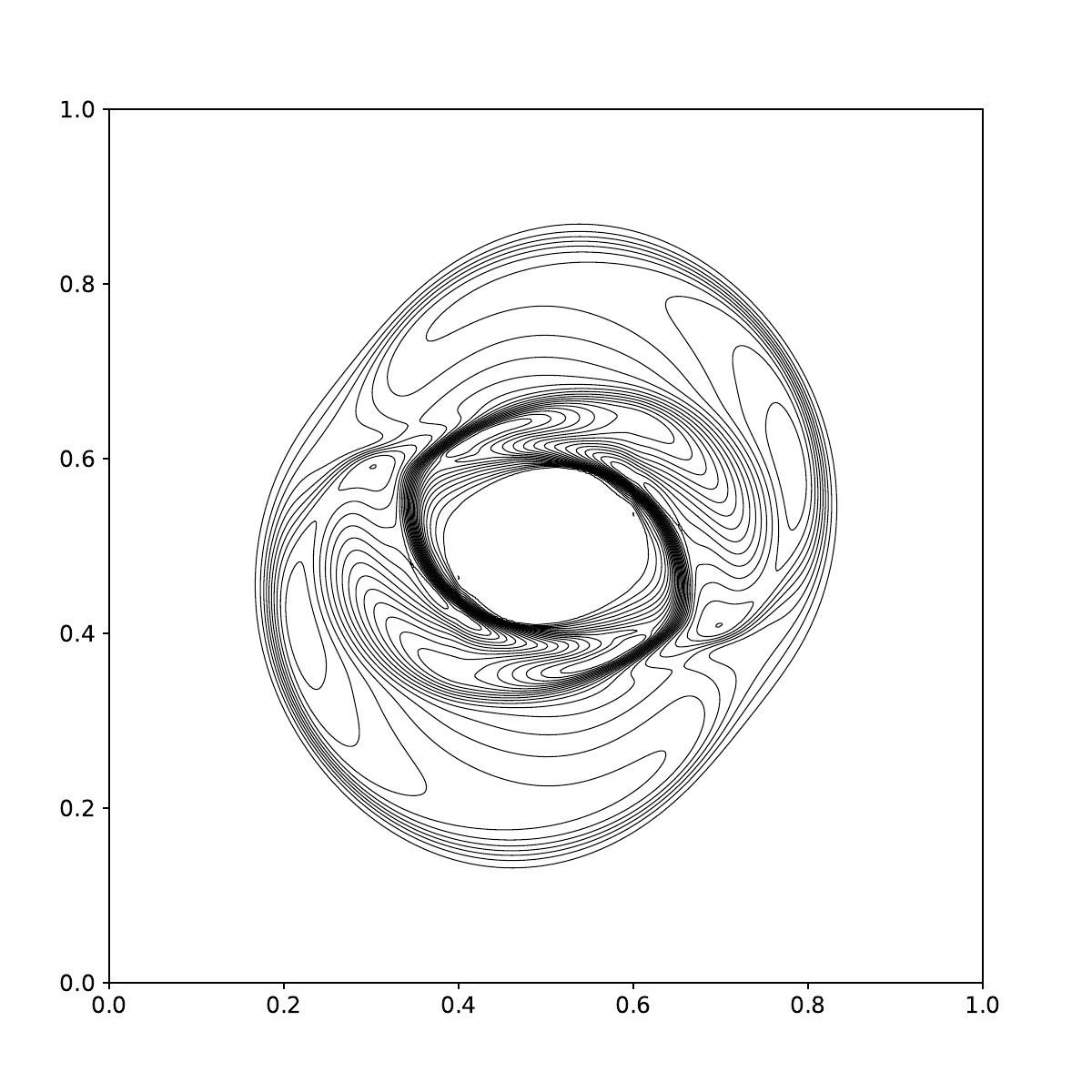}
 }
 \subfigure[\label{fig:HLLDPC_Rotor}{HLLD-ec}]{
 \includegraphics[width=0.48\textwidth]{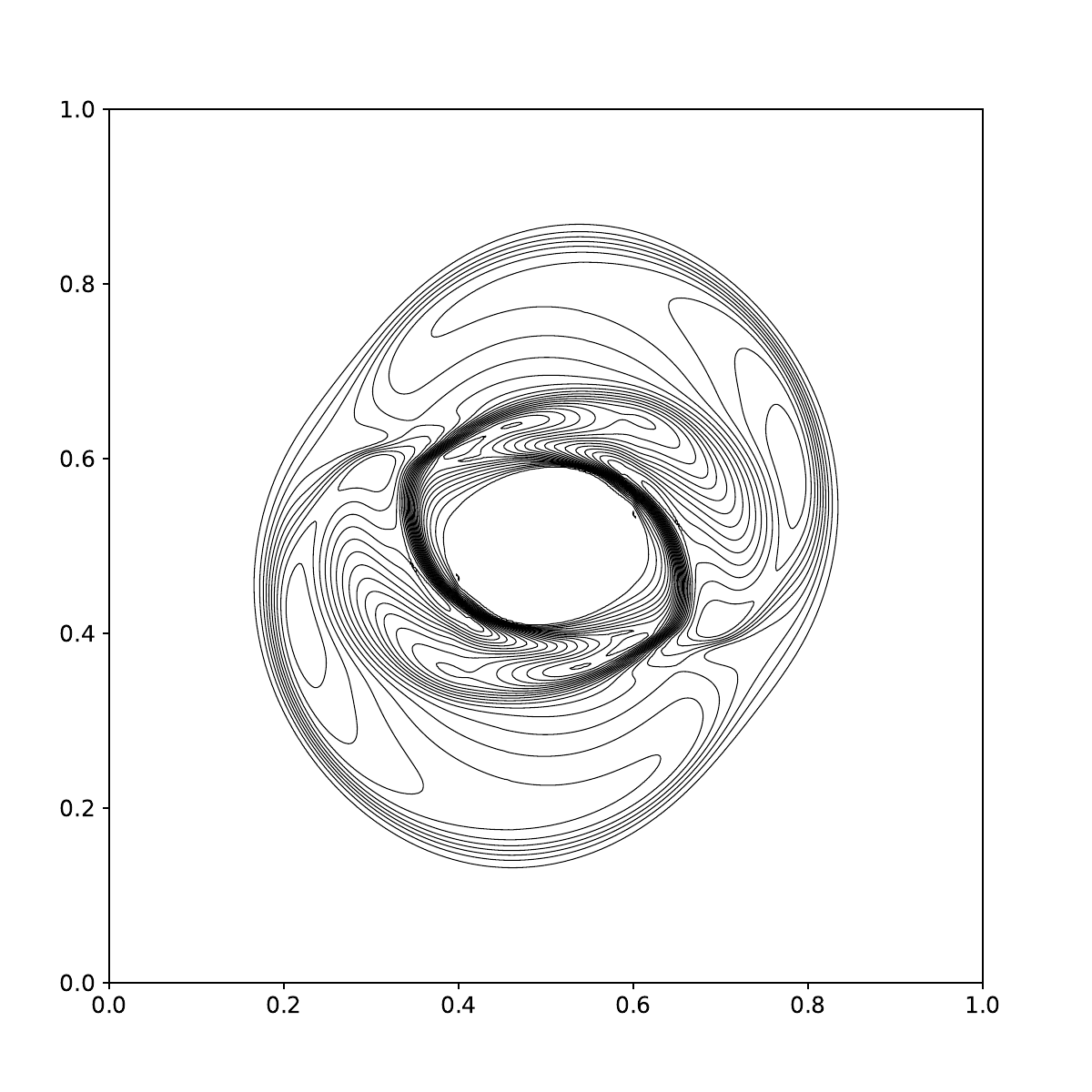}
 }  
 \caption{Internal energy at $t=0.2$ of the Rotor problem. The 30 contour lines are shown
for the range $\varrho e \in[0, 1.2]$.}
 \label{fig:Rotor}
\end{figure}

\subsection{The Rotor problem}

The two-dimensional Rotor problem was first introduced in Ref.~\cite{BALSARA1999}.  The computational domain is  a $[0, 1]\times [0,1]$ area, discretised by $400\times400$ cells.  The initial magnetic field and thermal pressure are both constant, which respectively are $\mathbf{B}=(B_x, 0, 0)^{\text{T}}$, where $B_x=2.5/\sqrt{4\pi}$, and $p=0.5$. The initial  ambient fluid at rest has $\varrho=1$ and $\mathbf{V}=\mathbf{0}$, for $r>0.115$, where $r=[(x-0.5)^2+(y-0.5)^2]^{1/2}$.  For the region where $r<r_0=0.1$, the initial status is $\varrho=10$, $u=-v_0(y-0.5)/r_0$, and $v=v_0(x-0.5)/r_0$, where $v_0=1$. For the region where $0.1<r<0.115$, the fluid quantities linearly vary, described by $\varrho=1+9f$, $u=-fv_0(y-0.5)/r_0$, and $v=fv_0(x-0.5)/r_0$, where $f=(0.115-r)/0.015$. The adiabatic index used here is $\gamma=5/3$.

Numerical solutions at $t=0.2$ are shown in Fig.~\ref{fig:Rotor}. The computations again use the first-order FV approximation, with $\Delta t = 0.001$ for the fully implicit time integration {\color{red}(thus CFL $\simeq 0.8$)}. The HDC is not imposed here to challenge the schemes, {\color{red} which has little effect on the (first-order) variables in the numerical results}. {\color{green}Another reason we do not use the HDC is that, as mentioned previously, Eq.~(\ref{eq:Bn}) is not equivalent to Eq.~(\ref{eq:divB}), and thus we try to minimise the contribution of the HDC.} While the results of the two HLLD-type schemes are sharper, the HLLC-ec scheme produces a more diffusive distribution due to the reasons above. Here, we only show the internal energy; the same conclusion can be drawn based on other quantities, which are therefore not shown.  

In the Brio-Wu shock-tube test, which is a one-dimensional problem simulated using a two-dimensional mesh, we have found that when the HDC is not imposed, the HLLD scheme may produce a significantly higher $B_x$ error in the low plasma $\beta$ area (equivalent to $\nabla \cdot\mathbf{B}$ error in a one-dimensional case, not shown), in comparison to the HLLC/D-ec schemes. However, in the present Rotor test, the $\nabla\cdot\mathbf{B}$ error of the HLLD scheme is slightly smaller than that of the HLLD-ec scheme. While the difference is slight, the original HLLD scheme may more accurately approximate fluid quantities, which dominate the solutions in this (relatively) high plasma $\beta$ case. Therefore, we introduce the following test case with a lower plasma $\beta$.

\subsection{A low plasma $\beta$ Rotor problem} \label{sec:lowbetarotor}

\begin{figure}[!ht]
 \centering
 \subfigure[\label{fig:HLLC_beta}{HLLC}]{
 \includegraphics[width=0.48\textwidth]{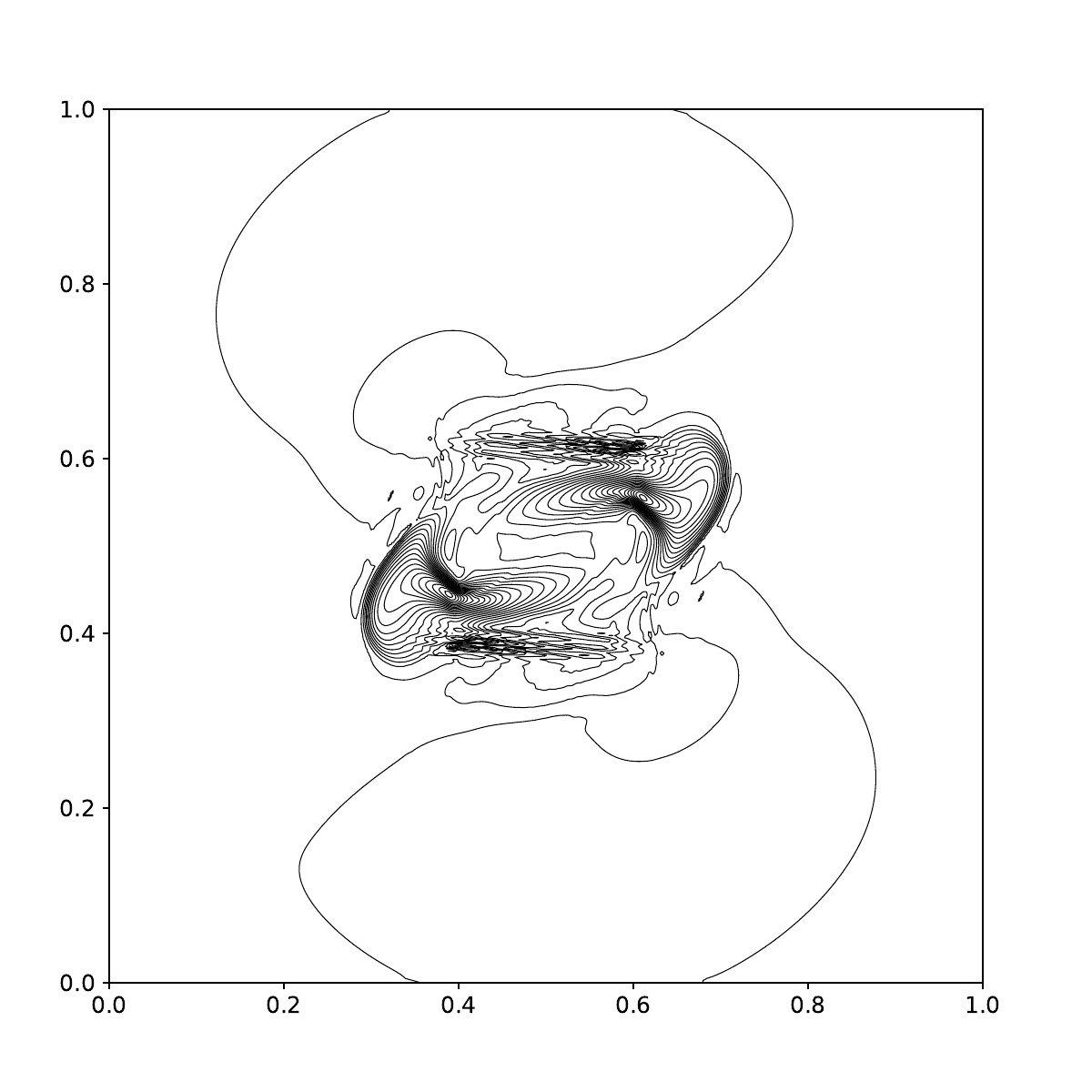}
 }
 \subfigure[\label{fig:HLLCPC_beta}{HLLC-ec}]{
 \includegraphics[width=0.48\textwidth]{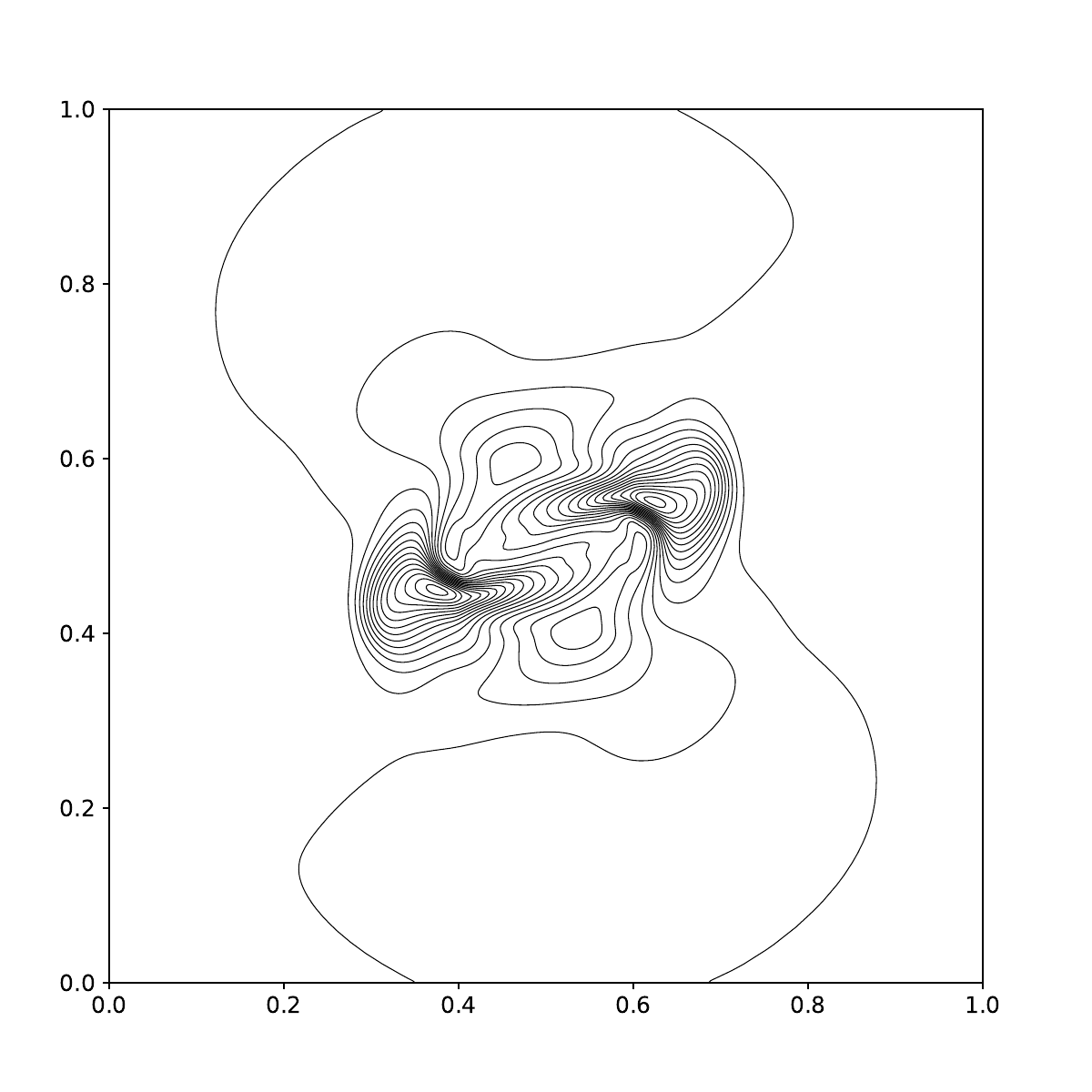}
 }  
  \subfigure[\label{fig:HLLD_beta}{HLLD}]{
 \includegraphics[width=0.48\textwidth]{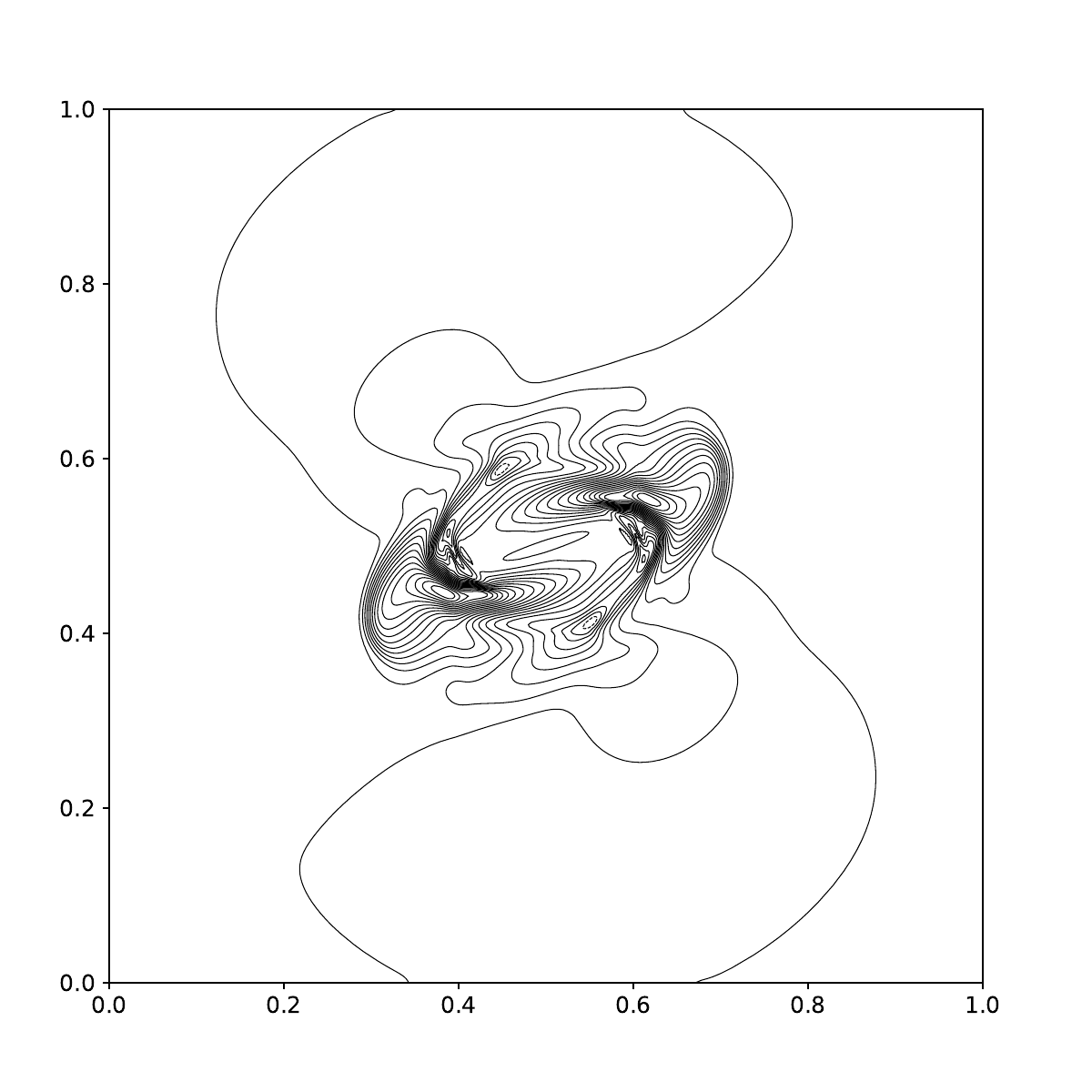}
 }
 \subfigure[\label{fig:HLLDPC_beta}{HLLD-ec}]{
 \includegraphics[width=0.48\textwidth]{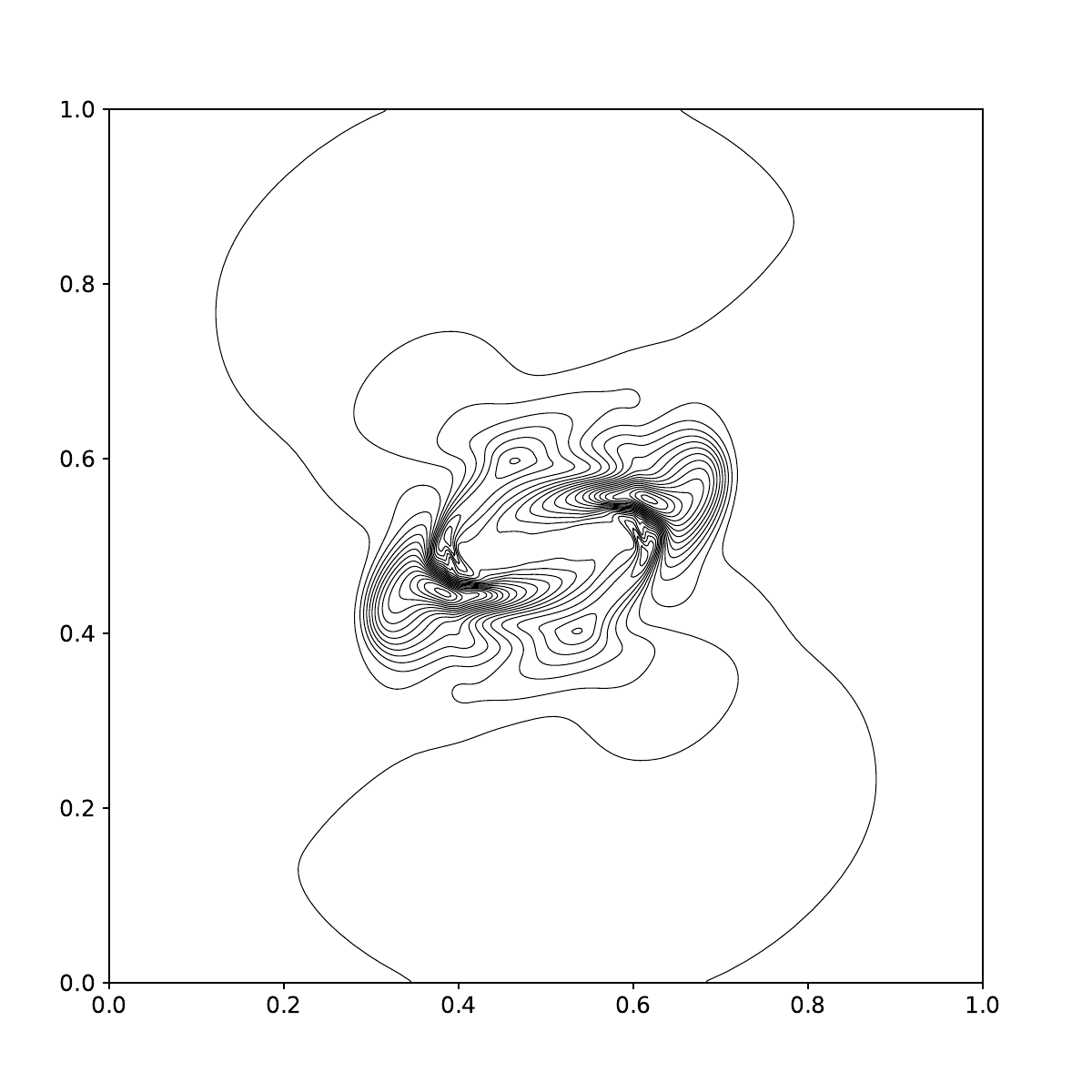}
 }  
 \caption{Internal energy at $t=0.12$ of the low plasma $\beta$ Rotor problem. The 40 contour lines are shown for the range $\varrho e \in[-0.15, 0.27]$. The HLLD scheme produces negative internal energy around $(x,y)=(0.45, 0.6)$ and $(x,y)=(0.55, 0.4)$, visualised by dashed contour lines.}
 \label{fig:Rotor_beta}
\end{figure}

\begin{figure}[htbp]
 \centering
 \includegraphics[width=0.6\textwidth]{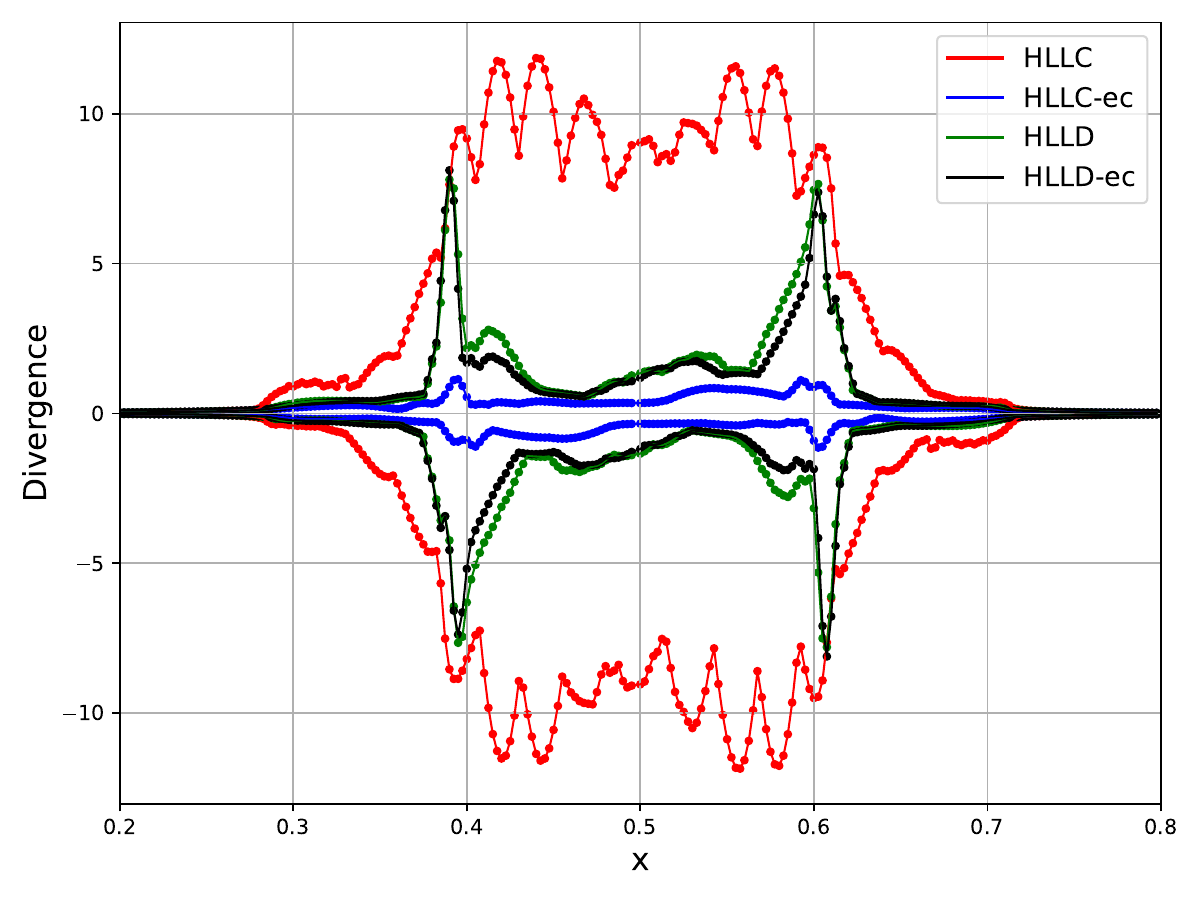}
 \caption{Envelopes (maximum and minimum values along each $y$-slice) of the divergence error $\nabla\cdot\mathbf{B}$.}
 \label{fig:Rotor_div}
\end{figure}

We propose a revised test case with lower plasma $\beta$ than the original Rotor problem. To do so, we reduced the density and plasma thermal pressure to $10\%$ of the original values in the Rotor problem, resulting in $0.1\le\varrho\le1$ and $p=0.05$. In addition, we add $B_y=2.5/\sqrt{4\pi}$, without changing $B_x$. The initial velocity and the adiabatic index both remain unchanged. The same numerical methods are used, and the HDC is again excluded to allow $\nabla\cdot\mathbf{B}$ to develop. Corresponding to the increased eigenwave speeds, we use $\Delta t = 0.0001$ for the same mesh here {\color{red}(thus CFL $\simeq 0.2$)}.

The internal energy contour lines at $t=0.12$ are shown in Fig.~\ref{fig:Rotor_beta}. Again, the difference between the HLLD and HLLD-ec schemes is relatively small, but the HLLC-ec scheme is clearly more diffusive, as it does not resolve contact discontinuity. However, without the HDC, the simulation using the HLLC(-L) scheme quickly develops oscillations and crashes. Although all the simulations eventually crash due to increasing divergence errors, the HLLC scheme is the most sensitive. The result of the HLLD scheme also has negative internal energy values at $t=0.12$.

We thus further present the divergence error envelopes in Fig.~\ref{fig:Rotor_div}, which show the maximum error along each $y$-slice. The divergence error produced by the HLLC scheme is significantly larger than those of the other schemes. This is likely because the HLLC scheme uses the HLL approximation for the magnetic field.  Note that, although a Riemann solver does (should) not directly cause an error in the longitudinal magnetic field component, the error in the tangential components will eventually affect the computations in other directions. They may also explain why the HLLD-type schemes exhibit larger errors than the HLLC-ec scheme, as they effectively reduce to the HLL approximation when the longitudinal magnetic field is substantial or dominant. Nonetheless, the present HLLD-ec scheme produces fewer errors than the HLLD scheme. 

Therefore, we may conclude that the present HLLD-ec scheme is less prone to producing divergence errors when plasma $\beta$ is relatively low, without sacrificing numerical resolution. The HLLC-ec scheme, on the other hand, exhibits a significant advantage in reducing the divergence error, {\color{red}albeit at the expense of sacrificing numerical solutions to the fluid equations.} Indeed, divergence-cleaning approaches may sufficiently remove the divergence error. Still, we should note that (I) the error should not occur in the first place, and (II) the divergence-cleaning process may diffuse the magnetic field, depending on the value of $c_{\text{h}}$ \cite{Perri2022}, which can be significantly different from the local physical eigenwave speeds. {\color{red}In the current case, the HLLC scheme can continue the simulation if the HDC is imposed,} {\color{green}which, however, brings extra numerical diffusion, since Eqs.~(\ref{eq:divB}) and (\ref{eq:Bn}) are not equivalent. We do not suggest not using the HDC procedure, but observe that improving the Riemann solvers can be beneficial for reducing the $\nabla\cdot\mathbf{B}$ error. Note that our schemes also produce smaller oscillations when the error exists.}

{\color{blue} We would like to mention that, there are more challenging test cases, which our HLLC/D-ec schemes may fail.  However, in an extremely low plasma $\beta$ ($\simeq 10^{-6}$) test   \cite{TREMBLIN2024}, for example, the present HLLD-ec scheme still performs better than the classic HLLD scheme.} {\color{red}Specifically,  the $\nabla\cdot\mathbf{B}$ error has to be effectively removed in this extreme test,  and then we found that the HLLD-ec scheme survives the test (using a smaller time step than in Ref.~\cite{TREMBLIN2024}) while the classic HLLD scheme still fails.} {\color{green}More comparison is not provided since significant advantages of the present schemes compared to the classic multi-state HLL-type schemes are already shown. However, further improvements are obviously needed in both the numerical schemes and the theoretical understanding for solving low plasma $\beta$ problems.}  
 
  \section{Conclusions} \label{sec:conclusions}

We have revised the multi-state HLL-type schemes to solve the ideal MHD equations. Specifically, we propose an energy consistency condition for calculating the intermediate energies within the Riemann fan, ensuring that numerically calculated energy terms are consistent with the (intermediate) solutions of other equations. As the intermediate states directly contribute to the dissipation terms of HLL-type schemes, ensuring the consistency condition also ensures the positivity of the intermediate internal energy. Furthermore, consistency leads to a lower error in the time-integrated energy equation solution, ultimately reducing spurious oscillations.  In addition, the multi-state assumptions of both the HLLC- and HLLD-type schemes are revisited in detail to achieve more accurate and robust solutions for problems with lower plasma $\beta$.  Analytical analysis and numerical tests show that the present schemes are less prone to spurious numerical behaviours in multi-dimensional, low plasma $\beta$ scenarios. In contrast, the numerical resolution is not necessarily sacrificed. Moreover, the revised schemes remain as simple as the original multi-state HLL-type schemes. {\color{green}On the other hand, we should note that the present effort is limited to solving the conservative MHD equations, with further extensions left for the future.}

%% Unnumbered versions of align and eqnarray
%\begin{align*}
% f(x) &= (x+a)(x+b) \\
%      &= x^2 + (a+b)x + ab
%\end{align*}

%\begin{eqnarray*}
% f(x)&=& (x+a)(x+b) \\
%     &=& x^2 + (a+b)x + ab
%\end{eqnarray*}

%% Refer following link for more details.
%% https://en.wikibooks.org/wiki/LaTeX/Mathematics
%% https://en.wikibooks.org/wiki/LaTeX/Advanced_Mathematics

%% Use a table environment to create tables.
%% Refer following link for more details.
%% https://en.wikibooks.org/wiki/LaTeX/Tables

%% Use figure environment to create figures
%% Refer following link for more details.
%% https://en.wikibooks.org/wiki/LaTeX/Floats,_Figures_and_Captions
%\begin{figure}[t]%% placement specifier 
%\centering%% For centre alignment of image.
%\includegraphics{example-image-a} 
%%\caption{Figure Caption}\label{fig1} 
%\end{figure}
%\begin{table}[t]
%\centering%% For centre alignment of tabular.
%\begin{tabular}{l c r}%% Table column specifiers
%% Tabular cells are separated by &
%  1 & 2 & 3 \\ %% A tabular row ends with \\
%  4 & 5 & 6 \\
 % 7 & 8 & 9 \\
%\end{tabular}
%% Use \caption command for table caption and label.
%\caption{Table Caption}\label{table1}
%\end{table}

\section*{Acknowledgments}
The authors thank K\=e X{\v{U}} for providing help with the exact Riemann solution, and thank Michaela Brchnelov\'a for helpful comments on the manuscript.
The results were obtained in the framework of the projects FA9550-18-1-0093 (AFOSR), C16/24/010  (C1 project Internal Funds KU Leuven), G0B5823N and G002523N (FWO-Vlaanderen), and 4000145223 (SIDC Data Exploitation, ESA Prodex). SP also acknowledges support from the Open SESAME project, which has received funding from the Horizon Europe programme (ERC-AdG agreement No.\ 101141362). The Research Council of Norway supports FZ  through its Centres of Excellence scheme, project number
262622. The computational resources and services used in this work were provided by the VSC (Flemish Supercomputer Centre),
funded by the Research Foundation - Flanders (FWO) and the Flemish Government - Department EWI. 
\appendix

\section{The numerical solver} \label{sec:CF}
We implement the numerical methods on \texttt{COOLFluiD} (Computational Object-Oriented Libraries for Fluid Dynamics) \cite{Lani2005, Lani2013}, which is an open-source component-based software framework\footnote{https://github.com/andrealani/COOLFluiD} for high-performance scientific and engineering computations. Specifically, the unstructured finite-volume MHD solver within \texttt{COOLFluiD} was initially developed by Yalim et al. \cite{Yalim2011} and has been further developed for modelling solar and space plasma \cite{Lani2014,Perri2022}. 

\texttt{COOLFluiD} features various numerical methods. Currently, the second-order-accurate unstructured finite-volume method is the leading approach for MHD simulations. Specifically, its second-order accuracy is achieved through (weighted) least-squares reconstructions, while a slope limiter may be activated to capture steep variations in physical quantities.  To improve the overall computational efficiency in large-scale simulations, the MHD solver of \texttt{COOLFluiD} employs fully implicit methods for temporal discretisation. Specifically, the backwards Euler scheme is used for steady problems, and the 2-step backwards difference formula is used for time-dependent problems.
The resulting sparse algebraic systems of the implicit methods are solved
 using the Generalised Minimal RESidual (GMRES) method \cite{Saad1986}  provided by \texttt{PETSc}  \cite{petsc-web-page},  with the restricted additive Schwarz preconditioner \cite{Widlund1987}.

\section{The HLLC scheme of Li} \label{sec:HLLC_Li}
Here, we introduce the HLLC scheme of Li (HLLC-L) \cite{Li2005} without going into a detailed explanation. We follow the original design as closely as possible, except that the discussions in Sections \ref{sec:MHDEqu} and \ref{sec:flux} are included, which means the current implementation is not precisely equivalent to Li's original one in multi-dimensional simulations. {\color{red}Although relatively minor, the changes improve overall robustness, particularly when time steps are large. Moreover, the changes are made to provide a fair comparison with energy-consistent schemes.}

A critical part of HLLC-type schemes is the estimation of the middle wave speed.  
 The middle wave speed is  given  by assuming a constant longitudinal velocity component within the whole Riemann fan, i.e., $V_{\parallel}^{\text{l*}}=S^{\text{m}}=V_{\parallel}^{\text{r*}}$, resulting in 
 \begin{equation}    \label{eq:SM}
 S^{\text{m}}=\frac{[\varrho V_{\parallel}(S-V_{\parallel})-P+
 {B}_{\parallel}^2]^{\text{r}}-[\varrho V_{\parallel}(S-V_{\parallel})-P+{B}_{\parallel}^2]^{\text{l}}}{[\varrho(S-V_{\parallel})]^{\text{r}}-[\varrho(S-V_{\parallel})]^{\text{l}}},
  \end{equation} 
\noindent {\color{green}where we use brackets $[~]$ only to simplify the use of the superscripts, without any mathematical implication.} As previously mentioned, the HLLC-L scheme uses the HLL average for calculating the (tangential components of) intermediate magnetic field, and thus we have 
 \begin{eqnarray}    \label{eq:consistentB}
 \mathbf{B}^{\text{l*,r*}}_{\perp}=\mathbf{B}^{\text{m}}_{\perp}=\frac{S^{\text{r}} \mathbf{B}^{\text{r}}_{\perp}-S^{\text{l}} \mathbf{B}^{\text{l}}_{\perp}-\left(\mathbf{F}^{\text{r}}-\mathbf{F}^{\text{l}}\right)}{S^{\text{r}}-S^{\text{l}}},
  \end{eqnarray} 
\noindent in which the flux terms $\mathbf{F}^{\text{l,r}}$ naturally only include the components of the induction equation in Eq.~(\ref{eq:ExactFlux}). Using Eq.~(\ref{eq:consistentB}) ensures that the magnetic field components in the Riemann fan satisfy the integral consistency condition; however, it obviously reduces the magnetic field's resolution. We should note again that in Eq.~(\ref{eq:HLLC}),  $\mathbf{U}^{\text{l*},\text{r*}}$ and $\mathbf{U}^{\text{l},\text{r}}$ do not include the longitudinal components of the magnetic field, corresponding to the zero longitudinal flux of the induction equation. 

Since the longitudinal component of the magnetic field is needed when calculating the other components of the numerical flux, it is always given as in Eq.~(\ref{eq:Ideal}). This option is different from those suggested in Refs. \cite{Li2005} and \cite{MIGNONE2010}. The rationale is that (I) the longitudinal component of the magnetic field should not be affected by the local physical wave speeds,  and (II) directly using the HDC solution \cite{MIGNONE2010} means that the intermediate state of the longitudinal component would be dependent on the value of $c_{\text{h}}$, which is typically a non-local and non-physical wave speed. Nonetheless, our solution is equivalent to using an HLL average for the hyperbolic system in Eq.~(\ref{eq:HDC}) alone. 

The HLLC-L scheme assumes that the total pressure is constant across the middle wave (allowing for a contact discontinuity), while plasma thermal pressure and magnetic pressure may vary. However, when assuming a constant intermediate magnetic field using the HLL average, the magnetic pressure remains effectively constant across the middle wave, as does the plasma pressure. By simply using Eq.~(\ref{eq:SM}) to replace  $S^{\text{m}}$ in 
 \begin{equation}    \label{eq:PM1}
P^{\text{l*,r*}}=\varrho^{\text{l,r}}(S^{\text{m}}-V_{\parallel}^{\text{l,r}})(S^{\text{l,r}}-V_{\parallel}^{\text{l,r}})+P^{\text{l,r}}-({B}_{\parallel}^{\text{l,r}})^2+({B}_{\parallel}^{\text{m}})^2, 
 \end{equation}
\noindent we obtain the intermediate total pressure 
\begin{eqnarray}  \label{eq:PM2} 
P^{\text{m}} = \frac{\left[\varrho(S-V_{\parallel})\right]^{\text{r}}(P-{B}_{\parallel}^2)^{\text{l}}-\left[\varrho(S-V_{\parallel})\right]^{\text{l}}(P-{B}_{\parallel}^2)^{\text{r}}+\left[\varrho(S-V_{\parallel})\right]^{\text{l}}\left[\varrho(S-V_{\parallel})\right]^{\text{r}}(V_{\parallel}^{\text{r}}-V_{\parallel}^{\text{l}})}{\left[\varrho(S-V_{\parallel})\right]^{\text{r}}-\left[\varrho(S-V_{\parallel})\right]^{\text{l}}}+({B}_{\parallel}^{\text{m}})^2.
  \end{eqnarray}

Then, the other variables can be given according to the RH relations across the fastest eigenwaves. Specifically, all the variables on the left side of the middle wave are given as
 \begin{eqnarray}   \label{eq:LMVars}
\left\{\begin{array}{ll}
 \varrho^{\text{l*}}=\varrho^{\text{l}}\frac{S^{\text{l}}-V_{\parallel}^{\text{l}}}{S^{\text{l}}- S^{\text{m}}},  \\ 
 (\varrho V_{\parallel})^{\text{l*}}=\varrho^{\text{l*}}S^{\text{m}},  \\ 
 (\varrho \mathbf{V}_{{\perp}})^{\text{l*}}=\frac{\varrho^{\text{l}} \mathbf{V}_{{\perp}}^{\text{l}}(S^{\text{l}}-V_{\parallel}^{\text{l}})-(B_{\parallel}^{\text{m}} \mathbf{B}_{{\perp}}^{\text{m}}-B_{\parallel}^{\text{l}} \mathbf{B}_{{\perp}}^{\text{l}})}{S^{\text{l}}- S^{\text{m}}}, \\
E^{\text{l*}}=\frac{E^{\text{l}}(S^{\text{l}}-V_{\parallel}^{\text{l}})+P^{\text{m}}S^{\text{m}}-P^{\text{l}}V_{\parallel}^{\text{l}}-\left[B_{\parallel}^{\text{m}}(\mathbf{B}^{\text{m}}\cdot \mathbf{V}^{\text{l*}})-B_{\parallel}^{\text{l}}(\mathbf{B}^{\text{l}}\cdot \mathbf{V}^{\text{l}})\right]}{S^{\text{l}}- S^{\text{m}}},  
 \end{array}\right.
 \end{eqnarray}
 \noindent where $\mathbf{V}_{\perp}^{\text{l*}}=(\varrho \mathbf{V}_{{\perp}})^{\text{l*}}/\varrho^{\text{l*}}$. 
 The variables on the right side of the middle wave can be given symmetrically. 
 Finally, Eq.~(\ref{eq:HLLC}) can be solved with all the conservative variables given above.  

The intermediate total energy includes ${B}_{\parallel}^2$. In section \ref{sec:HDC}, it is suggested to remove the contribution of ${B}_{\parallel}$ from the numerical diffusion terms for both the induction equation and energy equation. However, the jump relation cannot guarantee that the variation of magnetic energy is consistent with the HDC solution \textit{or} Eq.~(\ref{eq:Ideal}). Thus, it is not recommended to directly subtract ${B}_{\parallel}^2$ from the intermediate total energy. This is also the case for the HLLD scheme.

\section{The HLLD scheme} \label{sec:HLLD}

The HLLD scheme was initially proposed in Ref.~\cite{Miyoshi2005}. However, the formulae in Ref.~\cite{Miyoshi2005} are given under the assumption of a constant longitudinal magnetic field component ($B_{\parallel}$); therefore, they are reformulated below with a varying $B_{\parallel}$. In particular, Eq.~(\ref{eq:Ideal}) is used again for providing the intermediate state of  $B_{\parallel}$.

Nonetheless, some formulae can be reused from the HLLC-L scheme based on the given assumptions. Firstly, the intermediate pressure is also assumed to be constant over the whole Riemann fan, and thus, Eq.~(\ref{eq:PM2}) is also used in the HLLD scheme. The formula is essentially the same as the formula in Ref.~\cite{Miyoshi2005}, except that here, the longitudinal component of the magnetic field is not assumed to be constant from the left state to the right state (but constant between them). Moreover, the same assumption is used for the longitudinal component of the intermediate velocity, i.e., 
\begin{eqnarray}   \label{eq:HLLDSM}
V^{\text{l*,r*}}_{\parallel}=V^{\text{l**,r**}}_{\parallel}=S^{\text{m}},
\end{eqnarray} 
\noindent which means the plasma is incompressible across the intermediate waves. The density within the Riemann fan only changes across the middle linear wave (contact discontinuity), and thus we have 
 \begin{eqnarray}    \label{eq:HLLDRho}
\varrho^{\text{l*}}=\varrho^{\text{l**}}, \quad \text{and} \quad \varrho^{\text{r*}}=\varrho^{\text{r**}},
  \end{eqnarray} 
\noindent which are again the same as the HLLC-L scheme. On the contrary, in the HLLC-ec scheme, the fluid equations are solved by assuming a two-wave configuration; thus, the intermediate density remains constant throughout the entire Riemann fan.  

The other intermediate states behind the fast waves can be calculated according to the jump relations, which account for the varying $B_{\parallel}$.
Without going through the derivations, the tangential components of velocity and magnetic field are given as
 \begin{eqnarray}   \label{eq:V_parallel}
 \mathbf{V}_{{\perp}}^{\text{l*}}=\frac{(S^{\text{l}}-V_{\parallel}^{\text{l}})\varrho^{\text{l}}\mathbf{V}_{{\perp}}^{\text{l}}+B_{\parallel}^{\text{l}}\mathbf{B}_{\perp}^{\text{l}}-B_{\parallel}^{\text{m}}\frac{(S^{\text{l}}-V_{\parallel}^{\text{l}})\mathbf{B}_{\perp}^{\text{l}}+B_{\parallel}^{\text{l}}\mathbf{V}_{{\perp}}^{\text{l}}}{S^{\text{l}}-S^{\text{m}}}}{(S^{\text{l}}-V_{\parallel}^{\text{l}})\varrho^{\text{l}}-\frac{(B_{\parallel}^{\text{m}})^2}{S^{\text{l}}-S^{\text{m}}}}, 
  \end{eqnarray} 
\noindent and 
\begin{equation}  \label{eq:B_perp}  \mathbf{B}_{\perp}^{\text{l*}}=\frac{\mathbf{B}_{\perp}^{\text{l}}\left[S^{\text{l}}-V_{\parallel}^{\text{l}}-\frac{B_{\parallel}^{\text{m}}B_{\parallel}^{\text{l}}}{\varrho^{\text{l}}(S^{\text{l}}-V_{\parallel}^{\text{l}})}
 \right]-\mathbf{V}_{\perp}^{\text{l}}\left(B_{\parallel}^{\text{m}}-B_{\parallel}^{\text{l}}\right)}{S^{\text{l}}-S^{\text{m}}-\frac{(B_{\parallel}^{\text{m}})^2}{\varrho^{\text{l}}(S^{\text{l}}-V_{\parallel}^{\text{l}})}},
 \end{equation}
\noindent as $\mathbf{V}_{{\perp}}^{\text{r*}}$ and $\mathbf{B}_{\perp}^{\text{r*}}$ are given symmetrically. It is easy to find that when $B_{\parallel}$ is constant, these two formulae become the same as those in Ref.~\cite{Miyoshi2005}. {\color{green}Note that Ref.~\cite{Li2005} discussed similar formulae, but the HLLC scheme could not use both formulae, to satisfy the integral consistency condition \cite{Toro2009}.} Then, the intermediate total energy $E_{\perp}^{\text{l*},\text{r*}}$ can be given formally following Eq.~(\ref{eq:LMVars}), and of course the velocity and magnetic field components therein should be replaced by Eqs.~(\ref{eq:V_parallel}) and (\ref{eq:B_perp}). 

The intermediate states between the left and right propagating Alfv\'en waves can be given after having their propagation speeds provided, which are  
 \begin{eqnarray}   
S^{\text{l*}}=S^{\text{m}}-\frac{|B_{\parallel}^{\text{m}}|}{\sqrt{\varrho^{\text{l*}}}}, \quad \text{and} \quad S^{\text{r*}}=S^{\text{m}}+\frac{|B_{\parallel}^{\text{m}}|}{\sqrt{\varrho^{\text{r*}}}}.
  \end{eqnarray} 
  \noindent Note that when Alfv\'en speed is close to the fast wave speed, in Eqs.~(\ref{eq:V_parallel}) and (\ref{eq:B_perp}), the denominators may be close to zero, and thus a small value is needed in the denominators, which are not specifically discussed here.
  As has been explained in Ref.~\cite{Miyoshi2005}, the tangential components of the velocity and magnetic field should be constant between Alfv\'en waves, following the integral consistency formula of the four-state approximation
   \begin{equation} \label{eq:consistency4}
(S^{\text{r}}-S^{\text{r*}})\mathbf{U}^{\text{r*}}+(S^{\text{r*}}-S^{\text{m}})\mathbf{U}^{\text{r**}}+(S^{\text{m}}-S^{\text{l*}})\mathbf{U}^{\text{l**}}+(S^{\text{l*}}-S^{\text{l}})\mathbf{U}^{\text{l*}}=S^{\text{r}}\mathbf{U}^{\text{r}}-S^{\text{l}}\mathbf{U}^{\text{l}}-\mathbf{F}^{\text{r}}+\mathbf{F}^{\text{l}}.
 \end{equation}
\noindent Again, without going into details, the following formulae are used:
   \begin{eqnarray}   \label{eq:V_HLLD}
 \mathbf{V}_{{\perp}}^{\text{l**,r**}}=\frac{\sqrt{\varrho^{\text{l*}}}\mathbf{V}_{\perp}^{\text{l*}}+\sqrt{\varrho^{\text{r*}}}\mathbf{V}_{\perp}^{\text{r*}}+(\mathbf{B}_{{\perp}}^{\text{r*}}- \mathbf{B}_{{\perp}}^{\text{l*}})\text{sign}(B_{\parallel}^{\text{m}})}{\sqrt{\varrho^{\text{l*}}}+\sqrt{\varrho^{\text{r*}}}}, 
  \end{eqnarray} 
\noindent and 
\begin{equation}  \label{eq:B_HLLD}
\mathbf{B}_{\perp}^{\text{l**,r**}}=\frac{\sqrt{\varrho^{\text{r*}}}\mathbf{B}_{\perp}^{\text{l*}}+\sqrt{\varrho^{\text{l*}}}\mathbf{B}_{\perp}^{\text{r*}}+\sqrt{\varrho^{\text{l*}}\varrho^{\text{r*}}}( \mathbf{V}_{{\perp}}^{\text{r*}}- \mathbf{V}_{{\perp}}^{\text{l*}})\text{sign}(B_{\parallel}^{\text{m}})}{\sqrt{\varrho^{\text{l*}}}+\sqrt{\varrho^{\text{r*}}}}.
 \end{equation}

As density, pressure, and the longitudinal components of velocity and magnetic field are already given, the intermediate states of total energy between two Alfv\'en waves are
\begin{equation} 
E^{\alpha\text{**}}=E^{\alpha\text{*}}\mp\sqrt{\varrho^{\alpha\text{*}}}\left(\mathbf{V}^{\alpha\text{*}}\mathbf{B}^{\alpha\text{*}}-\mathbf{V}^{\alpha\text{**}}\mathbf{B}^{\alpha\text{**}}\right)\text{sign}(B_{\parallel}^{\text{m}}),
 \end{equation}
\noindent where the superscript $\alpha=\text{l}$ or $\alpha=\text{r}$ respectively correspond to the minus or the plus on the right-hand side. Since the longitudinal velocity and magnetic field components are constant throughout the entire Riemann fan, they do not directly contribute to the changes in total energy across Alfv\'en waves.

%%       \cite{<label>} ==> [1]

%% 




%\begin{thebibliography}{00}

%% For numbered reference style

%\bibitem{lamport94}
%  \textit{\LaTeX: a document preparation system},
%  Addison Wesley, Massachusetts,

%\end{thebibliography}
\bibliography{ref_MF}

\begin{thebibliography}{10}
\expandafter\ifx\csname url\endcsname\relax
  \def\url#1{\texttt{#1}}\fi
\expandafter\ifx\csname urlprefix\endcsname\relax\def\urlprefix{URL }\fi
\expandafter\ifx\csname href\endcsname\relax
  \def\href#1#2{#2} \def\path#1{#1}\fi

\bibitem{Godunov1959}
S.~K. Godunov, Finite difference method for numerical computation of discontinuous solutions of the equations of fluid dynamics, Matematicheskii Sbornik 47~(3) (1959) 271--306.

\bibitem{Toro2009}
E.~F. Toro, Riemann Solvers and Numerical Methods for Fluid Dynamics, 3rd Edition, Springer, 2009.

\bibitem{DAI1994}
W.~Dai, P.~R. Woodward, An approximate {R}iemann solver for ideal magnetohydrodynamics, Journal of Computational Physics 111~(2) (1994) 354--372.
\newblock \href {https://doi.org/10.1006/jcph.1994.1069} {\path{doi:10.1006/jcph.1994.1069}}.

\bibitem{Brackbill_1980}
J.~Brackbill, D.~Barnes, The effect of nonzero $\nabla$$\cdotp${B} on the numerical solution of the magnetohydrodynamic equations, Journal of Computational Physics 35~(3) (1980) 426--430.
\newblock \href {https://doi.org/10.1016/0021-9991(80)90079-0} {\path{doi:10.1016/0021-9991(80)90079-0}}.

\bibitem{Toth_2000}
G.~T{\'{o}}th, The $\nabla$$\cdotp${B}=0 constraint in shock-capturing magnetohydrodynamics codes, Journal of Computational Physics 161~(2) (2000) 605--652.
\newblock \href {https://doi.org/10.1006/jcph.2000.6519} {\path{doi:10.1006/jcph.2000.6519}}.

\bibitem{Dedner_2002}
A.~Dedner, F.~Kemm, D.~Kröner, C.-D. Munz, T.~Schnitzer, M.~Wesenberg, Hyperbolic divergence cleaning for the {MHD} equations, Journal of Computational Physics 175~(2) (2002) 645--673.
\newblock \href {https://doi.org/10.1006/jcph.2001.6961} {\path{doi:10.1006/jcph.2001.6961}}.

\bibitem{Wu_2018}
K.~Wu, Positivity-preserving analysis of numerical schemes for ideal magnetohydrodynamics, SIAM Journal on Numerical Analysis 56~(4) (2018) 2124--2147.
\newblock \href {https://doi.org/10.1137/18M1168017} {\path{doi:10.1137/18M1168017}}.

\bibitem{Wang2025}
H.~P. Wang, J.~H. Guo, L.~P. Yang, S.~Poedts, F.~Zhang, A.~Lani, T.~Baratashvili, L.~Linan, R.~Lin, Y.~Guo, {SIP}-{IFVM}: Efficient time-accurate magnetohydrodynamic model of the corona and coronal mass ejections, Astronomy \& Astrophysics 693 (2025) A257.
\newblock \href {https://doi.org/10.1051/0004-6361/202450771} {\path{doi:10.1051/0004-6361/202450771}}.

\bibitem{EINFELDT1991}
B.~Einfeldt, C.~Munz, P.~Roe, B.~Sjögreen, On {G}odunov-type methods near low densities, Journal of Computational Physics 92~(2) (1991) 273--295.
\newblock \href {https://doi.org/10.1016/0021-9991(91)90211-3} {\path{doi:10.1016/0021-9991(91)90211-3}}.

\bibitem{Batten_1997}
P.~Batten, N.~Clarke, C.~Lambert, D.~M. Causon, On the choice of wavespeeds for the {HLLC} {R}iemann solver, {SIAM} Journal on Scientific Computing 18~(6) (1997) 1553--1570.
\newblock \href {https://doi.org/10.1137/s1064827593260140} {\path{doi:10.1137/s1064827593260140}}.

\bibitem{Gallice2003}
G.~Gallice, Positive and entropy stable {G}odunov-type schemes for gas dynamics and {MHD} equations in {L}agrangian or {E}ulerian coordinates, Numerische Mathematik 94~(4) (2003) 673--713.
\newblock \href {https://doi.org/10.1007/s00211-002-0430-0} {\path{doi:10.1007/s00211-002-0430-0}}.

\bibitem{Harten_1983}
A.~Harten, P.~D. Lax, B.~van Leer, On upstream differencing and {G}odunov-type schemes for hyperbolic conservation laws, {SIAM} Review 25~(1) (1983) 35--61.
\newblock \href {https://doi.org/10.1137/1025002} {\path{doi:10.1137/1025002}}.

\bibitem{Gurski2004}
K.~F. Gurski, An {HLLC}-type approximate {R}iemann solver for ideal magnetohydrodynamics, SIAM Journal on Scientific Computing 25~(6) (2004) 2165--2187.
\newblock \href {https://doi.org/10.1137/S1064827502407962} {\path{doi:10.1137/S1064827502407962}}.

\bibitem{Li2005}
S.~Li, An {HLLC} {R}iemann solver for magneto-hydrodynamics, Journal of Computational Physics 203~(1) (2005) 344--357.
\newblock \href {https://doi.org/10.1016/j.jcp.2004.08.020} {\path{doi:10.1016/j.jcp.2004.08.020}}.

\bibitem{Miyoshi2005}
T.~Miyoshi, K.~Kusano, A multi-state {HLL} approximate {R}iemann solver for ideal magnetohydrodynamics, Journal of Computational Physics 208~(1) (2005) 315--344.
\newblock \href {https://doi.org/10.1016/j.jcp.2005.02.017} {\path{doi:10.1016/j.jcp.2005.02.017}}.

\bibitem{Toro1994}
E.~F. Toro, M.~Spruce, W.~Speares, Restoration of the contact surface in the {HLL}-{R}iemann solver, Shock Waves 4 (1994) 25--34.
\newblock \href {https://doi.org/10.1007/BF01414629} {\path{doi:10.1007/BF01414629}}.

\bibitem{Linde2002}
T.~Linde, A practical, general-purpose, two-state {HLL} {R}iemann solver for hyperbolic conservation laws, International Journal for Numerical Methods in Fluids 40~(3-4) (2002) 391--402.
\newblock \href {https://doi.org/10.1002/fld.312} {\path{doi:10.1002/fld.312}}.

\bibitem{Gary2001}
G.~A. Gary, Plasma beta above a solar active region: Rethinking the paradigm, Solar Physics 203 (2001) 71--86.
\newblock \href {https://doi.org/10.1023/A:1012722021820} {\path{doi:10.1023/A:1012722021820}}.

\bibitem{Brchnelova2023b}
M.~Brchnelova, B.~Ku{\'{z}}ma, F.~Zhang, A.~Lani, S.~Poedts, The role of plasma $\beta$ in global coronal models bringing balance back to the force, Astronomy \& Astrophysics 676 (2023) A83.
\newblock \href {https://doi.org/10.1051/0004-6361/202346788} {\path{doi:10.1051/0004-6361/202346788}}.

\bibitem{Ku_ma_2023}
B.~Ku{\'{z}}ma, M.~Brchnelova, B.~Perri, T.~Baratashvili, F.~Zhang, A.~Lani, S.~Poedts, {COCONUT}, a novel fast-converging {MHD} model for solar corona simulations. {III}. impact of the preprocessing of the magnetic map on the modeling of the solar cycle activity and comparison with observations, The Astrophysical Journal 942~(1) (2023) 31.
\newblock \href {https://doi.org/10.3847/1538-4357/aca483} {\path{doi:10.3847/1538-4357/aca483}}.

\bibitem{Zhang2018}
F.~Zhang, J.~Liu, B.~Chen, Modified multi-dimensional limiting process with enhanced shock stability on unstructured grids, Computers \& Fluids 161 (2018) 171--188.
\newblock \href {https://doi.org/10.1016/j.compfluid.2017.11.019} {\path{doi:10.1016/j.compfluid.2017.11.019}}.

\bibitem{Perri2022}
B.~Perri, P.~Leitner, M.~Brchnelova, T.~Baratashvili, B.~Ku{\'{z}}ma, F.~Zhang, A.~Lani, S.~Poedts, {COCONUT}, a novel fast-converging {MHD} model for solar corona simulations: {I}. benchmarking and optimization of polytropic solutions, The Astrophysical Journal 936~(1) (2022) 19.
\newblock \href {https://doi.org/10.3847/1538-4357/ac7237} {\path{doi:10.3847/1538-4357/ac7237}}.

\bibitem{TANAKA1994}
T.~Tanaka, Finite volume {TVD} scheme on an unstructured grid system for three-dimensional {MHD} simulation of inhomogeneous systems including strong background potential fields, Journal of Computational Physics 111~(2) (1994) 381--389.
\newblock \href {https://doi.org/10.1006/jcph.1994.1071} {\path{doi:10.1006/jcph.1994.1071}}.

\bibitem{Bouchut2007}
F.~Bouchut, C.~Klingenberg, K.~Waagan, A multiwave approximate {R}iemann solver for ideal {MHD} based on relaxation. {I}: theoretical framework, Numerische Mathematik 108 (2007) 7--42.
\newblock \href {https://doi.org/10.1007/s00211-007-0108-8} {\path{doi:10.1007/s00211-007-0108-8}}.

\bibitem{Bouchut2010}
F.~Bouchut, C.~Klingenberg, K.~Waagan, A multiwave approximate {R}iemann solver for ideal {MHD} based on relaxation {II}: numerical implementation with 3 and 5 waves, Numerische Mathematik 115 (2010) 647--679.
\newblock \href {https://doi.org/10.1007/s00211-010-0289-4} {\path{doi:10.1007/s00211-010-0289-4}}.

\bibitem{Chalons_2016}
C.~Chalons, M.~Girardin, S.~Kokh, An all-regime lagrange-projection like scheme for the gas dynamics equations on unstructured meshes, Communications in Computational Physics 20~(1) (2016) 188–233.
\newblock \href {https://doi.org/10.4208/cicp.260614.061115a} {\path{doi:10.4208/cicp.260614.061115a}}.

\bibitem{BOURGEOIS2024}
R.~Bourgeois, P.~Tremblin, S.~Kokh, T.~Padioleau, Recasting an operator splitting solver into a standard finite volume flux-based algorithm. the case of a lagrange-projection-type method for gas dynamics, Journal of Computational Physics 496 (2024) 112594.
\newblock \href {https://doi.org/10.1016/j.jcp.2023.112594} {\path{doi:10.1016/j.jcp.2023.112594}}.

\bibitem{TREMBLIN2024}
P.~Tremblin, R.~Bourgeois, S.~Bulteau, S.~Kokh, T.~Padioleau, M.~Delorme, A.~Strugarek, M.~Gonz\'alez, A.~S. Brun, A multi-dimensional, robust, and cell-centered finite-volume scheme for the ideal {MHD} equations, Journal of Computational Physics 519 (2024) 113455.
\newblock \href {https://doi.org/10.1016/j.jcp.2024.113455} {\path{doi:10.1016/j.jcp.2024.113455}}.

\bibitem{Goedbloed2004}
J.~P.~H. Goedbloed, S.~Poedts, Principles of Magnetohydrodynamics: With Applications to Laboratory and Astrophysical Plasmas, Cambridge University Press, 2004.
\newblock \href {https://doi.org/10.1017/CBO9780511616945} {\path{doi:10.1017/CBO9780511616945}}.

\bibitem{KULSRUD}
R.~M. Kulsrud, Plasma Physics for Astrophysics, Princeton University Press, 2005.

\bibitem{Fuchs2011a}
F.~G. Fuchs, A.~D. McMurry, S.~Mishra, N.~H. Risebro, K.~Waagan, Approximate {R}iemann solvers and robust high-order finite volume schemes for multi-dimensional ideal {MHD} equations, Communications in Computational Physics 9~(2) (2011) 324–362.
\newblock \href {https://doi.org/10.4208/cicp.171109.070510a} {\path{doi:10.4208/cicp.171109.070510a}}.

\bibitem{DAO2024}
T.~A. Dao, M.~Nazarov, I.~Tomas, A structure preserving numerical method for the ideal compressible {MHD} system, Journal of Computational Physics 508 (2024) 113009.
\newblock \href {https://doi.org/10.1016/j.jcp.2024.113009} {\path{doi:10.1016/j.jcp.2024.113009}}.

\bibitem{MIGNONE2010}
A.~Mignone, P.~Tzeferacos, A second-order unsplit {G}odunov scheme for cell-centered {MHD}: The {CTU}-{GLM} scheme, Journal of Computational Physics 229~(6) (2010) 2117--2138.
\newblock \href {https://doi.org/10.1016/j.jcp.2009.11.026} {\path{doi:10.1016/j.jcp.2009.11.026}}.

\bibitem{Yalim2011}
M.~Yalim, D.~{Vanden Abeele}, A.~Lani, T.~Quintino, H.~Deconinck, A finite volume implicit time integration method for solving the equations of ideal magnetohydrodynamics for the hyperbolic divergence cleaning approach, Journal of Computational Physics 230~(15) (2011) 6136--6154.
\newblock \href {https://doi.org/10.1016/j.jcp.2011.04.020} {\path{doi:10.1016/j.jcp.2011.04.020}}.

\bibitem{BALSARA1999}
D.~S. Balsara, D.~S. Spicer, A staggered mesh algorithm using high order {G}odunov fluxes to ensure solenoidal magnetic fields in magnetohydrodynamic simulations, Journal of Computational Physics 149~(2) (1999) 270--292.
\newblock \href {https://doi.org/10.1006/jcph.1998.6153} {\path{doi:10.1006/jcph.1998.6153}}.

\bibitem{GARDINER2005}
T.~A. Gardiner, J.~M. Stone, An unsplit {G}odunov method for ideal {MHD} via constrained transport, Journal of Computational Physics 205~(2) (2005) 509--539.
\newblock \href {https://doi.org/10.1016/j.jcp.2004.11.016} {\path{doi:10.1016/j.jcp.2004.11.016}}.

\bibitem{Einfeldt_1988}
B.~Einfeldt, On {G}odunov-type methods for gas dynamics, {SIAM} Journal on Numerical Analysis 25~(2) (1988) 294--318.
\newblock \href {https://doi.org/10.1137/0725021} {\path{doi:10.1137/0725021}}.

\bibitem{JANHUNEN2000}
P.~Janhunen, A positive conservative method for magnetohydrodynamics based on {HLL} and {Roe} methods, Journal of Computational Physics 160~(2) (2000) 649--661.
\newblock \href {https://doi.org/10.1006/jcph.2000.6479} {\path{doi:10.1006/jcph.2000.6479}}.

\bibitem{HARTEN1983}
A.~Harten, High resolution schemes for hyperbolic conservation laws, Journal of Computational Physics 135~(2) (1997) 260--278.
\newblock \href {https://doi.org/10.1006/jcph.1997.5713} {\path{doi:10.1006/jcph.1997.5713}}.

\bibitem{Harten1984}
A.~Harten, On a class of high resolution total-variation-stable finite-difference schemes, SIAM Journal on Numerical Analysis 21~(1) (1984) 1--23.
\newblock \href {https://doi.org/10.1137/0721001} {\path{doi:10.1137/0721001}}.

\bibitem{Balsara2000}
D.~S. Balsara, C.-W. Shu, Monotonicity preserving weighted essentially non-oscillatory schemes with increasingly high order of accuracy, Journal of Computational Physics 160~(2) (2000) 405--452.
\newblock \href {https://doi.org/10.1006/jcph.2000.6443} {\path{doi:10.1006/jcph.2000.6443}}.

\bibitem{Zhang2024}
F.~Zhang, Improving the quantification of overshooting shock-capturing oscillations, Progress in Computational Fluid Dynamics 24~(3) (2024) 135--142.
\newblock \href {https://doi.org/10.1504/PCFD.2024.138236} {\path{doi:10.1504/PCFD.2024.138236}}.

\bibitem{Ryu_1993}
D.~Ryu, J.~P. Ostriker, H.~Kang, R.~Cen, A cosmological hydrodynamic code based on the total variation diminishing scheme, The Astrophysical Journal 414 (1993) 1--19.
\newblock \href {https://doi.org/10.1086/173051} {\path{doi:10.1086/173051}}.

\bibitem{BALSARA1999a}
D.~S. Balsara, D.~Spicer, Maintaining pressure positivity in magnetohydrodynamic simulations, Journal of Computational Physics 148~(1) (1999) 133--148.
\newblock \href {https://doi.org/10.1006/jcph.1998.6108} {\path{doi:10.1006/jcph.1998.6108}}.

\bibitem{Popovas2025}
A.~Popovas, {DISPATCH} methods: an approximate, entropy-based {R}iemann solver for ideal magnetohydrodynamics, Astronomy \& Astrophysics 698 (2025) A69.
\newblock \href {https://doi.org/10.1051/0004-6361/202554028} {\path{doi:10.1051/0004-6361/202554028}}.

\bibitem{FUCHS2009}
F.~G. Fuchs, S.~Mishra, N.~H. Risebro, Splitting based finite volume schemes for ideal {MHD} equations, Journal of Computational Physics 228~(3) (2009) 641--660.
\newblock \href {https://doi.org/10.1016/j.jcp.2008.09.027} {\path{doi:10.1016/j.jcp.2008.09.027}}.

\bibitem{Qu2014}
F.~Qu, C.~Yan, J.~Yu, D.~Sun, A new flux splitting scheme for the {E}uler equations, Computers \& Fluids 102 (2014) 203--214.
\newblock \href {https://doi.org/10.1016/j.compfluid.2014.07.004} {\path{doi:10.1016/j.compfluid.2014.07.004}}.

\bibitem{Zhang2017}
F.~Zhang, J.~Liu, B.~Chen, W.~Zhong, A robust low-dissipation {AUSM}-family scheme for numerical shock stability on unstructured grids, International Journal for Numerical Methods in Fluids 84~(3) (2017) 135--151.
\newblock \href {https://doi.org/10.1002/fld.4341} {\path{doi:10.1002/fld.4341}}.

\bibitem{Xie2017}
W.~Xie, W.~Li, H.~Li, Z.~Tian, S.~Pan, On numerical instabilities of {G}odunov-type schemes for strong shocks, Journal of Computational Physics 350 (2017) 607--637.
\newblock \href {https://doi.org/10.1016/j.jcp.2017.08.063} {\path{doi:10.1016/j.jcp.2017.08.063}}.

\bibitem{Tadmor_2003}
E.~Tadmor, Entropy stability theory for difference approximations of nonlinear conservation laws and related time-dependent problems, Acta Numerica 12 (2003) 451–512.
\newblock \href {https://doi.org/10.1017/S0962492902000156} {\path{doi:10.1017/S0962492902000156}}.

\bibitem{Minoshima2020}
T.~Minoshima, K.~Kitamura, T.~Miyoshi, A multistate low-dissipation advection upstream splitting method for ideal magnetohydrodynamics, The Astrophysical Journal Supplement Series 248~(1) (2020) 12.
\newblock \href {https://doi.org/10.3847/1538-4365/ab8aee} {\path{doi:10.3847/1538-4365/ab8aee}}.

\bibitem{Minoshima2021}
T.~Minoshima, T.~Miyoshi, A low-dissipation {HLLD} approximate riemann solver for a very wide range of mach numbers, Journal of Computational Physics 446 (2021) 110639.
\newblock \href {https://doi.org/10.1016/j.jcp.2021.110639} {\path{doi:10.1016/j.jcp.2021.110639}}.

\bibitem{BRIO1988}
M.~Brio, C.~Wu, An upwind differencing scheme for the equations of ideal magnetohydrodynamics, Journal of Computational Physics 75~(2) (1988) 400--422.
\newblock \href {https://doi.org/10.1016/0021-9991(88)90120-9} {\path{doi:10.1016/0021-9991(88)90120-9}}.

\bibitem{Pinto_2017}
R.~F. Pinto, A.~P. Rouillard, A multiple flux-tube solar wind model, The Astrophysical Journal 838~(2) (2017) 89.
\newblock \href {https://doi.org/10.3847/1538-4357/aa6398} {\path{doi:10.3847/1538-4357/aa6398}}.

\bibitem{Sykora_2023}
J.~Mart\'inez-Sykora, B.~{De Pontieu}, V.~H. Hansteen, P.~Testa, Q.~M. Wargnier, M.~Szydlarski, The impact of multifluid effects in the solar chromosphere on the ponderomotive force under se and neq ionization conditions, The Astrophysical Journal 949~(2) (2023) 112.
\newblock \href {https://doi.org/10.3847/1538-4357/acc465} {\path{doi:10.3847/1538-4357/acc465}}.

\bibitem{Xu_2024}
K.~Xu, Z.~Gao, Z.~Qian, C.-H. Lee, Exact ideal magnetohydrodynamic {R}iemann solutions considering the strength of intermediate shocks, Physics of Fluids 36~(1) (2024) 016146.
\newblock \href {https://doi.org/10.1063/5.0185483} {\path{doi:10.1063/5.0185483}}.

\bibitem{Zhao2019}
G.~Zhao, M.~Sun, A.~Memmolo, S.~Pirozzoli, A general framework for the evaluation of shock-capturing schemes, Journal of Computational Physics 376 (2019) 924--936.
\newblock \href {https://doi.org/10.1016/j.jcp.2018.10.013} {\path{doi:10.1016/j.jcp.2018.10.013}}.

\bibitem{Lani2005}
A.~Lani, T.~Quintino, D.~Kimpe, H.~Deconinck, S.~Vandewalle, S.~Poedts, The {COOLFluiD} framework: Design solutions for high performance object oriented scientific computing software, in: V.~S. Sunderam, G.~D. van Albada, P.~M.~A. Sloot, J.~J. Dongarra (Eds.), Computational Science -- ICCS 2005, Springer Berlin Heidelberg, Berlin, Heidelberg, 2005, pp. 279--286.

\bibitem{Lani2013}
A.~Lani, N.~Villedie, K.~Bensassi, L.~Koloszar, M.~Vymazal, S.~M. Yalim, M.~Panesi, Coolfluid: an open computational platform for multi-physics simulation and research, in: 21st AIAA Computational Fluid Dynamics Conference, no. 2013-2589, AIAA, San Diego, CA, 2013.
\newblock \href {https://doi.org/10.2514/6.2013-2589} {\path{doi:10.2514/6.2013-2589}}.

\bibitem{Lani2014}
A.~Lani, M.~S. Yalim, S.~Poedts, A {GPU}-enabled finite volume solver for global magnetospheric simulations on unstructured grids, Computer Physics Communications 185~(10) (2014) 2538 -- 2557.
\newblock \href {https://doi.org/10.1016/j.cpc.2014.06.001} {\path{doi:10.1016/j.cpc.2014.06.001}}.

\bibitem{Saad1986}
Y.~Saad, M.~H. Schultz, {GMRES}: A generalized minimal residual algorithm for solving nonsymmetric linear systems, SIAM Journal on Scientific and Statistical Computing 7~(3) (1986) 856--869.
\newblock \href {https://doi.org/10.1137/0907058} {\path{doi:10.1137/0907058}}.

\bibitem{petsc-web-page}
S.~Balay, S.~Abhyankar, M.~F. Adams, S.~Benson, J.~Brown, P.~Brune, K.~Buschelman, E.~M. Constantinescu, L.~Dalcin, A.~Dener, V.~Eijkhout, J.~Faibussowitsch, W.~D. Gropp, V.~Hapla, T.~Isaac, P.~Jolivet, D.~Karpeev, D.~Kaushik, M.~G. Knepley, F.~Kong, S.~Kruger, D.~A. May, L.~C. McInnes, R.~T. Mills, L.~Mitchell, T.~Munson, J.~E. Roman, K.~Rupp, P.~Sanan, J.~Sarich, B.~F. Smith, S.~Zampini, H.~Zhang, H.~Zhang, J.~Zhang, {PETS}c {W}eb page, \url{https://petsc.org/} (2023).

\bibitem{Widlund1987}
O.~Widlund, M.~Dryja, An additive variant of the Schwarz alternating method for the case of many subregions, Technical Report 339, Ultracomputer Note 131, Department of Computer Science, Courant Institute, 1987.

\end{thebibliography}
\end{document}